\documentclass[mnsc]{informs3_no_remark} 

\OneAndAHalfSpacedXI 

\usepackage{natbib}
 \bibpunct[, ]{(}{)}{,}{a}{}{,}%
 \def\bibfont{\small}%

\usepackage{subfigure}
\usepackage{mwe}
\usepackage{lscape,booktabs,longtable}
\usepackage{amsmath}
\usepackage[pdftex,colorlinks=true,urlcolor=blue,citecolor=blue,pdfstartview=FitH]{hyperref}
\usepackage{comment}
\usepackage{enumerate}
\usepackage[ruled,vlined]{algorithm2e}
\usepackage[graphicx]{realboxes}
\newcommand{\edit}{}

\newcommand{\E}[1]{{\mathrm{E}\left[#1\right]}}

\newcommand{\PP}[1]{\mathrm{P}\left( #1 \right)}
\newcommand{\Var}[1]{{\mathrm{Var}\left(#1\right)}}
\newcommand{\Cov}[1]{{\mathrm{Cov}[#1]}}
\newcommand{\Corr}[1]{{\mathrm{Corr}\left(#1\right)}}
\newcommand{\T}{{\mathsf{T}}}
\newcommand{\naught}{{\scalebox{0.5}{$0$}}}

\usepackage{enotez}
\let\footnote=\endnote

\TheoremsNumberedThrough     
\ECRepeatTheorems

\EquationsNumberedThrough    

\MANUSCRIPTNO{MS-OPM-21-04060.R2}

\begin{document}



\RUNTITLE{Modeling Services using Hawkes Processes}

\TITLE{
The Co-Production of Service: Modeling Services in Contact Centers Using Hawkes Processes
}
\ARTICLEAUTHORS{%
\AUTHOR{Andrew Daw, Antonio Castellanos, Galit B. Yom-Tov, Jamol Pender, Leor Gruendlinger}
\AFF{University of Southern California Marshall School of Business, University of Chicago Booth School of Business, Technion---Israel Institute of Technology,  Cornell University School of Operations Research and Information Engineering, LivePerson Inc.} 
} 


\ABSTRACT{
In customer support contact centers, every service interaction involves a messaging dialogue between a customer and an agent; together, they exchange information, solve problems, and collectively \emph{co-produce} the service. Because the service progression is shaped by the history of conversation so far, we propose a bivariate, marked Hawkes process cluster model of the customer-agent interaction. 
To evaluate our stochastic model of service, we apply it to an industry contact center dataset containing nearly 5 million messages. Through both a novel residual analysis comparison and several Monte Carlo goodness-of-fit tests, we show that the Hawkes cluster model indeed captures dynamics at the heart of the service and also surpasses classic models that do not incorporate the service history. 
Furthermore, in an entirely data-driven simulation, we demonstrate how this  \edit{history-dependent model} can be leveraged operationally \edit{to inform a prediction-based routing policy. We} show that widely-used and well-studied customer routing policies can be outperformed with simple modifications according to the Hawkes model. Through analysis of a stylized model proposed in the contact center literature, we prove that service heterogeneity can cause this underperformance and, moreover, that such heterogeneity will occur if service closures are not carefully managed. 
}


\maketitle


%


\section{Introduction}

Most operations research assumes that the service duration is simply some random variable. Only a few attempts have been made to partition the service duration into its finer elements, and much of that work is focused on decomposing the service into a defined sequence of tasks, hence viewing the service as a part of a larger activity network \citep[e.g.,][]{Mandelbaum1998,RAN2022}. This is a good approach when tasks are clearly distinguished, for example, like steps in a loan approval process. But when considering a service conducted through a conversation, the tasks or phases can be ambiguous and hard to define, even more so in real time. This is certainly the case in the domain we consider here: customer support contact centers, in which the service interaction takes place through the exchange of written messages. Thus, instead, in this paper we will take a different approach and build a stochastic process model for the service interaction. 
We focus only on this interaction, restricting our attention to the ways that the customer and agent combine efforts and build the service from beginning to end.

The potential value of more granular and dynamic models of service interactions has actually been discussed in the operations research community for quite some time. In a both retrospective and prospective survey of queueing theory at its centenary, J.F.C. Kingman assessed the pitfalls of only considering independent and identically distributed random variables when constructing a queueing system. In \citet{kingman2009first}, he reasons that ``it is likely that the service mechanism, which may be very complicated, involving perhaps a network of queues with many servers, can be plausibly modelled as a random process, probably a (homogeneous) Markov process if enough variables are specified.'' Long before that, Jack Byrd listed ``undefined service times'' in which the ``type of service varies widely from one server to another'' as one of five challenges to the value of queueing theory in his provocative appraisal of the field \citep{byrd1978value}. These concerns are precisely what our modeling philosophy seeks to address. Here, our proposed model of the service interaction is stochastically driven by the history of the service so far, \`a la the self-exciting Hawkes process \citep{hawkes1971spectra}, allowing the scope of the exchange to evolve with each contribution.

By recognizing that service production is collaboratively achieved by the customer and the agent, our modeling philosophy also draws its inspiration from the literature about service \textit{co-production}. The fundamental role a customer plays in the service system is also well known; early definitions of the service economy noted that ``productivity in many service industries is dependent in part on the knowledge, experience, and motivation of the consumer'' \citep{fuchs1968service}.
We suggest that the co-produced service process should be modeled as a two dimensional stochastic process that captures the coupled interaction between the customer and the agent. 
We also go beyond proposing this stochastic model of co-produced service; we test the concept on real contact center data, showing the insights and value for operational decision making. 

The empirical literature shows that service pace is affected not just by the tasks to be completed but also by operational and behavioral elements \citep{Delasay2017LoadEffect}. An agent's response time has been shown to be impacted by factors like overall system load \citep{Kc2009ImpactOperations}, concurrency (or amount of multitasking) \citep{Kc2013DoesDepartment}, and customer-expressed sentiment \citep{Altman2019EmotionalLoad}. 
Of particular interest here, recent papers have shown that customers and agents impact each other's behavior simultaneously. For example, \citet{Altman2019EmotionalLoad} showed that the customer's expressed sentiment  influences agent response times and vice versa; \citet{Ashtar2021Reciprocal} showed that the two sides' sentiments impact one another within a service conversation. 
This notion of mutual impact fits our perspective of a service process characterized by two distinct roles --- customer and agent --- that need to cooperate with one another to  successfully co-produce the service. Hence, our proposed service models incorporate the reciprocal effect that an agent and customer have on one another through the pace of their responses, showing for the first time that not only will an agent's response time impact the customer's response time, but that the customer has the same impact on the agent.

Incorporating behavior within  service models is important due to its impact on fundamental operational decisions, such as staffing  \citep[see, e.g.,][]{Dong2015Service, Wu2019}.
We claim that an accurate dynamic model for the service progression, which captures the behavioral aspects of the service co-production, may also have a profound impact on the design of real-time optimal control algorithms for service systems.  Specifically, we both suggest and demonstrate that our model can be used to predict agent workload in real time and thus can be used to more appropriately balance load between agents. A theoretical analysis of our prediction-based routing policy suggests an interesting observation: the failure of current, concurrency-based \edit{routing} methods is related to mismanaged closure of service conversations. 
Our results coincide with previous findings showing that predictive information can improve service system operations, as was shown for healthcare operations by \citet{Xu2016}. 
This also coincides with the idea that the customer-agent service interaction should be considered in operational models of service systems, as was previously suggested by service design works such as  \citet{Roels2014,bellos2019should,bellos2021service} from a strategic view of the service and the division of labor.

\subsection{Contributions and Organization}

After further review of the service context and the relevant literature in Section~\ref{litReview}, the contributions of this paper are organized as follows:
\begin{itemize}
\item In Section~\ref{sec:model}, we define a general Hawkes service model for the co-produced conversation (Definition~\ref{modelDef}), which is a bivariate, marked, cluster stochastic process. We discuss the behavioral inspirations and interpretations for the model in a minimal assumption setting, and we establish a condition for the existence, uniqueness, and stability of the stochastic process (Theorem~\ref{stabilityThm}). \edit{We then define and contrast two manners of service closure, natural and systematic, and this formalizes the duration of the service.}
\item In Section~\ref{sec:case_study}, we apply the Hawkes service model to a large data set from a contact center in industry. To do so, we define three specific model forms that are amenable to estimation and analysis and also successively highlight the three main types of dependence in this paper: on the history, on the service relationships, and on the service system. We assess the performance relative to two benchmark models from the literature (Section~\ref{sec:quality}), and develop a novel residual analysis technique for goodness-of-fit evaluation in the univariate cluster case (Section~\ref{sec:residual}).
\item In Section~\ref{sec:insights}, we conduct an entirely data-driven simulation of the full service system in order to demonstrate the potential impact of this model on operational decision making. We propose a simple model-inspired change to the widely used lightest load routing policy, and demonstrate through the simulation study that it can achieve over 5\% and 10\% reduction in wait-during-service and wait-for-service\edit{, respectively}.
\item In Section~\ref{sec:routingA}, we analytically explore the origins of the observations from Section~\ref{sec:insights}. Through a static planning problem from the literature stylized to include salient elements \edit{from} our \edit{service} model, we prove that while the lightest load policy may be optimal if services are homogeneous, service heterogeneity can cause that policy to fail (Proposition~\ref{secondSPPprop}). Furthermore, we prove that the manner of service closure is what creates this heterogeneity, and that systematic closure according to our Hawkes service model can guarantee the desired homogeneity (Lemma~\ref{closureVarLemma}).
\end{itemize}
Finally, in Section~\ref{sec:conclusion}, we conclude, discuss these findings, and consider future work. All proofs are given in the appendix, as are complete computational details and other auxiliaries.

\section{Motivation, Background, and Further Literature Review}\label{litReview}

\subsection{Contact Center Context: Challenges for Modeling and Managing}

Currently,  asynchronous communication channels that serve customers via chat or messaging applications are steadily replacing call (voice-based) channels as the preferred form of customer-business communication. Indeed, a survey conducted by a cloud-based communications provider found that 78\% of respondents preferred to text with a company rather than call them \citep{ringCentral_2012}. Contact centers have also been recognized as important platforms for reaching potential customers and promoting sales \citep{Tan2019,YomTov2018Invitation}. Recently, data from contact centers has enabled researchers to observe detailed information about the conversational dependencies between customers and agents \citep{Rafaeli2020book}. The amount of available information about service encounters in contact center data is much more detailed than what is available for call centers. For example, while call center data generally only captures when an interaction started and ended, for contact centers one can know as much as what was written, when, and by which party.

The contact center environment has some important, but not necessarily unique, features which have attracted the attention of operations researchers in recent years. For example, customers can abandon a contact center's queue silently \citep{Castel2019}, which is similar to the unobserved abandonments in ``ticket queue'' models \citep{xu2007service,jennings2016comparisons} and the left-without-being-seen phenomena in emergency departments \citep{Batt2015}. Another important  aspect of contact centers is that agents can serve more than one customer concurrently \citep{Tezcan2014RoutingCustomersb,vanLeeuwaarden2017Restricted,Long2019}. Again, such multitasking is also apparent in settings such as hospital operations \citep{Kc2013DoesDepartment,Goes2017WhenMultitasking}, court systems \citep{bray2015multitasking}, and social welfare agencies \citep{campello2017}. This concurrency results in one of the key distinguishing factors between text-based contact centers and call centers: serving multiple customers at once alters the pace of the service interaction, which then directly affects the service length. This is because the communication is \textit{asynchronous}, meaning that its messages need not occur in quick succession or simultaneously. By comparison to a synchronous conversation in an in-person service or in a call center, dialogues in contact centers may have prolonged periods of inactivity.  
The level of this asynchrony can vary with the communication platform. For example, a web-chat communication is typically only mildly asynchronous, as service durations take place on the order of 10 minutes \citep{Castel2019}, while an email communication may be highly asynchronous, as durations can span from hours to even weeks  \citep{Halpin2013}. By comparison, a fully synchronous service interaction in a call center typically only lasts a few minutes \citep{Gans2010}. 

\begin{figure}[htb]
    \centering
    \includegraphics[width=0.8\textwidth]{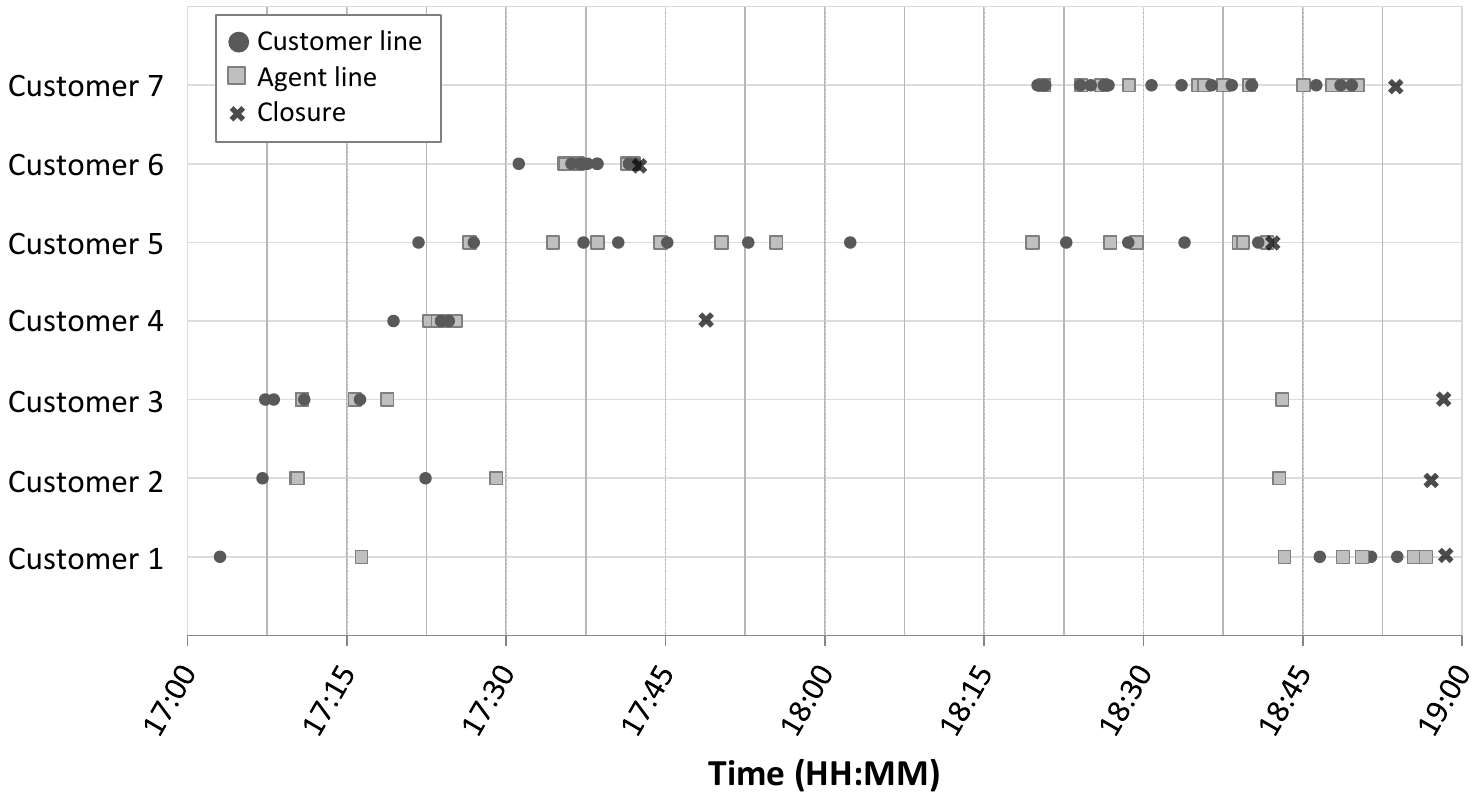}
    \caption{Sample  paths  of  seven  customers  served  by  the  same  agent  (May  4,  2017).  Circles (squares)  show when the customer (agent) sent a message\edit{, and $\times$ marks the conversation closure time}. }
    \label{fig:SamplePath7}
\end{figure}

In this paper, we use data from a moderately asynchronous setting:  message-based contact center service conversations of a telecommunications company. 
In this data, the service durations are on the order of minutes to hours; see Section \ref{sec:Data} for detailed summary statistics. It is important that we note that this long conversation duration need not mean that the agent is actively serving the customer the entire time. Indeed, in analyzing conversation data,  we notice long periods of times in which the conversation is inactive or ``on pause." Figure \ref{fig:SamplePath7} shows sample paths of seven conversations held concurrently by one agent. One can clearly see that, for some reason, the conversation with Customer 1 is inactive between 17:17 and 18:43. 
This phenomenon is not isolated to this example: If we discretize all conversations in our data into 5 minute intervals, we find that 57.5\% of those intervals are inactive, meaning they contain no messages at all.

\begin{figure}[htb]
\centering
\includegraphics[width=0.6\textwidth]{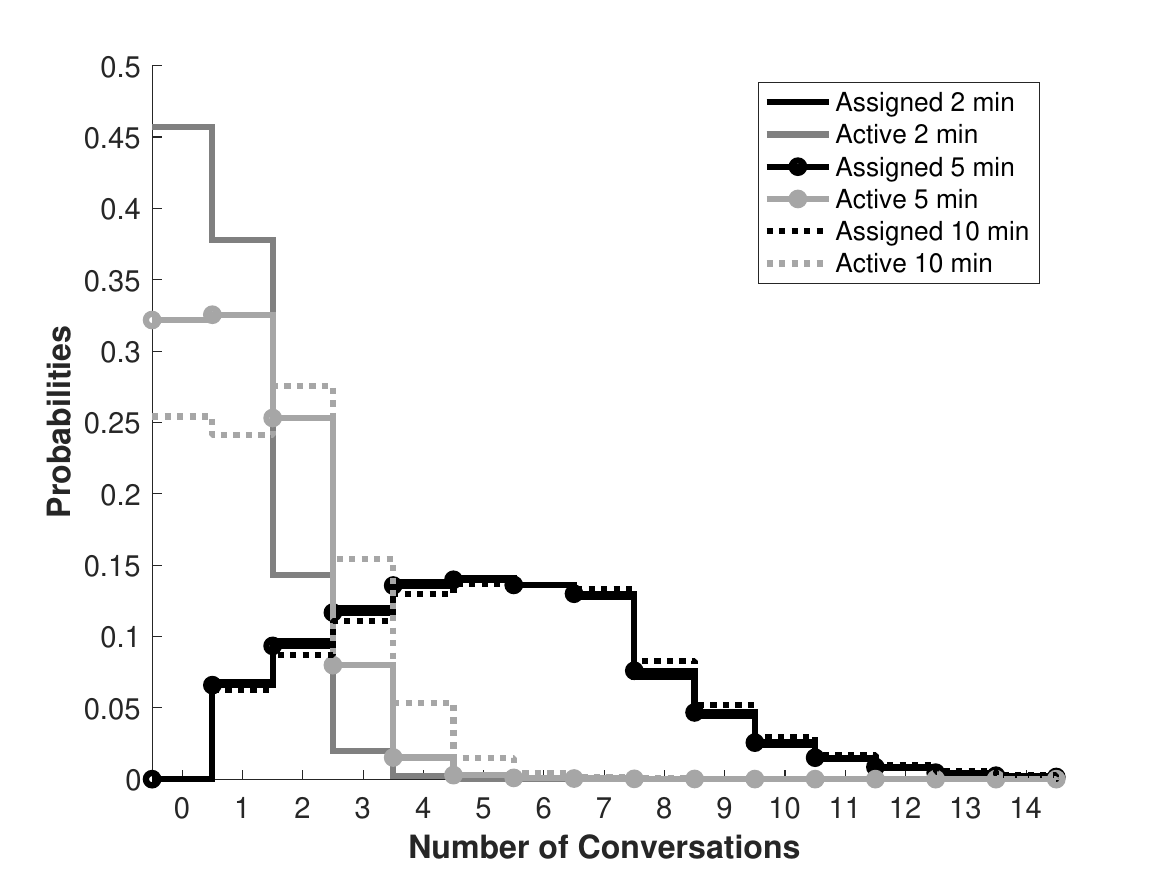}
\caption{Comparing the distributions of the number of active conversations and assigned conversations per agent. The probability indicates the proportion of intervals with observed activity. May 22--31, 2017.}\label{fig:HistogramAssigned}
\end{figure}

In this way, asynchronous communication offers benefits to both the customer and the agent. On the customer side, for example, this structure allows the \edit{guest} to take breaks during the conversation if they so desire. At any point, \edit{the customer} can choose to temporarily leave the conversation, perhaps to gather information related to the service interaction or simply to pursue other activities. On the agent side, these gaps offer the representative the flexibility to assist several customers in parallel. Rather than idly waiting for one particular customer's response, an agent can instead make use of the spare time and assist another person.
While this flexibility is beneficial individually, it actually creates challenges for system-level decisions. For example, these behaviors create a mismatch between the number of conversations assigned to an agent --- their concurrency --- and the number of those conversations that presently require work. This can be seen in Figure \ref{fig:HistogramAssigned}. 
Here, we discretize the agent's shifts into intervals \edit{of lengths 2, 5, and 10 minutes},\footnote{\edit{We use three interval lengths as a robustness check. Too long intervals may bundle conversations as concurrent when they were not actually served in parallel, while too short intervals may miss conversations that are inactive during that time. 
}} 
and, in each interval, we compare the number of conversations that are assigned to an agent and the number of those conversations that are active, meaning that they contain messages in that interval. As Figure \ref{fig:HistogramAssigned} shows, there is a large discrepancy between activity and assignment, and even between the measurements of activity \edit{over different interval lengths.}
These discrepancies can be problematic for operational decisions. Current contact center routing policies  \citep[e.g.,][]{Tezcan2014RoutingCustomersb,Long2019} assume that all assigned customers are equally active at any given point in time. One of this paper's goals is to \edit{develop a methodology to predict, in real time, the true level of upcoming conversation activity that an agent will handle, and to demonstrate the impact of such workload-measuring  methodology on routing policies and service levels.} 

\subsection{Hawkes Processes and Service Dynamics}
Taking a closer look at the dynamics depicted in Figure \ref{fig:SamplePath7}, we can observe that conversations progress in bursts of activity. Once one party sends a message, there is an increased chance that the other party will send a message too, leading to bursts of messages and a slight over-dispersion in the message arrival process  
($CV=1.07$). This could be related to the classic psychological concept of \textit{foot-in-the-door} \citep{Freedman1966}, since when one party succeeds in engaging their partner in the co-production of service, the service becomes more likely to continue. 
We therefore propose that the service model should depend on the conversation history, i.e., the sample path. This is different from classical  service models that have traditionally been based on Jackson networks \citep[e.g.,][]{Mandelbaum1998}.    

These bursty dynamics of contact center conversations are reminiscent of physical processes like earthquakes, where a sudden seismological event increases the probability of subsequent aftershocks. 
This type of stochastic behavior was classically modeled using a Hawkes process (HP), such as was done by \cite{Ogata1988}. Specifically, that paper showed that an earthquake can be viewed as a \textit{self-exciting point process} in which each arrival ``excites" the arrival rate, meaning it increases the probability of another arrival occurring soon afterwards. This stochastic intensity point process was originally defined by \citet{hawkes1971spectra}, and it has been used to model contagion and virality in a wide variety of applications, such as financial markets \citep[e.g.,][who adapt the process to daily stock market index data]{Embrechts2011}, social media \citep[e.g.,][who model retweet cascades]{Rizoiu2017}, 
 public health \citep[e.g.,][who connect HPs to epidemic models]{rizoiu2018sir, daw2018queue},
and queueing systems \citep[e.g.,][who study queues with HP arrivals]{gao2018functional,daw2017queues,koops2017infinite,chen2020perfect}.

There is also a rich stream of research that uses Hawkes processes to model various forms of communication \citep[e.g.,][]{malmgren2008poissonian,masuda2013self,Halpin2013,Rizoiu2017,rizoiu2018sir,Fox2016HawkesEmails,salehi2019learning}. We take this literature as both motivation and justification for the models we propose here. 
We build on these works in several important dimensions. First, while we share some similar goals of demonstrating the applicability of self- and mutually-exciting models for message-based communication, here we also study the effect of workload-level features (i.e., multitasking) and message-level features (i.e., word count and sentiment) on the conversational model, thus capturing both operational and behavioral elements that were not previously analyzed. There is also an important difference in scale. For example, the dataset in \citet{Halpin2013} is comprised of a little more than 400 email messages between the two individuals collected over a period of almost 18 months; here we have nearly five million messages across more than 300,000 separate conversations collected over one month in a contact center operating 24 hours \edit{each} day. The presence of many different conversations from many different customers and agents leads to another significant modeling departure, as we view each conversation as self-contained and driven solely by its own history (i.e., our model is one cluster), rather than also allowing Poisson-like baseline or exogenously generated messages, which are more natural in a longitudinal relationship that spans years rather than hours. This modeling change means that we develop new theory, employ different computational procedures, consider new behavioral dynamics, and generally seek out insights that were not explored in the literature before.

\section{Modeling the Service Co-Production through Stochastic Processes}
\label{sec:model}

As we have discussed in the introduction, the modeling scope in this paper is the customer-agent interaction within a single service exchange. Relative to traditional models of service that comprise the lion's share of the queueing theory literature, we seek a model for the service duration that is more granular than a single random variable, and more dynamic than phase- or stage-dependent models that rely upon exogenously defined service paths. For this, we turn to the self-exciting point process originally defined in~\citet{hawkes1971spectra}, and in particular to the cluster-based perspective originally identified by~\citet{hawkes1974cluster}. 
Definition~\ref{modelDef} establishes a bivariate, marked Hawkes cluster model of the service interaction. In the following sections, we will examine simplified forms of this model that are more amenable to analysis and estimation, but for now, let us describe the model with minimal assumptions.

\begin{definition}[Bivariate Hawkes Service Model]\label{modelDef}
Assuming that every service is initiated by a customer message at time $A_0^\mathsf{c} = 0$, let $N_t^\mathsf{c}$ and $N_t^\mathsf{a}$ be the respective point processes for the \emph{number of customer and agent messages} sent up to time $t \geq 0$ (excluding the initial), where this pair of point processes is driven by a corresponding pair of stochastic \emph{correspondence rate} intensities, defined with the customer correspondence rate given by 
\begin{align}
\lambda_{t}^{\mathsf{c}}
&=
\sum_{i=0}^{N_{t}^{\mathsf{c}}} f^\mathsf{c,c}(i) 
g^\mathsf{c,c}(t-A_i^\mathsf{c})
+
\sum_{j=1}^{N_{t}^{\mathsf{a}}} f^\mathsf{c,a}(j)
g^\mathsf{c,a}(t-A_j^\mathsf{a})
,
\label{SysBHPdefCustomer}
\end{align}
and with the agent correspondence rate given by
\begin{align}
\lambda_{t}^{\mathsf{a}}
&=
\sum_{i=0}^{N_{t}^{\mathsf{c}}}
f^\mathsf{a,c}(i)
g_{t}^\mathsf{a,c}(t-A_i^\mathsf{c})
+
\sum_{j=1}^{N_{t}^{\mathsf{a}}}
f^\mathsf{a,a}(j)
g_{t}^\mathsf{a,a}(t-A_j^\mathsf{a})
\label{SysBHPdefAgent}
,
\end{align}
where $A_\ell^x$ is the epoch for the $\ell$th message sent by party $x$ for all $\ell \in \mathbb{Z}_+$ and $x \in \{\mathsf{c}, \mathsf{a}\}$. That is,
\begin{align*}
\PP{N_{t+\delta}^x - N_t^x = n \mid \mathcal{F}_t} 
&
=
\begin{cases}
\lambda_t^x \delta + o(\delta) & n = 1,\\
1 - \lambda_t^x \delta + o(\delta) & n = 0,\\
o(\delta) & n > 1,
\end{cases}
\end{align*}
for each $x \in \{\mathsf{c,a}\}$, where $\mathcal{F}_t$ is the natural filtration of the bivariate stochastic process.
Here, $f^{x,y}: \mathbb{Z}_+ \to \mathbb{R}_+$ for each $x,y \in \{\mathsf{c},\mathsf{a}\}$ is the independent and identically distributed \emph{instantaneous impact} on the correspondence rate of party $x$ upon a new message sent by party $y$, and $g^{\mathsf{c},y}: \mathbb{R}_+ \to \mathbb{R}_+$ ($g_t^{\mathsf{a},y}: \mathbb{R}_+ \to \mathbb{R}_+$) for each $y \in \{\mathsf{c},\mathsf{a}\}$ is the customer's (agent's) correspondence rate \emph{decay function} adjusting the messages' impact on the correspondence rate as time passes. \hfill\Halmos
\end{definition}

Through three keywords in this definition, we can see how the Hawkes stochastic model parsimoniously captures the three foremost features of this service context. The model is \emph{bivariate} because it models the two parties within each service ($\mathsf{c}$: customer, $\mathsf{a}$: agent); it is \emph{marked} because it incorporates features that may vary randomly throughout the conversation; it is a \emph{cluster} because the present rate of activity in the service conversation is driven solely by the history of the conversation so far. The larger the correspondence rates become, the more message epochs there will be in the counting processes and the longer the service conversation will go on.

Let us elaborate on the terms within Definition~\ref{modelDef}, particularly the functions $f$ and $g$. First, \edit{although} the singleton superscripts may be unambiguous, we should clarify those written in pairs. The superscript $x,y$ is meant to be read in ``in-from'' form. For example, $\mathsf{c,a}$ captures the effects in the customer correspondence rate from messages sent by the agent. We will often refer to the four pairs as ``directions of the conversation,'' because the superscripts \edit{either} denote replies to the other party or \edit{denote} self-responses following up to one's own prior points. \edit{These directions capture the four \emph{relationships} ($(\mathsf{c,c})$, $(\mathsf{c,a})$, $(\mathsf{a,c})$ and $(\mathsf{a,a})$) between the two \emph{roles} ($\mathsf{c}$ and $\mathsf{a}$).} At times, we may also use these subscripts to refer to the correspondence rate for a specific direction of conversation, and the absence of superscripts will refer to all directions conglomerated. Now, let us look at the arguments indexing $f$ and $g$. The instantaneous impacts are enumerated by the indices of the message epoch sequences, meaning, for example, $f^\mathsf{c,a}(j)$ is the immediate effect in the customer correspondence rate of the $j$th message from the agent. By comparison to $f$, the decay function $g$ is dependent on a real-valued measure of time elapsed since a message, rather than a discrete message index.

The instantaneous impacts constitute the excitement effect within the Hawkes model; each new message increases the correspondence rates and thus drives the creation of future messages. Definition~\ref{modelDef} has provided that these impacts are independent and identically distributed (i.i.d.)~on each side of the conversation. This is the common assumption for marked Hawkes processes \citep[see, e.g.,][and references therein]{Embrechts2011}, but let us further specify the context of the randomness that we assume here. We will consider two particular features associated with each message that are known in the empirical literature to be drivers of behavior in conversational service: the message's sentiment and its number of words \citep[e.g.,][]{Goes2017WhenMultitasking,Altman2019EmotionalLoad}. For the word count, longer messages have been seen to prolong the conversation duration, where one may think of a message with more words as containing more information. On the other hand, for sentiment, messages with negative emotions have been observed to lead to more messages and longer conversations. While sentiment is of course a qualitative concept, we can leverage automated sentiment analysis engines to quantify the valence (positive or negative) and intensity of that emotion \citep[e.g.,][]{YomTov2018CustSent}. 

To incorporate these features in the Hawkes service model, we will let $S_\ell^x$ and $W_\ell^x$ be the sentiment score and word count for the $\ell$th message sent by party $x \in \{\mathsf{c,a}\}$. By construction in Definition~\ref{modelDef}, these sequences of random variables are identically distributed within each superscript, but \edit{the} distributions may vary across the two sides. Because Hawkes process intensities cannot be negative, we will assume that $S_\ell^x$ and $W_\ell^x$ are positive random variables with finite second moments. Without loss of generality, we will let each distribution have mean one. The prior literature then implies that a message with above-average word count will have $W_\ell^x > 1$, and a message with sentiment more negative than typical will have $S_\ell^x > 1$. With this in hand, we will suppose that the instantaneous impact $f^{x,y}(\ell)$ is a deterministic affine function of the underlying random variables $S_\ell^y$ and $W_\ell^y$, each with non-negative coefficients.

To balance the excitement generated by each instantaneous impact, the decay function regulates the influence of the history on the present. While we will place specific assumptions on the structure of these decay functions in later sections, $g$ will enforce that, generally speaking,  the longer it has been since a given message was sent, the less influence that message holds on the present correspondence rates. In addition to using the time elapsed since a message \edit{arrived}, in the agent's correspondence rate, we also allow $g$ to depend on the state of the broader service system, yielding the subscript of $t$. In particular, $g^{\mathsf{a},y}_t(\cdot)$ depends on the agent's concurrency at time $t$, meaning the number of customers assigned to the agent at time $t$. \edit{Let us} denote the concurrency at time $t$ as $K_t$. We will let $K_t \in \{1, \dots, \kappa\}$ for some maximal concurrency $\kappa \in \mathbb{Z}_+$, following the standard operational practice in industry to limit the maximum number of customers simultaneously assigned to one agent. Again, this stochastic modeling is behaviorally inspired. When the agent multitasks, waiting-within-service may increase due to the time spent serving other customers or from the cognitive load of managing several services at once \citep[e.g.,][]{Kc2013DoesDepartment,Tezcan2014RoutingCustomersb,bray2015multitasking}. Therefore, we expect that when $K_t$ is larger, the agent's individual response times should increase, decreasing the agent's correspondence rate. This observation was empirically validated in the contact center context by \citet{Altman2019EmotionalLoad}.

We will model these agent-side dynamics through the functional forms of $g_t^{\mathsf{a},y}(\cdot)$. Remarking briefly on nomenclature, we use $g_t^{\mathsf{a},y}(\cdot)$ rather than $g_{K_t}^{\mathsf{a},y}(\cdot)$ to condense notation, but we may use $g_{K_t=k}^{\mathsf{a},y}(\cdot)$ to specify the kernel function at a particular value of the agent's concurrency. Using this, let us assume that $\int_0^\infty g_{K_t=k}^{\mathsf{a},y}(t)\mathrm{d}t < \infty$ for every $1 \leq k \leq \kappa$ and for each $y \in \{\mathsf{c},\mathsf{a}\}$; we will similarly assume that the customer's decay functions satisfy $\int_0^\infty g^{\mathsf{c},y}(t)\mathrm{d}t < \infty$ for each $y \in \{\mathsf{c},\mathsf{a}\}$. Because $K_t$ takes on only finitely many values, one can think of the agent's decay functions as being one of finitely many deterministic functions, where the particular choice of function depends on the state of the agent's concurrency. Let us emphasize that we are defining the model with the assumption that the concurrency process $K_t$ is given. While $K_t$ is \edit{itself}, of course, a stochastically evolving process  in reality, our modeling scope is narrower than the total suite of an agent's assignment; recall that we are focusing specifically on the customer-agent interaction within a single service exchange. Hence, we will assume that an agent's conversations are mutually independent when given the history of $K_t$ up to time $t$, and we will not explicitly model $K_t$ outside of the assumptions that $K_t$ is a piecewise constant function of time, that it is deterministically known up to the current time throughout the conversation, and that it changes values only finitely many times.

\subsection{Establishing and Examining Stability of the Hawkes Conversational Model}\label{stabilitySubsec}

While Definition~\ref{modelDef} is sufficient to describe the process, we have not yet made claims about its existence or uniqueness, nor about its stability in the long-run. We do so now in this section through Theorem~\ref{stabilityThm}, leveraging both classic definitions for the Hawkes process \citep{hawkes1971spectra,hawkes1974cluster} and (slightly) more contemporary results \citep[e.g.,][]{massoulie1998stability,Embrechts2011}. However, first let us introduce two more pieces of notation that will help us provide a straightforward condition for stability.

Let $\bar \alpha^{x,y} = \E{f^{x,y}(i)} = \E{f^{x,y}(1)}$ be the expected instantaneous impact on party $x$'s correspondence rate upon a new message from party $y$. Naturally, the larger $\bar \alpha^{x,y}$ is, the more immediate attention a message from $y$ will demand from party $x$, on average. Considered across different $x,y$ pairs, this can serve as a measure of the immediate responsiveness in each direction of the conversation. However, $\bar \alpha^{x,y}$ is only the mean \emph{immediate} effect, and we know that the decay function dissipates these effects over time, so let us also quantify that regulation.

\edit{Let us define the effective long-run decay rate \edit{of each message}, $\beta^{x,y}$, as}
\begin{align}
\beta^{\mathsf{c},y} = 1\slash\left(\int_0^\infty g^{\mathsf{c},y}(t) \mathrm{d}t\right)
, 
\quad
\text{ and }
\quad
\beta^{\mathsf{a},y} = 1\slash\left(\int_0^\infty g_{K_t=k}^{\mathsf{a},y}(t) \mathrm{d}t\right)
,
\label{betaDefEq}
\end{align}
for every $k \in \{1, \dots, \kappa\}$. Notice that, on the agent side, this is more than simply defining notation: we are assuming that $\beta^{\mathsf{a},y}$ is the same across all possible values of the concurrency. This is part of the structure we are defining for the model\edit{, and this need not be restrictive. For example, $g_{K_t=k}^{\mathsf{a},y}(t) = g(t/k)/k$ for any $g(\cdot)$ with $\int_0^\infty g(u) \mathrm{d}u = 1/\beta^{\mathsf{a},y}$ will satisfy~\eqref{betaDefEq}, and this will ensure that conversations will be prolonged if the concurrency level increases}. Intuitively, this assumption says that the agent decay functions have the same regulatory power across concurrency values, but the precise timing and dynamics may still change with $K_t$.

Compared to the \edit{increase brought by the} instantaneous impact $\bar \alpha^{x,y}$, 
$\beta^{x,y}$ is a measure of \edit{continuous} regulation\edit{, or decrease in excitement over time,} in the $x,y$ direction of the conversation. The larger the $\beta^{x,y}$, the more quickly that messages from $y$ will lose their impact on the correspondence rate of party $x$. Hence, we can already see that $\bar \alpha^{x,y}$ and $\beta^{x,y}$ must constitute checks and balances within the evolution of the conversation. We can also notice that each quantity must be measured relative to time: The correspondence rates are themselves weighted sums of $f^{x,y}(i)$ terms, so $\bar \alpha^{x,y}$ is itself a rate. Similarly, we have said that $\beta^{x,y}$ measures how a message's impact fades through time, and so $\beta^{x,y}$ is a rate as well.

This recognition suggests considering the two quantities relative to one another. In fact, doing so will help us recognize that together these measure activity in each direction of the conversation. Let us define $\bar \alpha^{x,y} \slash \beta^{x,y}$ as the \textit{responsiveness ratio} in the $x,y$ direction of the conversation. Indeed, from the \citet{hawkes1974cluster} cluster-based definition of the Hawkes process, the $x,y$ responsiveness ratio is the mean number of messages that $x$ writes in response to a given message from $y$. The larger $\bar \alpha^{x,y} \slash \beta^{x,y}$ is, the stronger the influence of the $x,y$ conversation-direction will be on the overall service exchange. That is, larger responsiveness ratios imply greater likelihood of continued reply over the long run. By comparison to the rates $\bar \alpha^{x,y}$ or $\beta^{x,y}$ individually, the ratio $\bar \alpha^{x,y} \slash \beta^{x,y}$ is independent of time. In that lens, our assumption that $\beta^{\mathsf{a},y}$ is the same for each possible value of the concurrency implies that, while the agent's assignment load may change the pace at which they reply, it will not change their overall responsiveness --- it simply slows the responses down, or speeds them up.

In Theorem~\ref{stabilityThm}, we introduce a closed form condition for the stability of Definition~\ref{modelDef}'s Hawkes service model, implying that the correspondence rates in every conversation will eventually converge to 0 and that the total number of messages in any conversation will be finite. Moreover, this stability condition provides that Definition~\ref{modelDef} establishes a stochastic model that both exists and is unique; please see \citet{Embrechts2011} and references therein for precise definitions of existence, uniqueness, and stability for self-exciting stochastic models. \edit{Here}, the focus of our discussion around Theorem~\ref{stabilityThm} will be on what insights and interpretation the stability condition holds within the co-produced service operations context.

\begin{theorem}\label{stabilityThm}
If the responsiveness ratios satisfy
\begin{align}
\frac{\bar \alpha^{\mathsf{c,a}} }{\beta^{\mathsf{c,a}}}
\frac{\bar \alpha^{\mathsf{a,c}}}{\beta^{\mathsf{a,c}}}
<
\left(1 - \frac{\bar \alpha^{\mathsf{c,c}}}{\beta^{\mathsf{c,c}}}\right)
\left(1 - \frac{\bar \alpha^{\mathsf{a,a}}}{\beta^{\mathsf{a,a}}}\right)
,
\label{SysBHPStabilityEq}
\end{align}
\edit{with $\bar\alpha^\mathsf{c,c} < \beta^\mathsf{c,c}$ and $\bar\alpha^\mathsf{a,a} < \beta^\mathsf{a,a}$}, then the Hawkes service model exists, is unique, and is stable, and, moreover, 
$$
\lim_{t \to \infty} \lambda_t^\mathsf{c} = 0 
,
\qquad
\lim_{t \to \infty} \lambda_t^\mathsf{a} = 0 
,
\qquad
\lim_{t \to \infty} N_t^\mathsf{c} < \infty
,
\qquad
\text{and}
\qquad
\lim_{t \to \infty} N_t^\mathsf{a} < \infty
$$
almost surely.
\end{theorem}

Theorem~\ref{stabilityThm}, like the aforementioned results in the literature, is derived from a general Hawkes process stability condition based on the spectral radius of a matrix of excitation kernels \citep{massoulie1998stability}. The simple form in Equation~\eqref{SysBHPStabilityEq} is achieved for the deterministic marks case through manipulation of this matrix, and then extended to the random marks case through Doob's martingale convergence theorem; the full details of the proof are given in the appendix.

\edit{This stability condition provides more context for the assumptions around $\beta^{\mathsf{a},y}$ in Equation~\eqref{betaDefEq}. That is, by having $\int_0^\infty g^{\mathsf{a},y}_{K_t=k}(t) \mathrm{d}t$ be consistent across all values of $k$, Theorem~\ref{stabilityThm} reveals that we are supposing that the agent's concurrency level should not change whether or not the conversation will eventually terminate. Instead, our modeling assumption is that the concurrency changes only the {speed} at which the message excitement is regulated, and not the overall power of that regulation. By consequence, given that the stability condition is satisfied, the concurrency may alter the duration of the service, but it won't imply that the conversation could be infinitely long.}

Through Theorem~\ref{stabilityThm}, we can look into the ``space'' within a conversation and observe behavioral dynamics within co-produced service.
Let us start by inspecting the inequality in  Equation~\eqref{SysBHPStabilityEq}. The left-hand side of the inequality contains the \emph{co-production responsiveness ratios}, meaning those directions of the conversation in which one party responds to the other: agent-to-customer and customer-to-agent. On the other hand, the right-hand side is the product of the complements of the \emph{self-production responsiveness ratios}. Then, the stability condition inequality shows how these directions of the conversation must relate to one another. For example, if the left-hand side (the product of co-production responsiveness ratios) is close to 1, the self-production terms must both be close to 0; this means that the dominant service structure is in co-production, and the customer and the agent collaborate significantly. Conversely, if the left-hand side is close to 0, then at least one side-crossing relationship (i.e., agent-to-customer or customer-to-agent) is not very strong, and the service does not heavily depend on the two parties' interaction. In this case, both self-production terms may be close to 1, reflecting a service in which the two parties are loosely coupled --- each side can largely work independently and complete tasks on their own. 

Theorem~\ref{stabilityThm} also shows us limitations on how work can be structured and allocated between the customer and agent. \edit{First}, we can see that the self-production ratios must both be less than 1. Recalling the \citet{hawkes1974cluster} decomposition, this means that both parties must average less than one direct follow-up for each of their own messages. On the other hand, we can also see that it is possible for one of the co-production ratios to exceed 1 while the model remains stable. So, it may be possible in some services that, say, the agent is highly responsive to the customer, and so $\bar \alpha^{\mathsf{a,c}} \slash \beta^\mathsf{a,c} > 1$. For this to be a feasible design of service, Equation~\eqref{SysBHPStabilityEq} shows that the other directions of the exchange must be relatively inactive; suggesting that such arrangements may be best situated for high quality services in which the agent is expected to shoulder most of the load.

Finally, let us emphasize that all true contact center conversations should be stable, meaning they almost surely will contain only finitely many messages, like what Theorem~\ref{stabilityThm} implies. By comparison to a Hawkes \emph{arrival} process, this service model contains no baseline arrival rate; this is why we say it is a Hawkes cluster rather than a full Hawkes point process. That is, the stochastic process is initialized by the opening customer message at time 0, and this customer arrival is the only service activity that is exogenously driven. 
By comparison to Theorem~\ref{stabilityThm}, an arrival process stability condition would typically imply the existence of some (non-trivial) stationary distribution. Instead, here we have an almost surely finite sequence of points on the half-line, rather than an almost surely infinite sequence. Beyond simply matching our problem context, this distinction will be useful \edit{for fitting the model to data in Section \ref{sec:case_study}}.

\subsection{\edit{Formalizing the End of Service: Comparing Natural and Systematic Closure}}\label{closureDefSec}

\edit{Now, given Theorem~\ref{stabilityThm}'s implication that the service does end, let us consider how exactly it does. We will suppose that the conversation concludes through one of two styles: either \emph{natural} or \emph{systematic closure}.} 

\edit{By natural closure, we mean that the conversation endures its true duration up to the last written message, meaning $\hat \tau = \inf\{t\geq0 \mid N_t^\mathsf{c} = \lim_{u\to\infty} N_u^\mathsf{c}, N_t^\mathsf{a} = \lim_{u\to\infty} N_u^\mathsf{a}\}$ marks the end of service. That is, the service duration is the time from the first message until the last. Naturally, this occurs in conversations in which the final correspondence unambiguously declares that the service has concluded. This connotes a certain level of observability in the interaction, which may correspond to the quality or importance of the service \citep[e.g.,][]{ascarza2018some}. On the other hand, by systematic closure, we refer to the scenarios in which no message ever obviously announces the end of service, and thus the interaction must be closed by some policy or managerial decision.
This may be more likely associated with less formal or lower stakes services \citep[e.g.,][]{ascarza2018some,Castel2019}. For systematically closed services, the duration is the time from the first message until the time of automatic closure.} 

\edit{To motivate each of these in practice, let us revisit Figure~\ref{fig:SamplePath7} from the introduction, and, in particular, let us pay close attention to the ambiguity within service closure. Some conversations are closed soon after the last message (i.e., Customers 1, 5, 6, and 7) either by the customer or the agent, while in the other conversations the agent seems less sure whether the service was indeed concluded. The latter exhibit long gaps between the last message and closure time  (i.e., Customers 2, 3, and 4). This suggests that the former group's conversations may have closed naturally, while the later group's conversations were not. In the presence of these disparities, agents commonly seek to verify closure with the customer  (e.g., ``will you need anything else today?").} 

\edit{Consider the timelines for Customers 1, 2, and 3. For over an hour (roughly 17:30 to 18:30), all three of these conversations are entirely inactive. While such a quiet stretch might lead one to believe that the conversations are over, we can see that the context clues aren't enough to conclude this. Just before the 18:45 mark, the agent sends a message to each of the three customers, presumably to verify that the service is in fact over. At this point, Customer 1 responds and that conversation revives, but we do not see any activity from Customer 2 or 3 again (until 15 minutes later, when these two conversations were closed). 
This suggests that the inactivity itself is not a perfect indicator for a conversation's conclusion. }

\edit{The Customers 1, 2, and 3 example shows us that a conversation may be dormant but not yet complete (like for Customer 1), but, at the same time, a concluded conversation may not ever announce itself as such (like for Customers 2 and 3). Although Figure~\ref{fig:SamplePath7} is seen through the lens of a single agent, this challenge belongs as much to the system as it does to any agent, if not more so. To address this ambiguity in practice, firms deploy systematic closure policies or algorithms that determine when the service should be marked as complete. In contact centers, a simple policy may instruct the system to close conversations after some specified length of inactivity.\footnote{For example, \citet{Castel2019} observes systematic closure after 2 minutes of inactivity in a fast-paced chat setting.}}

\edit{To define this mathematically, we will treat the policy that systematically closes the service as a stopping time. Given the filtration of the Hawkes cluster stochastic process up to the current time, let a systematic closure rule be a stopping time that ends the interaction once some observable condition is met. More specifically, the system should decide upon an almost surely finite stopping time as its closure policy, so that the service system can be stable. To differentiate from natural closure, we will denote a systematic closure stopping time as $\tau$.}

\edit{For intuition's sake, let us notice that many natural rules through which  firms close services systematically are, in fact, stopping times. For example, we have already mentioned the simple (and, as we understand from our industry partners, widely used) practice to automatically close a service after some amount of time passes without a new message sent.
This closure rule is the stopping time $\tau = \inf\{t \geq \delta \mid N_{t} - N_{t-\delta} = 0\}$ for some $\delta > 0$. Also popular in practice is closure only after $n$ sufficiently-spaced agent messages without a customer response, such as $\tau = \inf\{t \geq n\delta' \mid N_{t-(\ell-1)\delta'}^\mathsf{a} - N_{t-\ell\delta'}^\mathsf{a} = 1 \,\forall\, 1 \leq \ell \leq n, N_{t}^\mathsf{c} - N_{t-n\delta'}^\mathsf{c} = 0\}$ for some $n \in \mathbb{Z}_+$ and $\delta' > 0$. }

\edit{Immediately, we can notice that natural closure and systematic closure are mutually exclusive concepts. That is, by definition, $\hat \tau$ uses future information about the stochastic process and thus cannot be a stopping time adapted to the natural filtration. Conversely, notice that the simple inactivity-based policy, $\tau = \inf\{t \geq \delta \mid N_t - N_{t-\delta} = 0\}$, is such that $\tau \ne \hat \tau$ almost surely  for any $\delta > 0$. More generally, this style of reasoning reveals that a stopping time $\tau$ can only be equal to the natural duration $\hat \tau$ with some probability strictly less than one.}

\edit{As Figure~\ref{fig:SamplePath7} suggests, we would expect that in reality, some conversations are closed naturally and others are closed systematically. In this paper, we will consider both styles. In particular, when fitting the model to data, we will take the simplifying assumption that each conversation is closed naturally, because our focus is on estimating the parameters of the service interaction's dynamics, rather than on inferring the mechanism through which it closed. However, in our analysis of routing managerial decisions, we will instead assume that all services are closed systematically; in fact, we will see that systematic closure may be an under-recognized, key determinant for the success (or failure) of routing policies. Of course, we find the interplay of natural and systematic closure to be highly interesting, but we reserve this system-level analysis for future research.}

\section{Evaluating the Conversational Models on Contact Center Data}\label{sec:case_study}

To evaluate this stochastic process representation of the conversation progression, we will now apply it to true contact center data from industry. This application will be at the heart of the remainder of the paper. To facilitate estimation and analysis of the model, let us introduce three particular forms of the Hawkes service model. In Table~\ref{modelTable}, we specify the three model forms in terms of parameter structures for the instantaneous impact $f$ and delay function $g$. We refer to these three respective cases as the Univariate Hawkes Process (UHP), Bivariate Hawkes Process (BHP), and System-Bivariate Hawkes Process (SysBHP). 

\begin{table}[htbp]
 \centering
  \caption{Summary of the Estimated and Evaluated Forms of the Hawkes Service Model.}\label{modelTable}
  \begin{scriptsize}
  \begin{tabular}{lll} \toprule
    \textbf{Model Form} & Parameter Structure & Modeled Dependence  \\
  \midrule
   Univariate (UHP) & $f^{x,y}(\cdot) = \alpha\slash2$,  & History  \tabularnewline 
   & $g^{\mathsf{c},y}(t) = g_t^{\mathsf{a},y}(t) = e^{-\beta t}$ & \tabularnewline
   Bivariate (BHP) & $f^{x,y}(\cdot) = \alpha^{x,y}$,  & History and relationship   \tabularnewline
   & $g^{\mathsf{c},y}(t) = e^{-\beta^{\mathsf{c},y}t}$, and $g_t^{\mathsf{a},y}(t) = e^{-\beta^{\mathsf{a},y} t}$ & \tabularnewline
   System Bivariate (SysBHP)  & $f^{x,y}(i) = \alpha^{x,y}_\mathsf{1} S_i^y + \alpha^{x,y}_\mathsf{2} W_i^y$,  
   & History, relationship, message features,
   \tabularnewline
   & $g^{\mathsf{c},y}(t) = e^{-\beta^{\mathsf{c},y}t}$, and $g_t^{\mathsf{a},y}(t) = \frac{1}{K_t}e^{-{\beta^{\mathsf{a},y} t} \slash {K_t}}$
   &
    and system state (concurrency) 
   \tabularnewline
  \bottomrule
  \end{tabular}
  \end{scriptsize}
\end{table}

As the names suggest and Table~\ref{modelTable} describes, these forms successively encapsulate each other in their parameter structures and, by consequence, in the dependence behavior that they model. In each case, the decay function $g$ is exponential, which is perhaps the most common decay kernel in the Hawkes literature. In the simplest case, the UHP, all messages have the same instantaneous impact ($\alpha$\edit{, because both processes jump by $\alpha/2$ at each message}) and all of these impacts decay at the same fixed rate ($\beta$). Hence, we can simply use $\lambda_t$ and $N_t$ instead of the superscripted versions. In this form, the service model primarily captures the conversation's dependence on its history. The BHP then extends this dependence to include relationship dependence as well. Here, the parameters remain constant and deterministic, but both the instantaneous impacts and decay rates vary across the directions of the conversation.

Finally, the SysBHP is the most fully-featured model form.\footnote{We also considered a SysBHP form with a deterministic shift in each instantaneous impact function. However, in practice there may be issues with statistical
stability of the estimated parameters, as the empirical distribution of $S^\mathsf{a}_1$ is tightly concentrated around 1. Because we have found that the performance of the form with $\alpha_0 + \alpha_1 S + \alpha_2 W$ is essentially the same as $\alpha_1 S + \alpha_2 W$, we have opted for the latter for the sake of reproducibility.} Relative to the BHP, the SysBHP adds dependence on the sentiment and word count random variables within the immediate impacts, and the agent's decay function is adapted to depend on the agent's concurrency. In particular, the concurrency divides the decay rate and divides the exponential function overall. These divisions can be thought of as modulating the time scale of the agent's response process. Because the instantaneous impacts are multiplied by the decay functions in Definition~\ref{modelDef}, the SysBHP both softens the immediate effect of each message and slows each message's rate of decay. That is, the mean jumps in the SysBHP agent correspondence rate are $\bar \alpha^{\mathsf{a},y}\slash K_t$ and the correspondending decay rates are $\beta^{\mathsf{a},y} \slash K_t$. One can quickly verify that $\beta^{\mathsf{a},y} = 1\slash\left(\int_0^\infty g_{K_t=k}^{\mathsf{a},y}(t)\mathrm{d}t\right)$ for every $k$ as we have assumed, confirming that although the agent's concurrency hampers their pace, it does not alter the expected number of messages that they will send. 
This is similar in nature to the philosophy of processor-sharing queueing models \citep[e.g.,][]{Borst2005}, which have been used in many prior queueing models of contact centers \citep[e.g.,][]{Tezcan2014RoutingCustomersb}.

To evaluate the Hawkes service model relative to the classic queueing literature, we will compare the three model forms to two benchmark models: the sum-of-exponential static (SES) and the sum-of-exponentials dynamics (SED). As the names may suggest, these are absorbing Markov chain models where the times between messages are exponentially distributed. In the SES, these exponential distributions are identical, but they need not be in the SED. Similarly, the absorption (end-of-conversation) probability is the same after every message in the SES, while it may vary from one message to the next in the SED. For brevity's sake, precise definitions of the SES and SED models are reserved for the appendix. Instead, here let us emphasize that, relative to the Hawkes model in any form, the salient difference of these benchmarks is that they are memoryless and independent from the history of the service timestamps, message features, and system states.

\subsection{Contact Center Background, Summary Statistics, and Parameter Estimates}\label{sec:Data}

Before reviewing the parameter estimates, let us first provide summary-level statistics describing our data source.
From a communications company's contact center observed during the month of May 2017 (31 days), we have acquired data containing 337,224 service conversations and a total of 4,964,895 messages. 
This center operates 24 hours per day, 7 days per week. The average number of new conversations is 602.68  per hour ($\textit{standard deviation }[SD]=83.59$). 
The mean number of online agents is 134.69 ($SD=31.06$), and all can serve any customer (i.e., the company does not employ any skill-based routing schema). The mean agent concurrency is 4.79 customers per agent ($SD=2.49$), and  99.91\% of conversations are handled by the same agent for their full duration (only 291 of the 337,224 conversations are transferred).
The average conversation duration is 53.48 minutes ($SD=65.15$). 
The average time between customer  and agent messages respectively is 2.58 and 4.26 minutes ($SD=9.63, 16.38$, respectively). Each conversation contains an average of 14.72 messages ($SD=15.02$), out of which 27.9\% were written by the customer and 72.1\% by the agent. Before normalization to  unit mean, each customer message averages 13.14 words ($SD = 16.02$) and each agent message averages 23.0 words ($SD = 22.74$). 
Likewise, before normalization and mapping to the positive reals, the sentiment scores in the data range from 24 (extremely positive) to -14 (highly negative), with the customer having a relatively neutral mean of $0.102$ ($SD = 0.794$) and the agent having a truly neutral mean of $0.00$ ($SD = 0.015$).

\edit{In each model form, the estimation algorithm uses the assumption that the service closes naturally (see Appendix~\ref{app:Comp} for the computational details).} For a realistic assessment, we estimate the parameters on training data and evaluate using out-of-sample data. We split the data chronologically, because our modeling assumption that conversations are dependent through the concurrency (and conditionally independent given the concurrency) implies that concurrent conversations should be held together. We use the first 23 days (approx.~75\% of conversations) as training data for estimation and the last 8 days (approx.~25\%) for evaluation.\footnote{We have also performed in-sample tests, but the performance was similar and thus we omit it for brevity's sake.} We use 10-fold cross validation to verify robustness, and we use bootstrapping (400 samples of 10\% of the data set size) to estimate standard errors for each parameter estimate. Relative to the estimates, these standard errors are quite small.  The estimated parameters for the Hawkes model forms and the literature benchmark models are available in Table~\ref{tbl:ParametersTest}.

\begin{table}[htb]
 \centering
  \caption{Estimation of Parameters for Each Model Form from Training Set (May, 1--23, 2017)}
  \begin{scriptsize}
  \begin{tabular}{lll} \toprule
  Model & Form & Parameters$^*$  \\
  \midrule
  Benchmark & SES & Exp.~mean $1/\mu=0.068$ (\emph{$<$0.001}), Absorption prob.~$p=0.065$ ($<$\emph{0.001}) \tabularnewline
  & SED & Exp.~means in App.~\ref{app:outof},  Abs.~prob.~$p_k={k+r \choose k-1}(1-\varrho)^{k-1}\varrho^r$ for $k \geq 1$, $r=1.3$ (\emph{0.011}), $\varrho=0.09$ (\emph{0.001}) 
 \\ \midrule
   Hawkes & UHP & $\alpha = 7.821$ (\textit{0.066}),\quad  $\beta = 8.395$ (\textit{0.070}) \tabularnewline
   & BHP & $ \alpha^{\mathsf{c,c}} = 0.885$ (\textit{0.026}),\quad  $\alpha^{\mathsf{c,a}}=14.630$ (\textit{0.171}),\quad  $\alpha^{\mathsf{a,c}}=3.678$ (\textit{0.067}),\quad  $\alpha^{\mathsf{a,a}}=20.048$ (\textit{0.264}),
  \tabularnewline
  &  &
    $ \beta^{\mathsf{c,c}}=3.710$  (\textit{0.067}),\quad  $\beta^{\mathsf{c,a}}=38.356$  (\textit{0.454}),\quad $\beta^{\mathsf{a,c}}=4.136$  (\textit{0.057}),\quad  $\beta^{\mathsf{a,a}}=47.559$   (\textit{0.782})  \tabularnewline
&
  {SysBHP} &  $ \alpha^{\mathsf{c,c}}_\mathsf{1} = 0.667$ (\textit{0.029}),\quad  $\alpha^{\mathsf{c,a}}_\mathsf{1}=14.053$ (\textit{0.192}),\quad  $\alpha^{\mathsf{a,c}}_\mathsf{1}=14.829$, (\textit{0.530}),\quad  $\alpha^{\mathsf{a,a}}_\mathsf{1}=113.650$ (\textit{1.904}),
 \tabularnewline
 &   &
   $\alpha^{\mathsf{c,c}}_\mathsf{2}=0.176$ (\textit{0.017}),\quad  $\alpha^{\mathsf{c,a}}_\mathsf{2}=0.030$ (\textit{0.002}),\quad\,  $\alpha^{\mathsf{a,c}}_\mathsf{2}=2.273$ (\textit{0.232}),\quad\,\,\,  $\alpha^{\mathsf{a,a}}_\mathsf{2}=0.069$ (\textit{0.009}), 
     \tabularnewline
 &  &
   $\beta^{\mathsf{c,c}}=3.640$ (\textit{0.017}),\quad  $\beta^{\mathsf{c,a}}=38.388$ (\textit{0.438}),\quad  $\beta^{\mathsf{a,c}}=20.374$ (\textit{0.434}),\quad\,  
   $\beta^{\mathsf{a,a}}=260.100$ (\textit{5.543})  
  \tabularnewline
  \bottomrule
  \multicolumn{3}{l}{$^*$The data is measured in hours; hence the parameters are hourly rates. (\textit{Standard errors are given in parenthesis.})}
  \end{tabular}
  \end{scriptsize}
 \label{tbl:ParametersTest}
\end{table}

\edit{These estimates immediately reveal some structures and behaviors in the conversational service. For example, the respective $\alpha$ parameters show how the model reacts to each customer and agent message. Both the BHP and SysBHP model forms feature strong instantaneous impact upon agent messages (large $\alpha^\mathsf{c,a}$ and $\alpha^\mathsf{a,a}$), and in the SysBHP we see that these jumps are more closely tied to sentiment than to word count. This implies that the instantaneous impacts do not vary too much, given the size of the word count standard deviation relative to that of the sentiment scores. However, this is less true when focusing on responses to customer messages, where we see that the size of $\alpha_2^\mathsf{c,c}$ and $\alpha_2^\mathsf{a,c}$ shows that both the customer and agent have a greater sensitivity to the number of words in customer messages.}

\edit{To further contextualize the parameters in Table~\ref{tbl:ParametersTest}, let us connect them to a measure of conversation activity and workload, the expected number of messages. In Proposition~\ref{totalMsgPropShort}, we provide an expression for the remaining number of messages given the current correspondence rates.}

\begin{proposition}\label{totalMsgPropShort}
\edit{Excluding the messages already sent in the observation period up to time $t_0 \geq 0$, the total expected number of messages until the natural end of the conversation in the SysBHP is 
\begin{align*}
\E{N_\infty - N_{t_0} \mid \boldsymbol{\lambda}_{t_0}}
&=
\frac{
\left(
1 + \frac{\bar \alpha^\mathsf{a,c}}{\beta^\mathsf{a,c}} - \frac{\bar\alpha^\mathsf{a,a}}{\beta^\mathsf{a,a}}
\right)
\left(
\frac{\lambda_{t_0}^{\mathsf{c,c}}}{\beta^{\mathsf{c,c}}} 
+
\frac{\lambda_{t_0}^{\mathsf{c,a}}}{\beta^{\mathsf{c,a}}} 
\right)
+
\left(
1
+
\frac{\bar\alpha^{\mathsf{c,a}}}{\beta^{\mathsf{c,a}}}
-
\frac{\bar \alpha^\mathsf{c,c}}{\beta^\mathsf{c,c}}
\right)
\left(
\frac{\bar \lambda_{t_0}^{\mathsf{a,a}}}{\beta^{\mathsf{a,a}}} 
+
\frac{\bar \lambda_{t_0}^{\mathsf{a,c}}}{\beta^{\mathsf{a,c}}} 
\right)
}{
\left(
1 
- 
\frac{\bar\alpha^\mathsf{c,c}}{\beta^\mathsf{c,c}}
\right)
\left(
1 - \frac{\bar\alpha^\mathsf{a,a}}{\beta^\mathsf{a,a}}
\right)
-
\frac{\bar\alpha^{\mathsf{c,a}}}{\beta^{\mathsf{c,a}}}
\frac{\bar\alpha^{\mathsf{a,c}}}{\beta^{\mathsf{a,c}}}
}
,
\end{align*}
where $\bar \lambda_{t_0}^{\mathsf{a},y} = \sum_{\ell: A_\ell^y \leq t_\naught}(\alpha^{\mathsf{a},y}_\mathsf{1} S_\ell^y + \alpha^{\mathsf{a},y}_\mathsf{2} W_\ell^y) e^{-\beta^{\mathsf{a},y} (t_\naught - A_\ell^y) \slash K_{t_\naught}}$ for $y \in \{\mathsf{c},\mathsf{a}\}$.}
\end{proposition}

\edit{Through total expectation over the initial sentiment and word count random variables, we can also obtain a simple expression for the mean number of messages in any given conversation only in terms of the $\bar \alpha$ and $\beta$ parameters.}

\begin{corollary}\label{totalMsgCor}
\edit{Including the initial customer message, the total expected number of messages in a conversation is} 
\begin{align*}
\edit{
\E{N_\infty + 1}
=
\frac{{
1
- 
\frac{\bar\alpha^\mathsf{a,a}}{\beta^\mathsf{a,a}}
+
\frac{\bar \alpha^{\mathsf{a,c}}}{\beta^{\mathsf{a,c}}} 
}
}{{
\left(
1 
- 
\frac{\bar\alpha^\mathsf{c,c}}{\beta^\mathsf{c,c}}
\right)
\left(
1 - \frac{\bar\alpha^\mathsf{a,a}}{\beta^\mathsf{a,a}}
\right)
-
\frac{\bar\alpha^{\mathsf{c,a}}}{\beta^{\mathsf{c,a}}}
\frac{\bar\alpha^{\mathsf{a,c}}}{\beta^{\mathsf{a,c}}}
}}
.}
\end{align*}
\end{corollary}

\edit{Corollary~\ref{totalMsgCor} gives us a way to reason about the division of labor in the expected number of tasks (meaning, messages) that arise in follow-up to a prior task. That is, recall from Definition~\ref{modelDef} that the $x,y$ responsiveness ratio $\frac{\bar \alpha^{x,y}}{\beta^{x,y}}$ is the mean number of messages party $x$ sends in response to a given message from party $y$.  Returning to the estimates in Table~\ref{tbl:ParametersTest}, we can observe some of the task dynamics in the data's context. In this service setting, we can see that, although we have already remarked that the agent messages have strong instantaneous impact, it is customer activity that drives the heart of the service over the long run. The largest responsiveness ratio is the agent responding to the customer, $\frac{\bar \alpha^\mathsf{a,c}}{\beta^\mathsf{a,c}} = 0.839$ ($\textit{standard error [SE]} = 0.008$), and a customer message begets more responses on average than an agent message, $\frac{\bar \alpha^\mathsf{c,c}}{\beta^\mathsf{c,c}} + \frac{\bar \alpha^\mathsf{a,c} }{ \beta^\mathsf{a,c}} = 1.071$ ($\textit{SE} = 0.009$) compared to $\frac{\bar \alpha^\mathsf{c,a}}{\beta^\mathsf{c,a}} + \frac{\bar \alpha^\mathsf{a,a} }{ \beta^\mathsf{a,a}} = 0.804$ ($\textit{SE} = 0.006$). Moreover, these estimates show us that the service could also be described as \emph{moderately co-productive}, as the co-production responsiveness ratios dominate the self-production ratios for each message type, i.e.~$\frac{\bar \alpha^\mathsf{a,c}}{\beta^\mathsf{a,c}} = 0.839 > \frac{\bar \alpha^\mathsf{a,a}}{\beta^\mathsf{a,a}} = 0.437 \,(\textit{SE} = 0.005)$ and $\frac{\bar \alpha^\mathsf{c,a}}{\beta^\mathsf{c,a}} = 0.367 \,(\textit{SE} = 0.003) > \frac{\bar \alpha^\mathsf{c,c}}{\beta^\mathsf{c,c}} = 0.232 \,(\textit{SE} = 0.004)$, but it is not the case that the two co-production ratios are the two largest overall.}

\edit{Given this parameter context, let us now evaluate the fit of the model.} We conduct two primary assessments of the models: a Monte Carlo comparison of two key performance distributions for all model forms (Section~\ref{sec:quality}) and a novel extension of residual analysis techniques for the univariate cluster case (Section~\ref{sec:residual}). Details of the estimation and Monte Carlo simulation methodologies are available in the appendix.

\subsection{Monte Carlo Goodness-of-Fit and Benchmark Performance Evaluation}\label{sec:quality}

We evaluate the accuracy on two main empirical distributions from the service exchange: the \textit{conversation duration} (meaning the time from first to last message) and the \textit{gap times} (meaning the time between successive messages). 
Because these distributions may be difficult to access analytically for all of the model forms, we compare the data to model distributions generated by simulating 100,000 synthetic conversations with the estimated parameters.

\begin{figure}[tbh]
\centering
\subfigure[Conversation Duration]{
\includegraphics[width=0.48\textwidth]{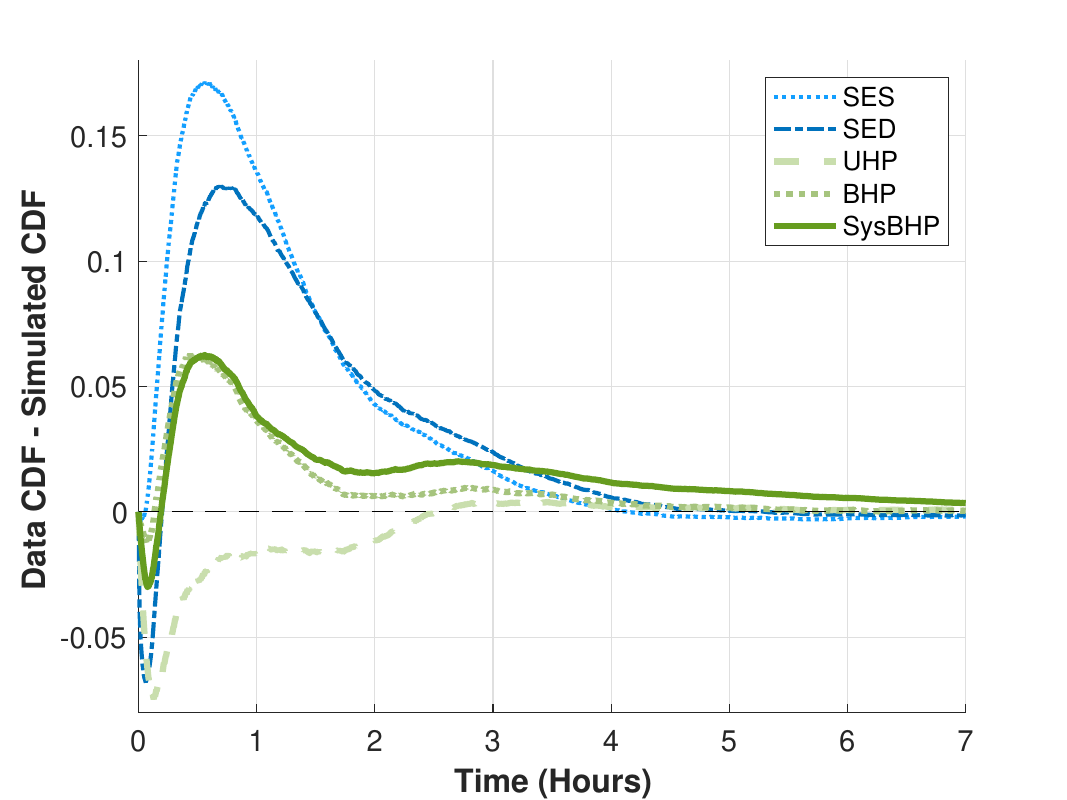} \label{fig:CDFLOSTest}
} 
\subfigure[Gap Time]{
\includegraphics[width=0.48\textwidth]{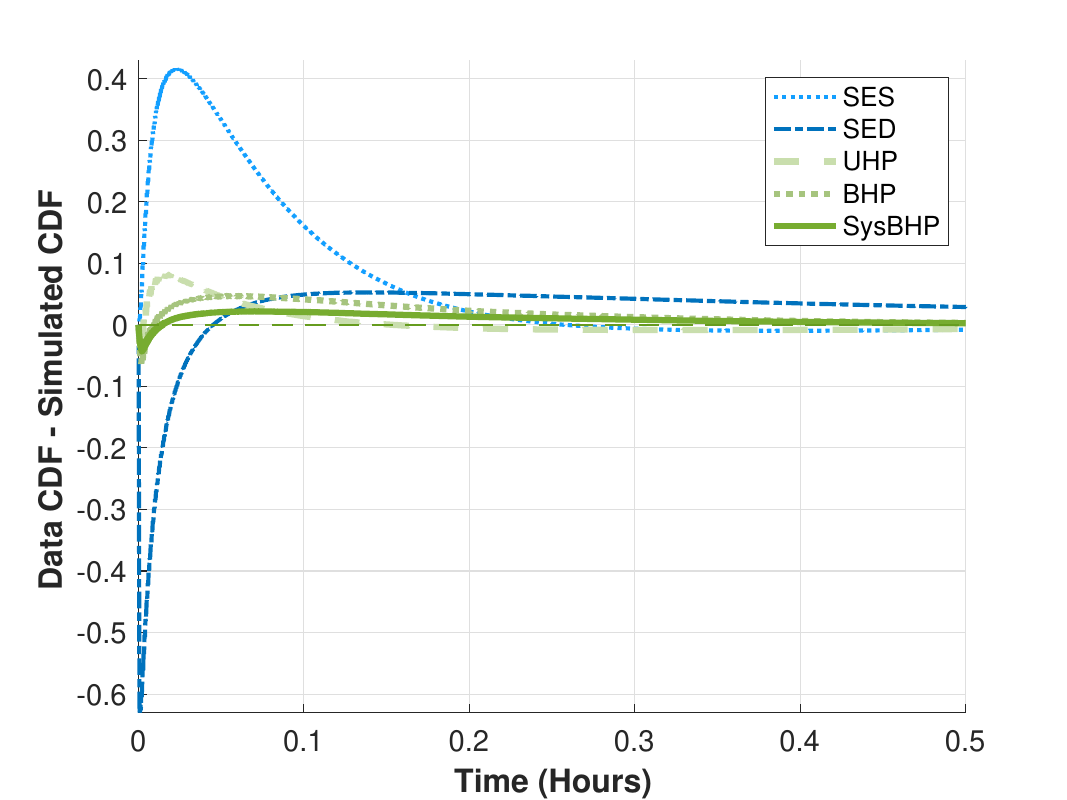} \label{fig:CDFgapsTest}
} 
\caption{Difference of empirical CDFs for the data and simulated conversation models. Out-of-sample test. }
\label{fig:CDFsTest}
\end{figure}

To provide a comprehensive perspective on the accuracy of the stochastic models,  let us offer both visual and metric comparisons. Figure \ref{fig:CDFsTest} plots the differences between the empirical cumulative distribution function (CDF) of the data ($\hat F_D$) and the empirical CDF of each simulated stochastic model ($\hat F_M$). Specifically, we consider the differences in CDFs of the conversation duration in Figure~\ref{fig:CDFLOSTest} and for the gap times in Figure~\ref{fig:CDFgapsTest}. The benchmarks are in thinner blue curves, and the Hawkes model forms are in wider green curves.
Then, Table~\ref{tbl:CompAll} contains the Kolmogorov-Smirnov ($\mathsf{KS} = \max_x |\hat F_D(x) - \hat F_M(x)|$) and 1-Wasserstein ($\mathsf{W1} = \int_0^\infty | \hat F_D(x) - \hat F_M(x) | \mathrm{d}x$, not to be confused with the word count $W_i$) distances for each of the Hawkes model forms. Table~\ref{tbl:CompAll} denotes the best $\mathsf{KS}$ and $\mathsf{W1}$ performance in bold font. 
For relative scales, it can be directly seen that the $\mathsf{KS}$ distance is no more than 1, and by the triangle inequality, $\mathsf{W1}$ is at most the sum of the means of the data and the simulation. 
Naturally, both these distances are intimately related to the curves in Figure~\ref{fig:CDFsTest}. For a given model, the $\mathsf{KS}$ distance will be the global maximum or minimum of the CDF difference, and $\mathsf{W1}$ will be the total area between the curve and the dashed line at 0.

\begin{table}[htbp]
 \centering
  \caption{Evaluation of Model Fit. Out-of-sample test.
  }
  \begin{scriptsize}
  \begin{tabular}{llcccccc} \toprule
   &&& \multicolumn{2}{c}{Duration}  & & \multicolumn{2}{c}{Gap }  \\ 
   \cmidrule{4-5} \cmidrule{7-8}
    {Model} & Form && $\mathsf{KS}$  & $\mathsf{W1}$ & & $\mathsf{KS}$  & $\mathsf{W1}$   \\
  \midrule
  Benchmark & SES && 0.171 & 0.254 &   & 0.417 & 0.041 \tabularnewline
   & SED && 0.130 & 0.228 &   & 0.633 & 0.046 \tabularnewline
  \midrule
   Hawkes & UHP && 0.074 & \textbf{0.061} &   & 0.096 & 0.031 \tabularnewline
    & {BHP} && \textbf{0.058} & 0.078 &   & {0.059} & 0.020 \tabularnewline
    & {SysBHP}  && 0.063 & 0.137 &   & \textbf{0.043} & \textbf{0.006}
   \tabularnewline
  \bottomrule
  \end{tabular}
  \end{scriptsize}
 \label{tbl:CompAll}
\end{table}

In reviewing Figure~\ref{fig:CDFsTest} and Table \ref{tbl:CompAll}, we see three leaps in performance. First, the plots and distances both show that moving from non-behavioral benchmark models, i.e., the static (SES) and time-varying exponential (SED) models, to the history-dependent UHP offers improvement in both the duration fit and the gap-time fit. 
Particularly in comparing the UHP and the benchmarks, this improvement is important in that it shows that the conversation's pace and sequencing are valuable service modeling components, and that merely tracking the number of messages  does not offer the same level of performance.\footnote{Let us point out a cautionary tale. In reviewing the duration fits, the benchmarks perhaps seem not so far off of the Hawkes counterparts, at least relative to the disparity in the gaps. However, this can be seen as an empirical example of the law of large numbers limits that self-exciting processes are known to satisfy, even though the gap times are dependent \citep[e.g.,][]{bacry2013some,fierro2015hawkes,gao2018limit,daw2018queue}. Because the conversational duration is a sum of the gap times, this mutual closeness of the conversation-duration curves not only should be expected, but also demonstrates the pitfalls of not inspecting and modeling with finer granularity.}

We can see a second improvement when moving from the UHP to the  BHP, particularly so for the gap times. 
Just as the SES and SED to UHP improvement shows the modeling value of the conversation's history, this improvement when moving to the BHP demonstrates the value of distinguishing the roles within the service co-production. We believe this is a particularly important takeaway for the goals of modeling and analyzing customer and agent behavior, as it shows (a) the different contribution that each role has in the conversation's pace and (b) the influence that the interaction between the two roles can have on the service exchange. 
Finally, our third leap comes when moving from the BHP to the SysBHP, as seen in the gap time distributions. Although the SysBHP has clear limitations in representing the conversation's duration from start to finish, it has by far the best performance in the gaps. 
This shows us that when using these Hawkes conversational models for decision making over short intervals of time, factoring in the agent's caseload with customer and agent behavior can improve model accuracy.

\subsection{Goodness-of-Fit through Residual Analysis for Univariate Hawkes Clusters}\label{sec:residual}

When applying point processes to data, we are typically granted \emph{residual analysis} techniques to assess goodness-of-fit; see \citet{Ogata1988} for an early application to Hawkes processes in seismology, Chapter 9 of \citet{laub2021elements} for a recent Hawkes-specific review, or, e.g.,~\citet{brown2005statistical,kim2014call} for application to non-stationary Poisson processes in operations management. The idea of this methodology is that, through a transformation according to the random time change theorem, the data would become a unit-rate Poisson process if it was truly a realization from the point process model. However, the random time change theorem critically relies upon an assumption that the point process is unceasing, almost surely yielding an infinite sequence of points on the positive reals. 
This assumption will not hold for these cluster-based models. Hence, the Poisson process comparison is not available for our setting. Instead, in the univariate setting, a similar transformation has recently been seen to yield a connection to a jointly uniform distribution \citep{daw2023conditional}. However, \citet{daw2023conditional} did not consider inference applications, so let us now describe how to extend this idea to evaluate model fit for the UHP.

Following the same transformation function as the classic random time change theorem, let us define the \emph{compensator points}, meaning the integral of the $\lambda_t$ evaluated at the UHP timestamps:
\begin{align}
\Lambda_\ell = \sum_{i=0}^{\ell-1}\left(1 - e^{-\beta(A_\ell - A_i)}\right) = \ell - \sum_{i=0}^{\ell-1} e^{-\beta(A_\ell - A_i)}
\label{compEq}
.
\end{align}
We will let $\Lambda_0 = A_0 = 0$ without loss of generality. Equation~\eqref{compEq} actually defines a normalized compensator, meaning the true integral of the correspondence rate divided by $\frac{\alpha}{\beta}$; this saves a scaling step in the residual analysis. We will also let $N = \lim_{t\to\infty} N_t$ be the number of messages in the conversation. By definition, the compensator function is increasing, and it is straightforward to verify that we have $\Lambda_{\ell-1} < \Lambda_\ell < \ell$ for every $1 \leq \ell \leq N-1$. In a point process where $\PP{N = \infty} = 1$, the difference between compensator points would be exponentially distributed. Instead, in this case where $\PP{N < \infty} = 1$, \citet{daw2023conditional} finds that the compensator points are jointly uniform on a particular convex polytope. In \edit{Theorem}~\ref{dialplotprop}, we show that these points can be further transformed to reveal i.i.d.~standard uniform random variables.

\begin{theorem}\label{dialplotprop}
For a UHP model with $N \in \mathbb{Z}_+$ total points, let $\Lambda_1 < \dots \Lambda_{N-1}$ be the compensator transforms of the epochs via Equation~\eqref{compEq}. Additionally, let $\mathcal{I}_\ell = \{i : \ell - 1 < \Lambda_i \leq \ell\}$ and let $\sigma : \mathcal{I}_\ell \to \mathcal{I}_\ell$ be a uniformly random permutation of the $\mathcal{I}_\ell$ indices. Then, 
\begin{align*}
\Lambda_{\sigma(i)} - \lfloor\Lambda_{\sigma(i)}\rfloor \stackrel{\mathsf{iid}}{\sim} \mathsf{Uni}(0,1)
,
\end{align*}
for every $i \in \mathcal{I}_\ell$ and every $1 \leq \ell \leq N-1$.
\end{theorem}

Because $\Lambda_0 = 0$ is excluded from \edit{Theorem}~\ref{dialplotprop}, the result is vacuous at $N=1$. By consequence of \edit{Theorem}~\ref{dialplotprop}, we can evaluate the fit of the UHP estimated on the contact center data and, moreover, we have the ability for a deeper level of analysis than the previous Monte Carlo comparisons. Empirical CDFs of the transformations according to Equation~\eqref{compEq} should be uniform on $(0,1)$ when shuffled according to \edit{Theorem}~\ref{dialplotprop}, so the comparison now is between data and theory, rather than data and simulation. In this way, we are extending the idea of residual analysis to this finite cluster setting.

\begin{figure}[tbh]
\centering
\subfigure[$N=5$]{
\includegraphics[width=0.31\textwidth]{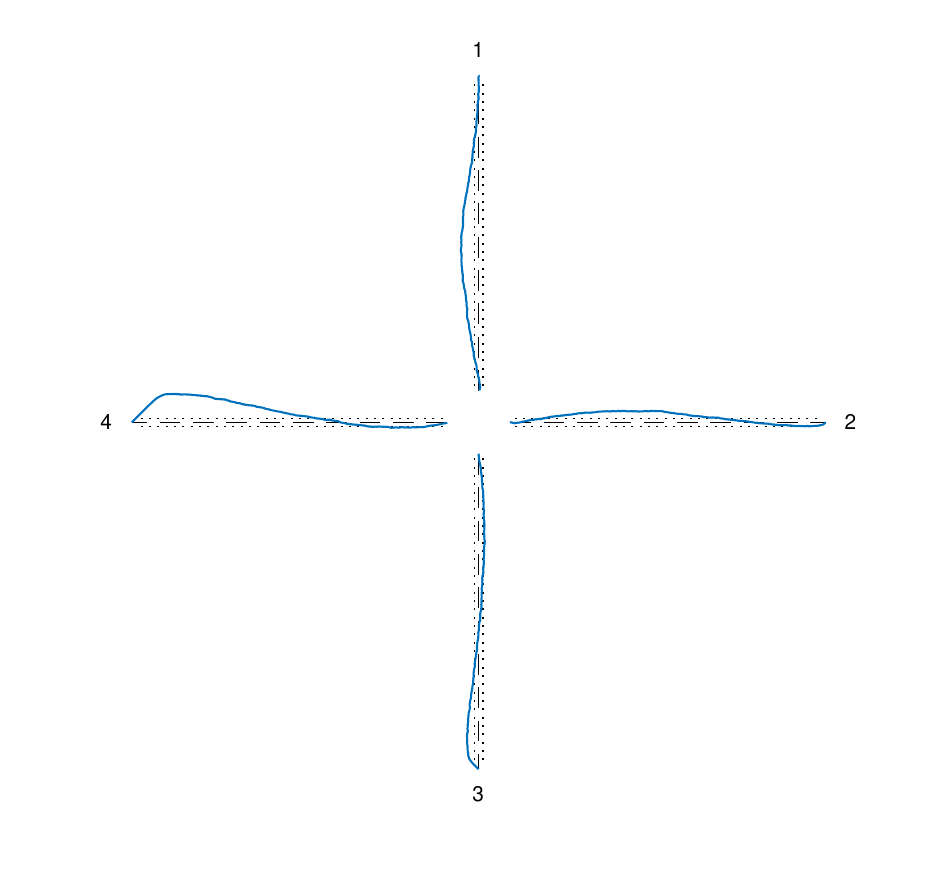} 
} 
\subfigure[$N=10$]{
\includegraphics[width=0.31\textwidth]{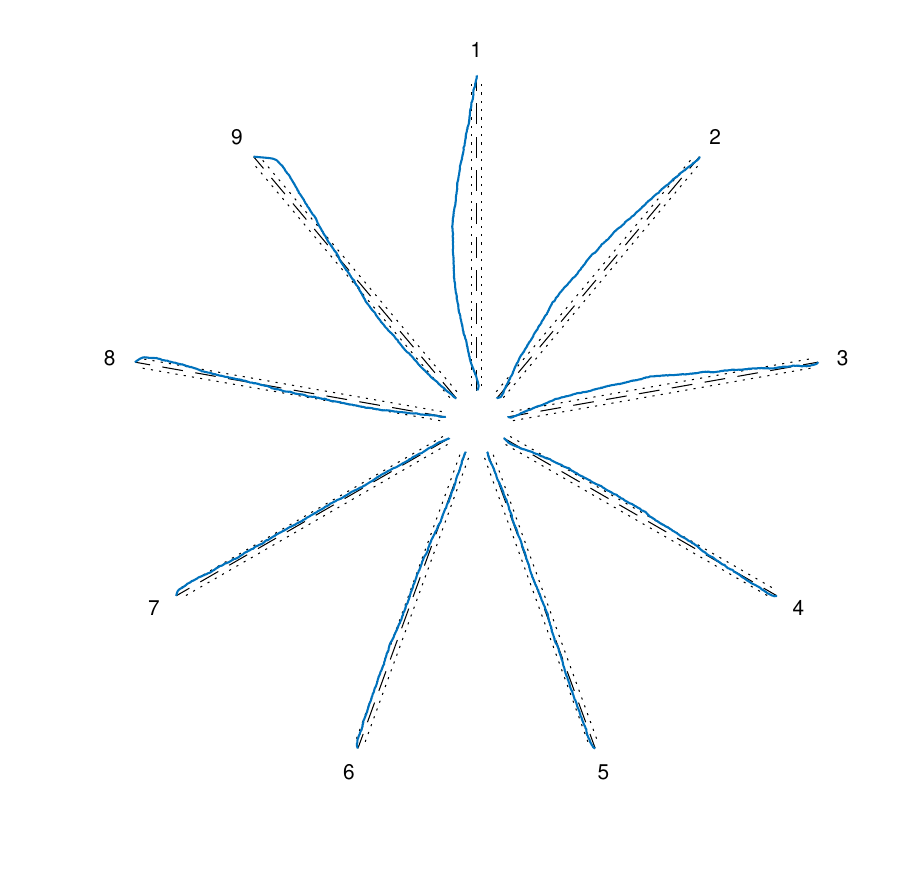} 
} 
\subfigure[$N=15$]{
\includegraphics[width=0.31\textwidth]{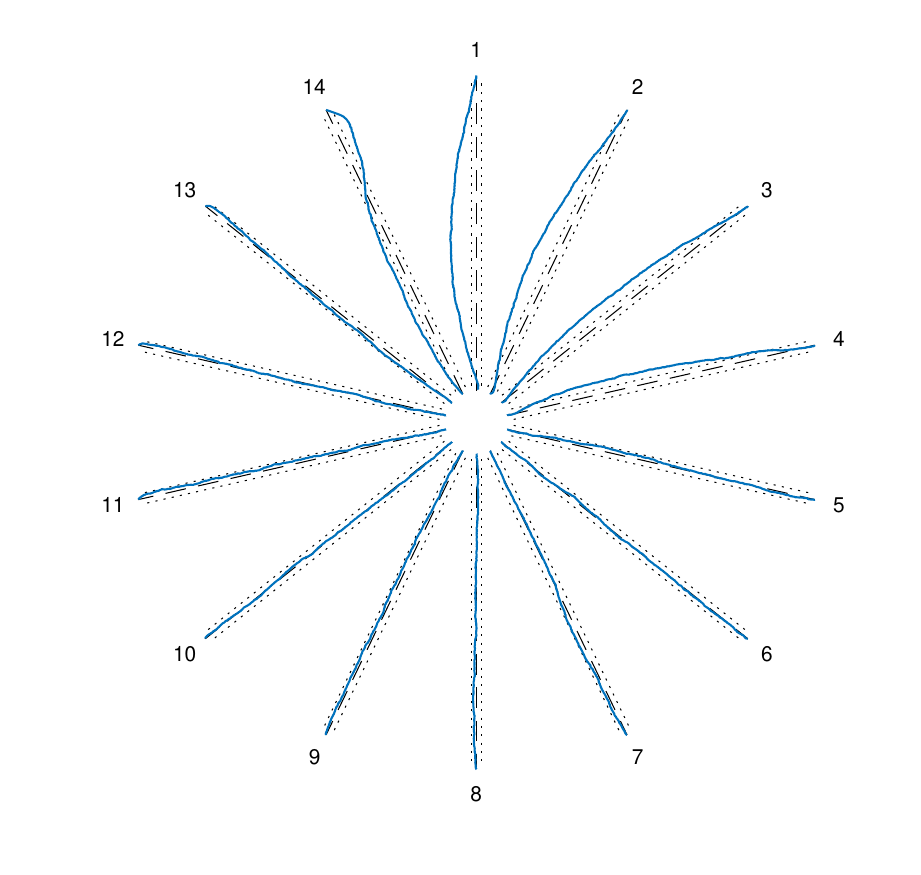} 
} 
\subfigure[$N=20$]{
\includegraphics[width=0.31\textwidth]{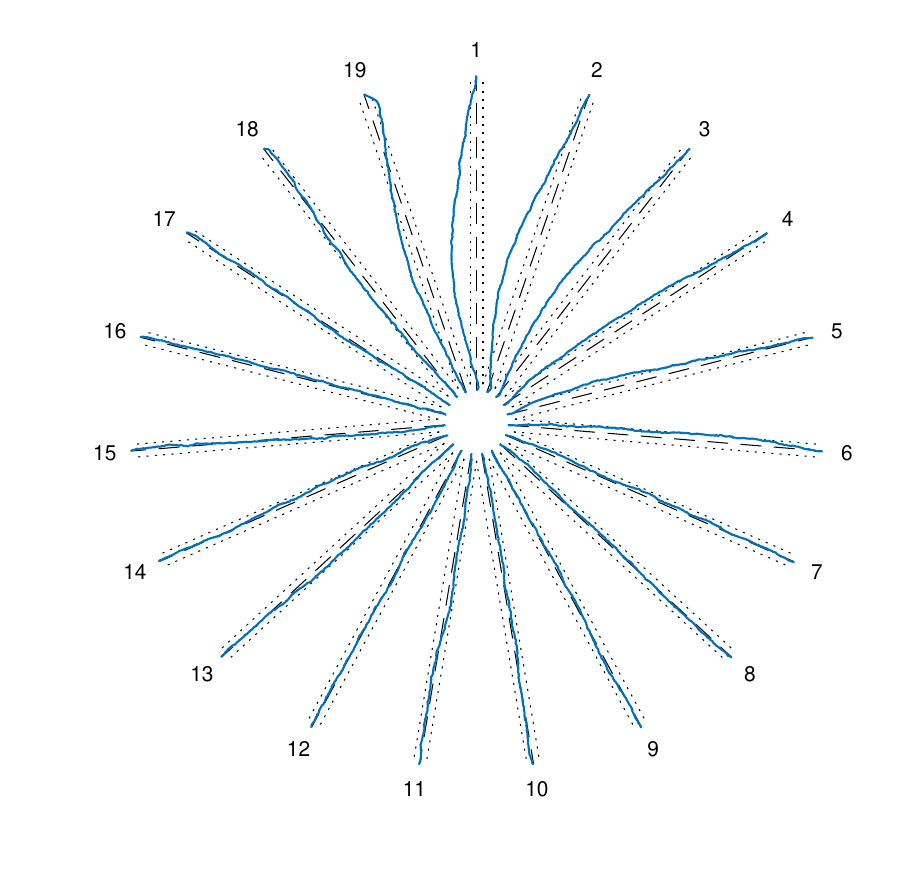} 
} 
\subfigure[$N=25$]{
\includegraphics[width=0.31\textwidth]{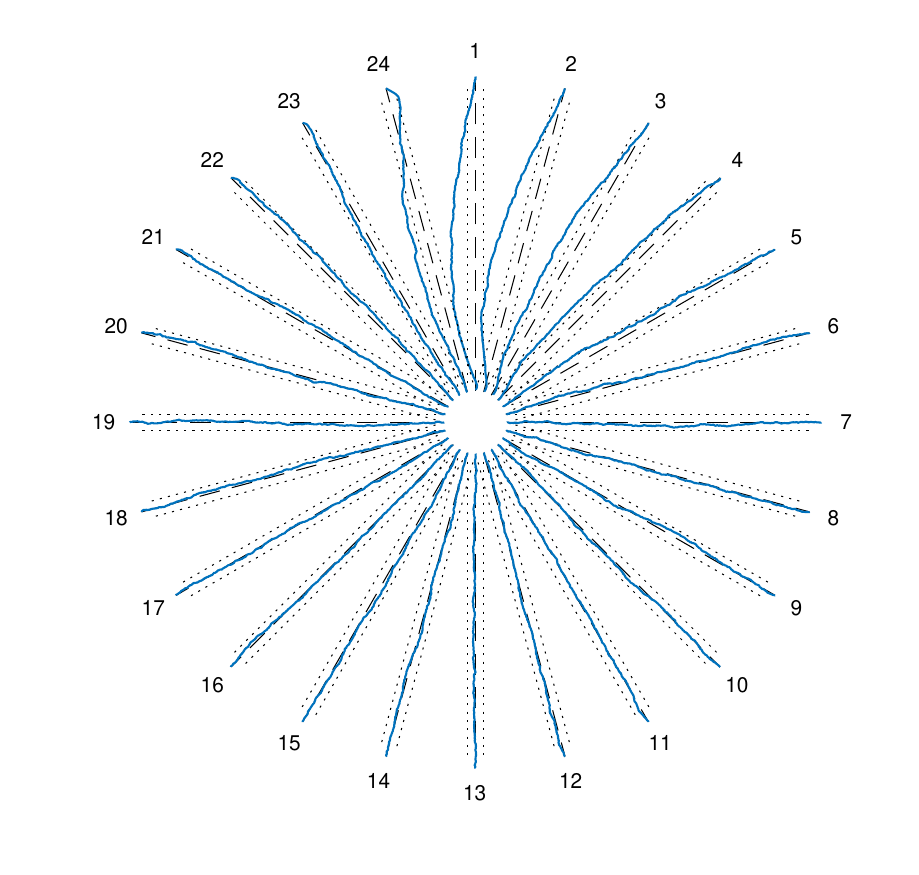} 
} 
\subfigure[$N=30$]{
\includegraphics[width=0.31\textwidth]{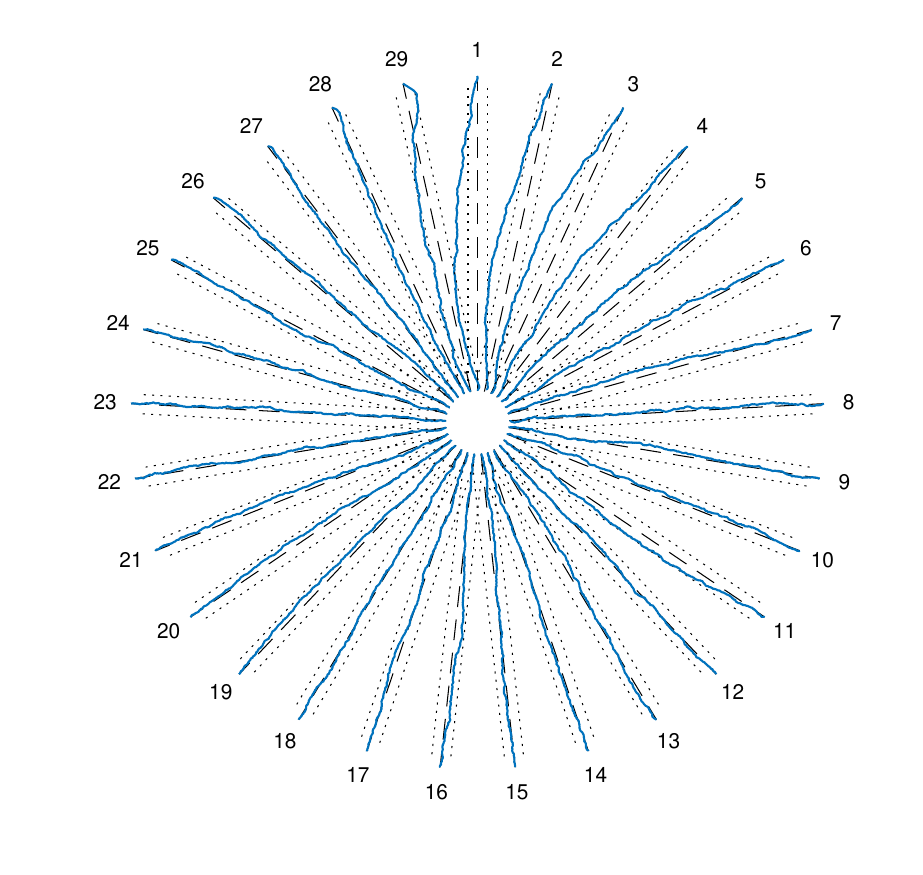} 
} 
\caption{Residual analysis ``dial plots'' of conditionally uniform UHP transformed empirical CDFs compared with true standard uniform CDFs and 99\% Kolmogorov-Smirnov error bounds. Out-of-sample test.}
\label{fig:dialplots}
\end{figure}

In Figure~\ref{fig:dialplots}, let us introduce ``dial plots'' for residual analysis of the UHP. Conditioned on there being a total of $N$ messages in the conversation, we have $N-1$ compensator points. After shifting and randomly reordering those that share an integer interval as in \edit{Theorem}~\ref{dialplotprop}, these plots compare each of the $N-1$ empirical CDFs with both standard uniform CDFs (dashed lines) and associated Kolmogorov-Smirnov error bounds (dotted lines). While each true CDF would of course be a $45^\circ$ line from $(0,0)$ to $(1,1)$, here, for the sake of space, the collection of lines is arranged clockwise with equidistant spacing, starting with the first (shuffled) compensator transformation at the twelve o'clock position of an analog dial.

Figure~\ref{fig:dialplots} contains the dial plots for conversations with a multiple of 5 messages, from 5 to 30. A larger collection of plots for $N = 3$ to $33$ is available in the appendix, but this sample is enough for the full insights. If the data was truly from the UHP, we would expect to see the empirical CDFs universally aligned with the dashed true CDFs. However, following Table~\ref{tbl:CompAll} and Figure~\ref{fig:CDFsTest}, we should not expect the fit to be perfect.  Instead, these dials show us where the UHP performs well and where it does not. Although the random de-ordering means that the plotted sequence need not be the true message sequence, the shuffling is only among those with the same integer ceiling, so earlier hands of the dial are still more likely to correspond to earlier messages in the conversation. Hence, we can still inspect these plots with a loose interpretation of the original service sequence. Certainly, the model performance suffers from what may be transient effects at the start and end of conversations; the first and final few of the hands clearly exceed the error bounds in each case of $N$. It may be the case that automated messages are more common in practice during such nascent stages, and these would not align with the model's behavioral assumptions. \edit{Furthermore, Figure~\ref{fig:SamplePath7} shows that some conversations are closed systematically, and that possible truncation is not covered in this estimation.} However, outside of these idiosyncrasies \edit{at the first and last few messages}, the UHP fit in the interior of conversations is actually quite good, with many hands entirely contained within the bounds. This is particularly true when we move beyond the short conversations (recall that $\E{N} = 14.72$ in this data), where a majority of the hands are accurately aligned in the dial. 

While the theory is not yet available to conduct this analysis for the other Hawkes model forms, anecdotally, we have seen similar performance when simulating BHP data and applying these UHP dial plot tests to it. This may suggest that relationship and role heterogeneity could capture some of the early and late stage conversation effects that escape the UHP. Regardless, we are encouraged by the UHP results, as the middle of the conversation is perhaps the most relevant operationally. That is, conversations that are not clearly just beginning nor clearly over are those that impact assessments of how busy agents are and decisions of who should receive the next customer. 
Most importantly, though, these dial plots help us formalize what \edit{we} believe is a key takeaway from this paper: There is naturally  still room for these Hawkes service models to be improved; nevertheless, the data demonstrates that there is fundamental merit to the dependence structure proposed in these stochastic processes.

\section{Exploring Routing Policies through Data-Driven Simulation}\label{sec:insights}

Given the dynamic and history-dependent nature of the Hawkes service model, it feels naturally applicable to real-time decisions. For the contact center context, new customer routing is perhaps the foremost of such managerial problems. In this section, we will explore that decision through the data, and in the sequel we will analyze our observations with theoretical tools. To explore these Hawkes-based routing decisions in an environment credibly connected to practice, we will now develop an entirely data-driven contact center simulation. In particular, our stochastic models are not at all involved in the randomness of the simulation. Rather, the customer arrival times, the messaging sequence within the conversation, the agent's read-to-reply time (i.e., the time from opening the message to when they send a response), the customer's receive-to-reply time (i.e., the time from receiving the message to sending a response), and the time from last message to the formal conversation closure are all sampled directly from the original data. The only role our models will play is in calculations for deciding to which agent a new conversation should be routed, and, consequently, the altered wait times in the conversation duration.

At a summary level, this data-driven simulation model treats the contact center as a collection of parallel re-entrant Erlang-R queueing systems with dedicated servers, with similar structure to \citet{campello2017}. There are two upfront managerial inputs that establish the synthetic service system environment: the number of agents, $\mathcal{S} \in \mathbb{Z}_+$, and the maximal concurrency for each agent, $\kappa \in \mathbb{Z}_+$. The simulation progresses over the true customer arrival process from the data, moving from one arrival epoch to the next. 
At each arrival, we first check if there is at least one agent with an available concurrency slot. If there is, the new conversation will be assigned to an agent according to the system's routing rule (stated below). If all agents have $\kappa$ conversations assigned, the simulation will instead assign the new customer to the first agent to become available. Hence, the conversation start time will be the next conversation completion time, and a queue of arriving customers may build during this delay. Between arrivals, each conversation progresses according to the data's true reply times. We assume that each agent responds to customer messages in a first-come-first-serve (FCFS) fashion across all their assigned conversations. Hence, the full conversation duration and, more precisely, the pauses between the sampled customer and agent message times, may depart from reality due to the simulation's assignment: the FCFS discipline means that the waiting time depends on the particular collection of conversations assigned to each agent.\footnote{Connecting to the prior literature on contact center operations, the simulation model we use is effectively a collection of parallel re-entrant Erlang-R queueing systems with dedicated servers, with similar structure to \citet{campello2017} (see Figure 1 therein for a visualization of the system with particular distributional assumptions, although here we sample directly from data instead). That is, each customer assigned to an agent can be thought of as being in that agent's re-entrant queueing system. 
When the customer sends a message to the agent, they enter the agent's message queue. The customer enters the service phase when the agent begins responding to their message, and exits the service phase upon receipt of the agent's response. If the customer eventually sends a message in response, they will re-enter the agent's message queue once they send the next message. 
}

We measure system performance through two types of waiting metrics: the \textit{outer wait}, meaning the wait for agent assignment, and the \textit{inner wait}, meaning the wait during service due to the agent multitasking across their concurrency. In our experiment, the  inner wait is reported per agent response and the outer wait is reported per conversation. 
To maintain rigor, the experiment will be conducted on the test data set from Section~\ref{sec:case_study}, so that the parameters of the Hawkes models are estimated from ``historical data'' in the eyes of the routing simulation.

Naturally, the operational focal point within these experiments is the manner in which new customers are routed to agents. The number of agents and the maximal concurrency are thus not just parameters of the experiment, but also parameters of the policy. 
The routing policies we consider are modified from the ``lightest load (LL),'' ``least assigned,'' or ``join the smallest caseload (JSC)'' policy,  which has been shown to be asymptotically optimal for some sets of assumptions \citep[see, e.g.,][]{luo2013staffing,Tezcan2014RoutingCustomersb,campello2017,Long2019}. From our industry partners, we also understand that this routing algorithm is very popular in practice. Re-phrasing the description in \citet{luo2013staffing}, this policy is as follows:

\hrulefill\textbf{ Lightest Load Policy (LL) }\hrulefill
{\it
\begin{itemize}
\item If there are agents available, route to the one with the lightest load: $i^L \in \argmin_{i \leq \mathcal{S}} K_{t,i}$
\begin{itemize}
\item If multiple agents have the lightest load, break ties randomly. 
 \end{itemize}
\end{itemize}}
\hrulefill

We propose a modest modification of this policy. Instead of breaking ties randomly between agents with the lightest load, we compute the expected number of messages in the next $\delta > 0$ units of time in each of lightest-loaded agents' conversations according to the Hawkes service models, and then select the agent with the lowest expectation. This estimation can be readily evaluated for each of the estimated three model forms; in Proposition~\ref{transMean} we provide it for the SysBHP.

\begin{proposition}\label{transMean}
Given the state of the SysBHP correspondence rates at time $t_0$, $\boldsymbol{\lambda}_{t_0}$,
and assuming that the concurrency is constant from $t_0$ to time $t \geq t_0$, the expected number of upcoming messages sent from time $t_0$ until time $t$ is 
\begin{align*}
\E{N_t - N_{t_0} \mid \boldsymbol{\lambda}_{t_0}}
= 
-
\mathbf{v}^\mathrm{T}
\mathbf{M}^{-1}
(\mathbf{I} - e^{\mathbf{M} (t - t_\naught)})
 \boldsymbol{\lambda}_{t_0}
\end{align*}
where $\mathbf{v}$ is an all-ones column vector and 
\begin{align*}
\mathbf{M}
=
\begin{bmatrix}
-(\beta^\mathsf{c,c} - \bar \alpha^\mathsf{c,c})\phantom{-}  & \bar \alpha^\mathsf{c,c}  & 0 & 0 \\
 0 & -\beta^\mathsf{c,a}\phantom{-} & \bar \alpha^\mathsf{c,a}  & \bar \alpha^\mathsf{c,a} \\
\frac{\bar \alpha^\mathsf{a,c} }{ K_{t_0}} & \frac{\bar \alpha^\mathsf{a,c}} {K_{t_0}}  & -\frac{\beta^\mathsf{a,c}}{K_{t_0}}\phantom{-}   & 0 \\
 0 & 0 & \frac{\bar \alpha^\mathsf{a,a}}{K_{t_0}} & -\frac{\beta^\mathsf{a,a} - \bar \alpha^\mathsf{a,a} }{K_{t_0}}\phantom{-}  \\
\end{bmatrix}
.
\end{align*}
\end{proposition}

Inherently, Proposition~\ref{transMean} is a Hawkes-based workload measurement for each conversation, because it scores the likelihood of upcoming activity to which the agent must respond. We can define a LL-style policy with these tiebreakers as follows:

\hrulefill\textbf{ Lightest Load with Hawkes-Based Projections Policy (LL$+$HP) }\hrulefill
{\it
\begin{itemize}
\item If there are agents available, route to the one with the lightest load: $i^L \in \argmin_{i \leq \mathcal{S}} K_{t,i}$
\begin{itemize}
\item If multiple agents have the lightest load, select $i^* = \argmin_{i:K_{t,i}=K_{t,i^L}} \E{N_{t+\delta,i} - N_{t,i} \mid \boldsymbol{\lambda}_{t,i}}$.
 \end{itemize}
\end{itemize}}
\hrulefill

This change from the LL policy is small, and it only invokes the Hawkes process models when multiple agents have the lowest assigned concurrency. However, this simple change targets the exact discrepancy seen in Figure~\ref{fig:HistogramAssigned} in the introduction: assignment is not necessarily a measure of present activity, and thus not necessarily a measure of workload.

Using these two policies, we now compare the capabilities of the UHP, BHP, and SysBHP. We evaluate the projections over short, medium, and long interval sizes: $\delta =$ 30 seconds, $\delta =$ 5 minutes, and $\delta = \infty$. We run the simulation in two different staffing settings: $\mathcal{S}=$ 125 and 135 servers. The maximum concurrency each agent may have, $\kappa$, takes on values 5, 10, 15, and 20. 
The comparison between the routing performance delivered by each model (and by the unmodified LL policy) is summarized in Figure \ref{fig:improvement} and detailed in full in Tables \ref{tbl:CompAllMIwait}--\ref{tbl:CompAllSDOwait} in the Appendix. Each of the performance metrics --- mean inner wait, mean outer wait, and probability of outer wait --- are shown here as percent improvements of the various LL+HP forms relative to the baseline LL policy.\footnote{Due to the memoryless property, random tie-breaking in the LL policy is equivalent to projecting future activity according to the SES benchmark. Hence, these improvements can also be thought of as relative to this classic model.} 

Relative to the BHP and SysBHP, the UHP stands out in these experiments for its invariance across $\delta$. In this case, the expected number of messages given in Proposition~\ref{transMean} becomes
$
\E{N_{t+\delta} - N_{t} \mid {\lambda}_{t}}
=
\frac{\lambda_t}{\beta - \alpha}
\left(
1 - e^{-(\beta-\alpha)\delta}
\right)
,
$
where $\lambda_t$ is calculated using the timestamps of the conversation so far. While the value of these projections will change with $\delta$, the rankings are insensitive to $\delta$. 
In this way, the UHP tie-breaking reduces from a ranking of projections to a ranking of histories: $i^* = \argmin_{i:K_{t,i}=K_{t,i^L}} \E{N_{t+\delta,i} - N_{t,i} \mid {\lambda}_{t,i}} = \argmin_{i:K_{t,i}=K_{t,i^L}} {\lambda}_{t,i}$. This reduction does not hold in general for the BHP or SysBHP models, as the weight of the projections can shift with different $\delta$'s because of the disparities in impact \edit{for} different directions of the conversation. 

\begin{figure}[tbh]
\centering
\subfigure[Mean Outer Wait ($\mathcal{S} = 125$)]{
\includegraphics[width=0.31\textwidth]{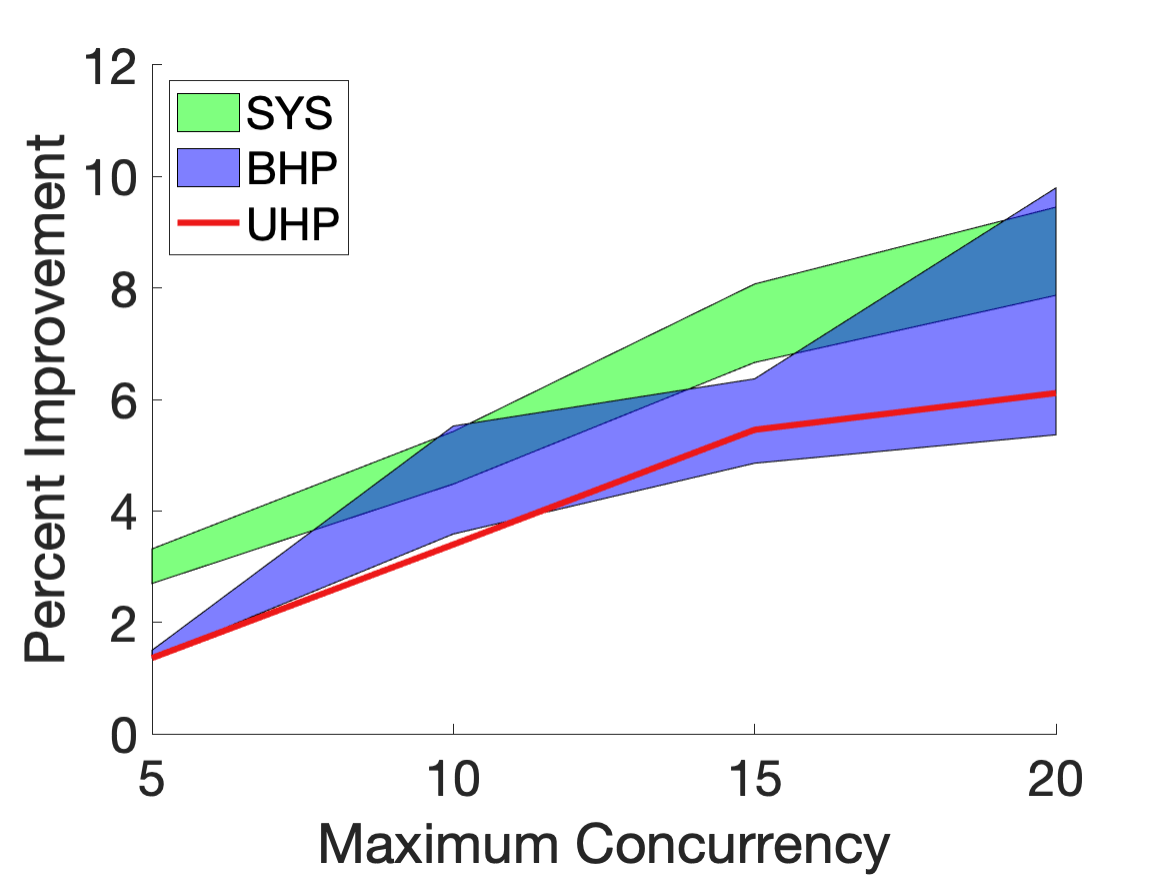} \label{fig:ow125}
}
\subfigure[Mean Inner Wait ($\mathcal{S} = 125$)]{
\includegraphics[width=0.31\textwidth]{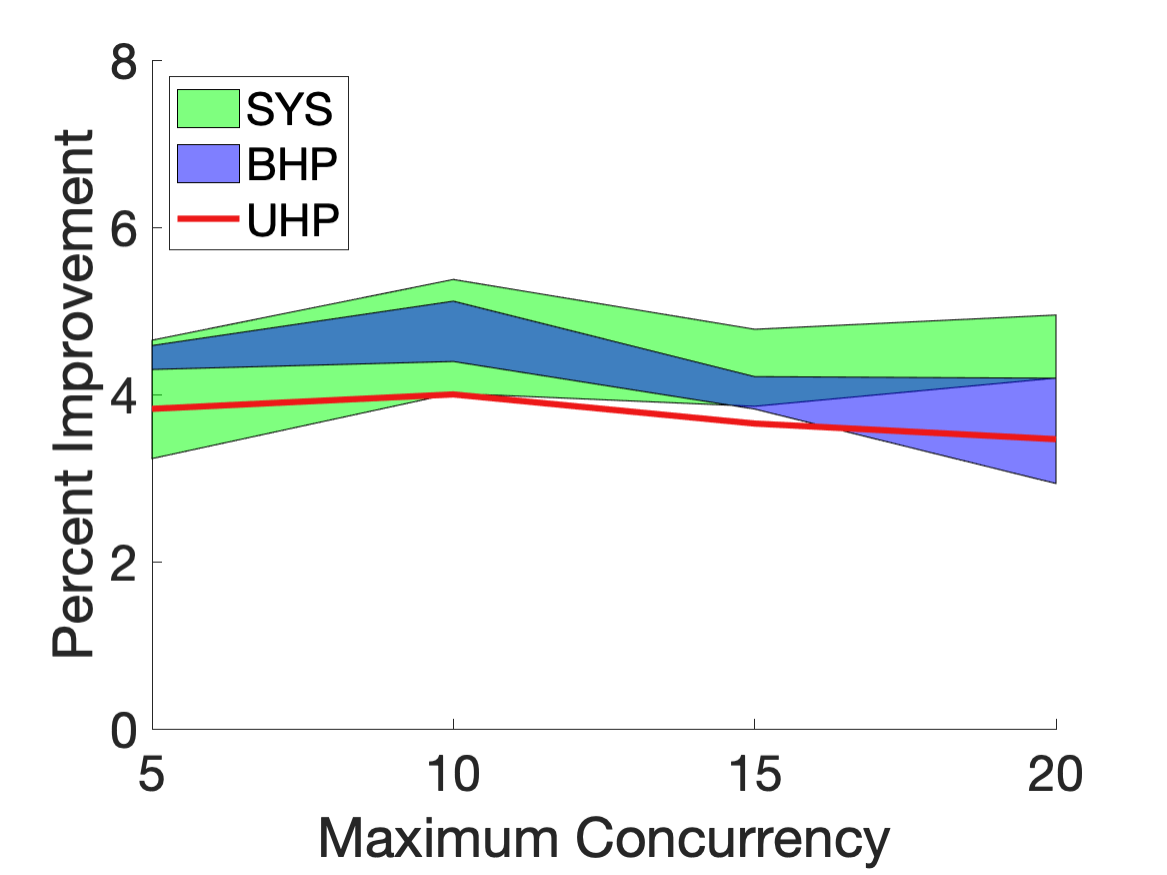} \label{fig:iw125}
} 
\subfigure[Prob. of Outer Wait ($\mathcal{S} = 125$)]{
\includegraphics[width=0.31\textwidth]{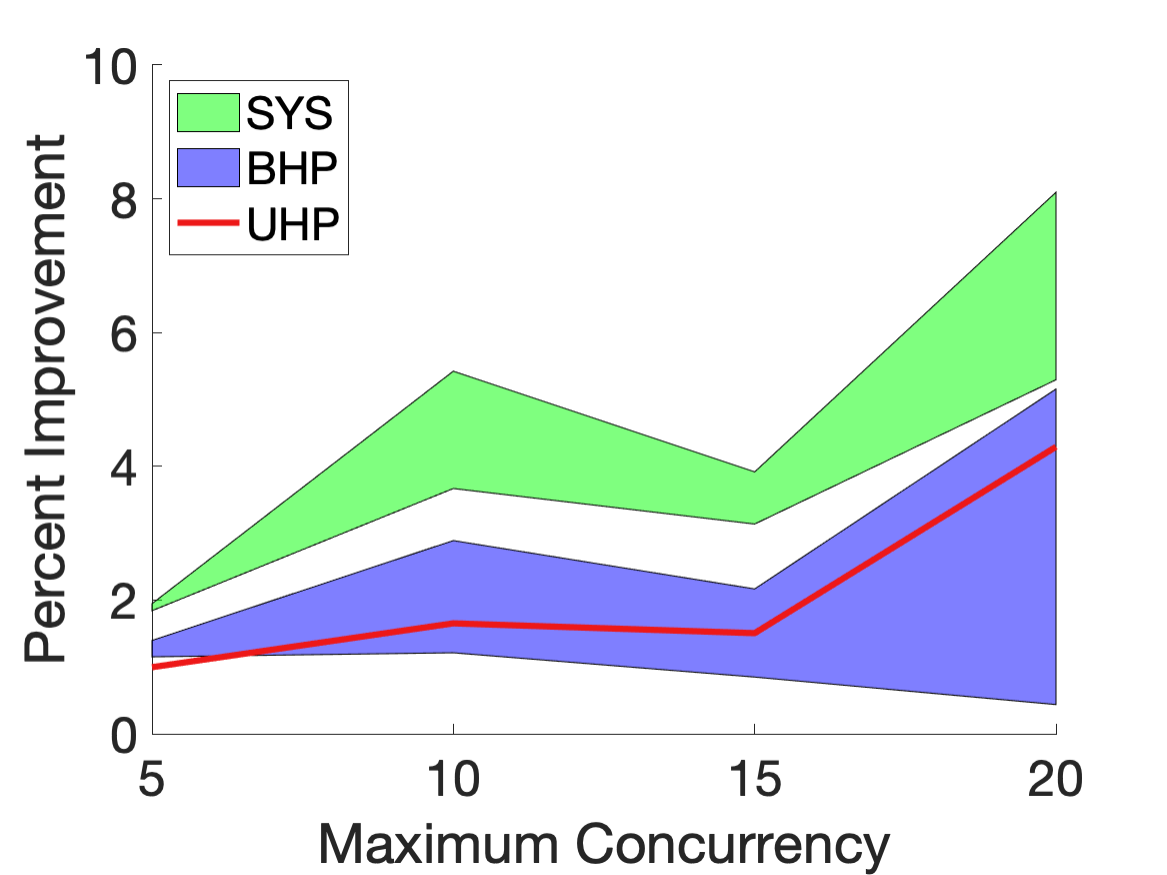} \label{fig:pw125}
} 
\subfigure[Mean Outer Wait ($\mathcal{S} = 135$)]{
\includegraphics[width=0.31\textwidth]{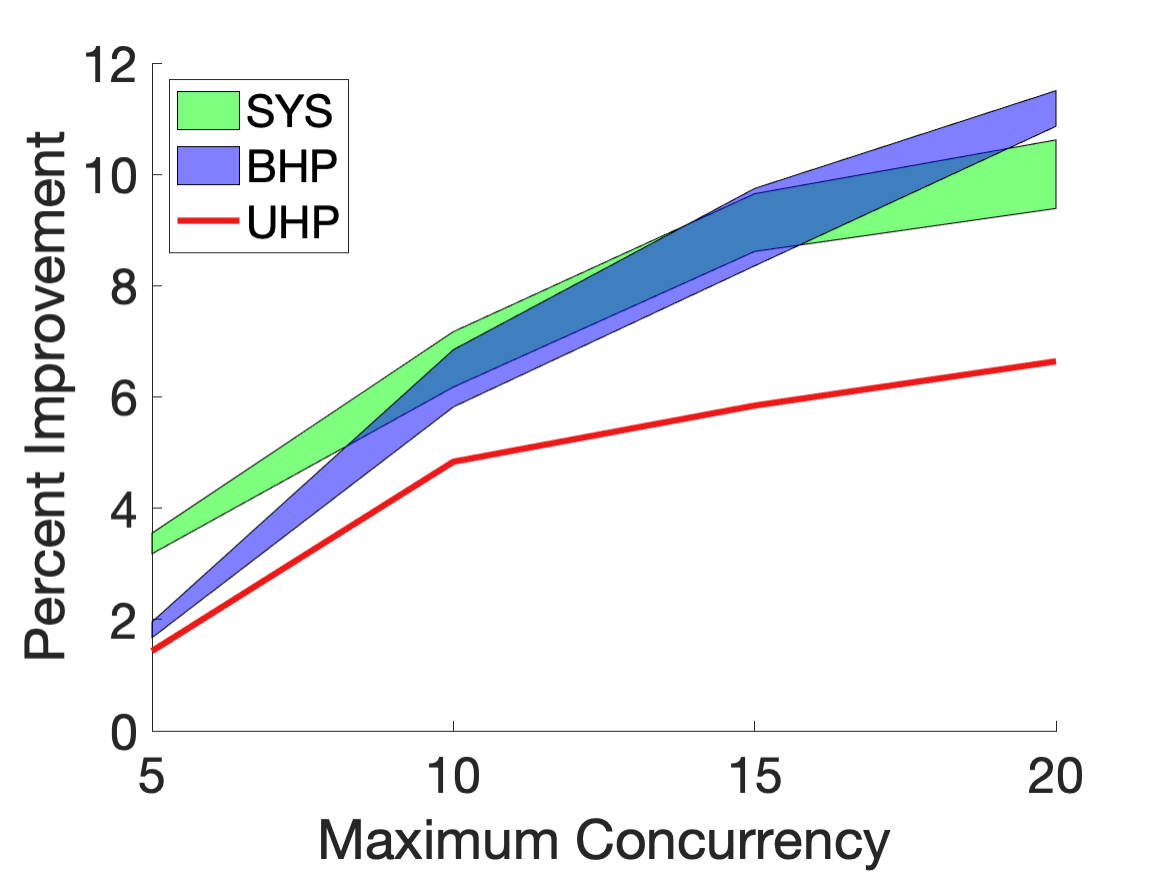} \label{fig:ow135}
} 
\subfigure[Mean Inner Wait ($\mathcal{S} = 135$)]{
\includegraphics[width=0.31\textwidth]{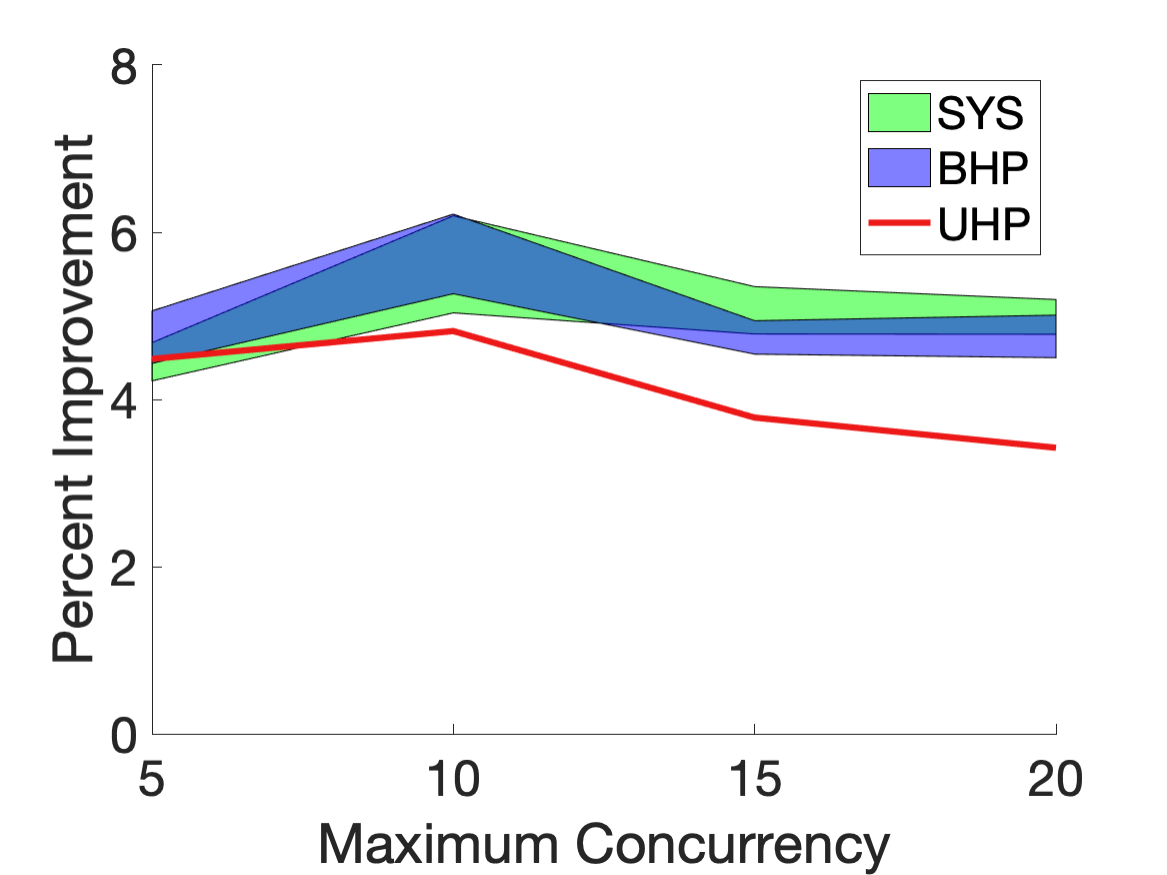} \label{fig:iw135}
} 
\subfigure[Prob. of Outer Wait ($\mathcal{S} = 135$)]{
\includegraphics[width=0.31\textwidth]{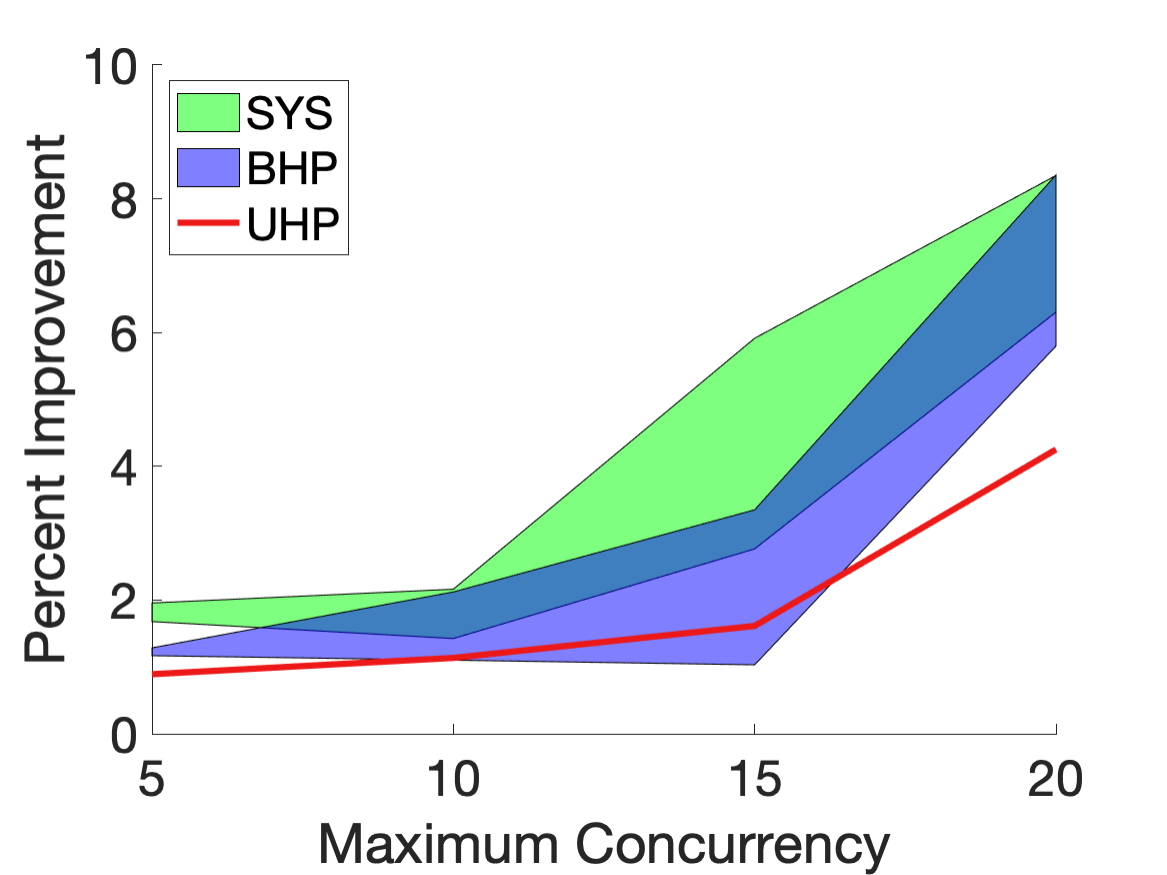} \label{fig:pw135}
} 
\caption{Percent improvement of wait time metrics for the LL$+$HP policy across a range of the projection interval $\delta$, plotted as an area of the minimum to maximum improvements at each maximum concurrency value.}
\label{fig:improvement}
\end{figure}

This constitutes a first takeaway from this experiment: accounting for history dependence alone can improve operational decision making, and expanding to more nuanced dependence can offer even more improvement still.
As can be seen throughout these plots and tables, each of the Hawkes model forms offers substantial improvement over the LL policy, with the most general SysBHP typically leading the pack. Indeed, the SysBHP improves the inner and outer wait times by over 5\% and 10\%, respectively. 
Supporting our assertion that the customer and agent relationships are vital components of service progression models, the BHP policies also perform quite well across the board; in some cases, it can be seen to rival or surpass the SysBHP.

Let us note that these policies are not just decreasing the average wait, but reducing system variability as well. Tables~\ref{tbl:CompAllSDIwait} and~\ref{tbl:CompAllSDOwait} show that the Hawkes projection policies can also reduce the standard deviation of the waits, often by 3-6\% of the standard deviations under the unadjusted LL policy. Furthermore, all differences in mean between the standard LL routing and the model projection policies are statistically significant. 
Let us also remark that either through inspection of Tables \ref{tbl:CompAllMIwait}--\ref{tbl:CompAllSDOwait} or through the heights of the ranges in Figure~\ref{fig:improvement}, it is clear that the percent improvement is fairly robust to choice of short, medium, or long projection intervals.\footnote{We also conducted a localized search over $\delta \in \{10, 20, \dots, 180\}$ seconds, and found similar robustness.} Identification of the best  $\delta$ may be quite context specific, but in the very least our results suggest that any choice of $\delta$ is poised to improve performance --- as is demonstrated by the UHP.

\section{Analyzing Routing Policies and the Impact of Service Closure}\label{sec:routingA}

Although Section~\ref{sec:insights}'s data-driven simulation found that LL+HP outperformed LL with statistical significance, we also reviewed that the LL policy has been shown to be asymptotically optimal in certain conditions. Let us offer a closer investigation of this discrepancy by leveraging the results of \citet{Tezcan2014RoutingCustomersb} and stylizing that analysis for comparison to our model's focal features. \edit{In particular, we claim that the interplay between the routing policy and the systematic closure policy is at the heart of this gap in performance.}

For this illustration of the potential pitfalls of LL routing and the interplay of algorithmic closure, let us suppose that all services must be closed systematically. We will conduct this analysis in two parts. First, we will construct a service-side extension of the static planning problem (SPP) from \citet{Tezcan2014RoutingCustomersb} that is motivated by properties we observe in the Hawkes service model. Then, in the second part, we will prove that the manner of systematic closure can create heterogeneity in the service performance metrics, and, accordingly, we will  further extend the SPP to the heterogeneous setting. \edit{In the contrast between} these two parts, we will observe the impact of the system's choice of closure algorithm. 
By comparison to \citet{Tezcan2014RoutingCustomersb}, our analysis is meant to be illustrative rather than exhaustive.

\subsection{Motivating and Analyzing a Service Success Static Planning Problem}\label{firstStage}

To start the first stage of routing analysis, let us begin by motivating the service-side performance measure, the probability of service success. With the stylized assumption that conversations must be closed systematically, we are implicitly assuming that neither the customer nor agent indicate service closure explicitly in their messages, like we saw in the vignette in Figure~\ref{fig:SamplePath7}. Instead, when to close the service must be determined simply by the history of conversation timestamps (and possibly also word count and sentiment score random variables). Intuitively, a conversation should be closed if the system is confident that it has truly concluded. Hence, we will call a service closure successful if in fact no more messages would occur after the closure time. By contrast, an unsuccessful closure would end the conversation prematurely. 
Because the Hawkes service model is history-dependent, we can leverage the sample path of the conversation to quantify the probability that a closure would be successful, or, equivalently, the risk that closing now would be too early. 
Given the full history of the Hawkes service model \edit{contained} in its natural filtration, we can provide the probability that further conversation activity will (or will not) occur soon.\footnote{In fact, in the case of the three specific model forms we introduced in Section~\ref{sec:case_study}, the exponential decay functions can enable us to use only the present state of the correspondence rates.}

\begin{proposition}\label{nextMsgDist}
Let $X$ be the time until the next message occurs after time $t_0 \geq 0$ in the SysBHP. Assuming that the concurrency is constant from $t_0$ onward, the probability that no message is sent in the next $x \geq 0$ time units is
\begin{align*}
\PP{X \geq x \mid \boldsymbol{\lambda}_{t_0}}
&=
e^{
-
\frac{\lambda_{t_\naught}^\mathsf{c,c}}{\beta^\mathsf{c,c}}
\left(
1 - e^{-\beta^\mathsf{c,c} x}
\right)
-
\frac{\lambda_{t_\naught}^\mathsf{c,a}}{\beta^\mathsf{c,a}}
\left(
1 - e^{-\beta^\mathsf{c,a} x}
\right)
-
\frac{\bar \lambda_{t_\naught}^\mathsf{a,c}}{\beta^\mathsf{a,c}}
\left(
1 - e^{-{\beta^\mathsf{a,c} x}\slash{K_{t_\naught}} }
\right)
-
\frac{\bar \lambda_{t_\naught}^\mathsf{a,a}}{\beta^\mathsf{a,a}}
\left(
1 - e^{-{\beta^\mathsf{a,a} x}\slash{K_{t_\naught}} }
\right)
}
,
\end{align*}
where $\bar \lambda_{t_0}^{\mathsf{a},y} = \sum_{\ell: A_\ell^y \leq t_\naught}(\alpha^{\mathsf{a},y}_\mathsf{1} S_\ell^y + \alpha^{\mathsf{a},y}_\mathsf{2} W_\ell^y) e^{-\beta^{\mathsf{a},y} (t_\naught - A_\ell^y) \slash K_{t_\naught}}$ for $y \in \{\mathsf{c},\mathsf{a}\}$.
\end{proposition}

Of course, through the limit as $x$ goes to infinity, Proposition~\ref{nextMsgDist} also provides the probability that the conversation is truly over, $\PP{X=\infty \mid \boldsymbol{\lambda}_{t_0}} = \exp\left(-
{\lambda_{t_0}^\mathsf{c,c}}\slash{\beta^\mathsf{c,c}}
-
{\lambda_{t_0}^\mathsf{c,a}}\slash{\beta^\mathsf{c,a}}
-
{\bar \lambda_{t_0}^\mathsf{a,c}}\slash{\beta^\mathsf{a,c}}
-
{\bar \lambda_{t_0}^\mathsf{a,a}}\slash{\beta^\mathsf{a,a}}
\right)$.
Viewing $\PP{X = \infty \mid \boldsymbol{\lambda}_{t_0}}$ as a function of $K_{t_0}$, we can see that this probability decreases as the agent's concurrency increases. However, we also know by Definition~\ref{modelDef} that the service pace slows as $K_{t_0}$ increases.
Informally, this suggests that LL routing may indeed be a good idea: Performance and pace both worsen as the agent's concurrency rises, so there is a natural desire to prioritize routing to agents with the fewest customers on hand.

To solidify the intuition from this pair of observations, let us introduce a SPP stylized and extended from that in \citet{Tezcan2014RoutingCustomersb}, where that paper justifies the use of this linear program (LP) through the fluid limit of a Markov chain model of a contact center. Let us define $\mu_k$ as the mean service closure rate per customer when an agent has concurrency level $k \in \{1, \dots, \kappa\}$. Similarly, let us also define $\eta_k$ as the probability of satisfactorily completed service upon closure at concurrency level $k$. Equivalently, $1-\eta_k$ is the fraction of conversations that are prematurely closed. As inspired by Proposition~\ref{nextMsgDist}, these service success probabilities will constitute the SPP's metrics of performance. 
Like in \citet{Tezcan2014RoutingCustomersb}, the decision variables $\{\theta_k : 1 \leq k \leq \kappa\}$ are the \emph{idealized arrival rate} of customers to each concurrency level $k$. Naturally, these decision variables constitute an idealized routing policy in the fluid sense. We describe the SPP through the following LP.
\begin{align}
\max\quad &\sum_{k=1}^\kappa \eta_k \theta_k 
 \nonumber \\
\mathrm{s.t.}\quad & \sum_{k=1}^\kappa \frac{\theta_k}{k \mu_k} \leq \mathcal{S}
 \nonumber \\
\quad& \sum_{k=1}^{\kappa} \theta_k = \Theta
 \nonumber \\
\quad& \theta_k \geq 0 \quad \forall\, k \in \{1, \dots, \kappa\} \nonumber
\end{align}

Also like in \citet{Tezcan2014RoutingCustomersb}, the first constraint enforces that the capacity of the total number of servers $\mathcal{S}$ is sufficient to handle the idealized offered load, and the second constraint assures that all incoming traffic is routed, where  $\Theta$ is the overall customer arrival rate to the system.  We will assume $\Theta < \mathcal{S} \max_{1\leq k\leq \kappa} k \mu_k$ for stability of the service system; otherwise this routing LP will be infeasible.
However, this SPP departs from its predecessor through its objective function. In \citet{Tezcan2014RoutingCustomersb}, the goal was to minimize customer abandonment, where customer impatience could occur both while waiting for service to begin (outer abandonment, if you will) and while waiting for responses within service (inner abandonment). 
Instead, our objective is to maximize the rate of service success. 
Through the framework of systematic closure, we are extending to a setting that mirrors or complements \citet{Tezcan2014RoutingCustomersb}'s case on the server or system side. That is, abandonment occurs when the customer's internal patience clock cuts off a slow-moving agent within service; by contrast, unsuccessful systematic closure cuts off a dormant but not yet satisfied customer. Hence, rather than minimizing how often customers leave the service before it ends, this SPP maximizes how often the system correctly concludes the service.

Informed by \citet{Tezcan2014RoutingCustomersb}'s analysis and the present endeavor to understand the pitfalls of LL routing, let us suppose a simple and stylized structure for each agent's total service rate, $k \mu_k$. We will assume that this agent-level service rate is linear in the concurrency up to some point, but constant after. Again, this is simplistic by definition, but it matches qualitative intuition: it is easy to multitask and remain productive, so long as the load doesn't get too high. (See Figures~\ref{fig:durations} and~\ref{fig:gaptimes} in the appendix for empirical support of this style.) Furthermore, we will assume that the successful rate, $\eta_k$, is decreasing with $k$, like we saw in Proposition~\ref{nextMsgDist}. With this stylized structure, we can characterize the routing SPP's optimal solution.

\begin{proposition}\label{firstSPPprop}
Suppose that there exist $c_0, c_1 \in \mathbb{R}_+$ such that $k \mu_k = \left( k c_0 \wedge c_1\right)$ with $\Theta < \left(\kappa c_0 \wedge c_1\right) \mathcal{S}$. If
the service success probabilities $\eta_1 \geq \dots \geq \eta_\kappa$ satisfy
\begin{align}
\eta_j - \eta_{k^*-1}  \leq \frac{\Delta_{k^*}}{j}\left(k^*-1-j\right)\left(\eta_{k^*-1} - \eta_{k^*}\right),
\label{successProbAssump}
\end{align}
for each $1 \leq j \leq \kappa$, where $\Delta_k = k \mu_k / (k \mu_k - (k-1)\mu_{k-1})$ and where $k^* \leq \kappa$ is such that $(k^*-1)c_0 \mathcal{S} \leq \Theta < k^* c_0 \mathcal{S}$, then the solution vector $\boldsymbol{\theta}^* \in \mathbb{R}_+^\kappa$ defined by
\begin{align}
\theta_k^* 
=
\begin{cases}
\frac{(k^*-1)c_\naught}{c_\naught + (c_1 - k^* c_\naught)^+}
\left(\left(k^* c_0 \wedge c_1\right) \mathcal{S} - \Theta\right)  & \text{ if } k = k^* - 1
,\vspace{0.1 in} \\ 
\frac{(k^* c_\naught \wedge c_1)}{c_\naught + (c_1 - k^* c_\naught)^+}
\left(\Theta - (k^*-1) c_0 \mathcal{S}\right) & \text{ if } k = k^* 
,\vspace{0.1 in} \\
0 & \text{ otherwise}
,
\end{cases}
\label{thetaStarDef}
\end{align}
is an optimal solution to the routing static planning problem.
\end{proposition}

Through Proposition~\ref{firstSPPprop}, we have found service rate and success probability conditions under which LL routing is optimal in the SPP.
In addition to the non-increasing success probabilities and the simple piecewise multitasking structure of the service rates, informally, Equation~\eqref{successProbAssump} says that the success probabilities (as a function of $k$) should not change too drastically relative to the change around the lowest concurrency level that can maintain system stability, $k^*$. Given this, the optimal solution identified in~\eqref{thetaStarDef} constitutes an idealized routing that prioritizes this lowest stable level; \citet{Tezcan2014RoutingCustomersb} identified this as the LL policy. Hence, Proposition~\ref{firstSPPprop} confirms our earlier intuition. We saw that both service pace and performance worsen as concurrency rises, and now we have an SPP setting that reproduces these dynamics and prescribes LL routing as optimal. 

\subsection{Systematic Closure and the Emergence of Success Heterogeneity}\label{secondStage}

\edit{In formalizing systematic closure in Section~\ref{closureDefSec}, we reviewed some popular systematic closure policies that can be viewed as stopping times in the Hawkes service model, such as closing a service after some pre-defined time of inactivity, or upon a certain number of agent messages without a customer response.} Drawing inspiration from our stochastic model, one could also think of a policy that closes the service when the correspondence rate becomes sufficiently low. This would be the hitting time of some level in the Hawkes cluster. 
Our claim in this second stage of analysis is that the structure of this stopping time is actually a cornerstone operational choice, and the manner of systematic closure will dictate the nature of the service success probabilities and, by consequence, alter the performance of broader operational decisions upstream.

To see this, let us consider a family of level-based stopping times for the BHP service model form. For each $0 < \zeta \leq \alpha^\mathsf{c,c}\slash\beta^\mathsf{c,c} + \alpha^\mathsf{a,c}\slash\beta^\mathsf{a,c}$, define the correspondence rate hyperplane, $L_\zeta = \{\boldsymbol{\lambda} \in \mathbb{R}_+^4 : {\lambda_1}\slash{\beta^\mathsf{c,c}}
+
\lambda_2\slash{\beta^\mathsf{c,a}}
+
\lambda_3\slash{\beta^\mathsf{a,c}}
+
{\lambda_4}\slash{\beta^\mathsf{a,a}}
=
\zeta
\}$. Then, let $\mathcal{T} = \{\tau_\zeta = \inf\{t \geq 0 \mid \boldsymbol{\lambda}_{t} \in L_\zeta\} : 0 < \zeta \leq  \alpha^\mathsf{c,c}\slash\beta^\mathsf{c,c} + \alpha^\mathsf{a,c}\slash\beta^\mathsf{a,c}\}$ be the family of hitting-time stopping rules for each possible $\zeta$. Here, the upper bound on $\zeta$ ensures that it will be below the initial correspondence rate, eliminating the risk that the process will always be below the hyperplane and thus ensuring that the stopping time will be finite. Through Proposition~\ref{nextMsgDist}, we can see that the closure policies in $\mathcal{T}$ hold more managerial meaning than simply closing upon the first time some hyperplane is reached. That is, these hyperplanes orient the systematic closure precisely around the service success. In Lemma~\ref{closureVarLemma}, we prove that this family contains the only closure rules that both are almost surely finite and specify the closing success probability exactly.

\begin{lemma}\label{closureVarLemma}
Let $\tau$ be a stopping time for the BHP service model. Then, if and only if $\tau \in \mathcal{T}$, $\Var{\PP{X = \infty \mid \boldsymbol{\lambda}_{\tau}}} = 0$ with $\PP{\tau < \infty} = 1$.
\end{lemma}

Through the BHP, Lemma~\ref{closureVarLemma} reveals the structure of SysBHP performance within each possible concurrency level. That is, even \edit{if the concurrency is restricted to be some fixed level},  Lemma~\ref{closureVarLemma} provides conditions under which the service success probabilities will vary. Connecting Lemma~\ref{closureVarLemma} to the SPP, this means that if the systematic closure exactly prescribes the success probability (i.e., $\tau \in \mathcal{T}$), then it is sufficient to model the SPP with one success probability $\eta_k$ at each concurrency level $k$, which yields Proposition~\ref{firstSPPprop}. On the other hand, if $\tau \not \in \mathcal{T}$, then the SPP should include more variety than just one success probability per level.

To see that this heterogeneity is itself consequential, we return to the SPP and stylize on the salient details. We will incorporate the variety of success probabilities by expanding the LP coefficients simply, yet meaningfully. Let us suppose that at every concurrency level there are two types of close-of-service activity conditions: high success (Type \textsf{1}) and low success (Type \textsf{2}). Let these have corresponding probabilities $\eta_k^\mathsf{1} > \eta_k^\mathsf{2}$ for every concurrency level $k$. At each $k$, we will let the index $h \leq k$ track the number of Type \textsf{1} services. Then, for every $k$ let $p_{k,h}$ with $0 = p_{k,0} \leq p_{k,1} \leq \dots \leq p_{k,k-1} \leq p_{k,k} = 1$ be the probability that a Type \textsf{1} service is the next to conclude for an agent with $h$ Type \textsf{1} services among $k$ overall. If Types \textsf{1} and \textsf{2} conclude at equal rates, we would have $p_{k,h} = h \slash k$, but we need not make this assumption.

Relative to the original SPP, here we will expand the formulation to account for the success heterogeneity. Let the decision variables $\{\theta_{k,h} : 1 \leq k \leq \kappa, 0 \leq h \leq k\}$ again be the {idealized arrival rates}, where this desired routing now considers both the concurrency level $k$ and the high type index $h$. Using these, let us define the heterogeneous SPP through the following LP. 
\begin{align*}
\max\quad &\sum_{k=1}^\kappa \sum_{h=0}^k
\left(p_{k,h} \eta_k^\mathsf{1} + (1-p_{k,h})\eta_k^\mathsf{2} \right) \theta_{k,h} 
\\
\mathrm{s.t.}\quad & \sum_{k=1}^\kappa \sum_{h=0}^k \frac{\theta_{k,h}}{k \mu_k} \leq \mathcal{S}
\\
\quad&  \sum_{k=1}^{\kappa} \sum_{h=0}^k \theta_{k,h} = \Theta
\\
\quad& \theta_{k,h} \geq 0 \quad \forall\, 1 \leq k \leq \kappa, 0 \leq h \leq k
\end{align*}
This optimization problem is essentially the same as the homogeneous SPP in terms of constraints, which once again enforce stability and conservation of arrivals. However, like the decision variables, the objective function now reflects the heterogeneity using the simple coefficients we have just defined. We will now see that this heterogeneity does indeed impact the routing policy performance, even when the SPP assumptions are otherwise the same as Proposition~\ref{firstSPPprop}.

To contrast with Proposition~\ref{firstSPPprop}, let us define two candidate solutions, $\boldsymbol{\theta}^\mathsf{LL}$ and $\boldsymbol{\theta}^\mathsf{HP}$, which each adapt the homogeneous optimal LL solution to this new setting. First, we will let $\boldsymbol{\theta}^\mathsf{LL}$ be any policy that prioritizes the lowest stable levels but disregards the collection of activity types within those levels. Then, $\boldsymbol{\theta}^\mathsf{HP}$ similarly prioritizes the lowest levels of concurrency; however, much like how the LL$+$HP policy from Section~\ref{sec:insights} routes to agents with the lowest projected activity within the lowest concurrency levels, $\boldsymbol{\theta}^\mathsf{HP}$ prioritizes those with the largest amount of the high success rate types within the lowest concurrency levels. In Proposition~\ref{secondSPPprop}, we prove that non-increasing success probabilities and a piecewise multitasking structure of the service rates (like in Proposition~\ref{firstSPPprop}) are enough to guarantee that the activity-aware $\boldsymbol{\theta}^\mathsf{HP}$ will strictly outperform any other lightest load policy, $\boldsymbol{\theta}^\mathsf{LL}$. If we also assume that the high success probabilities have a rate of change structure like in \eqref{successProbAssump}, then $\boldsymbol{\theta}^\mathsf{HP}$ is optimal.

\begin{proposition}\label{secondSPPprop}
Suppose that there exist $c_0, c_1 \in \mathbb{R}_+$ such that $k \mu_k = \left( k c_0 \wedge c_1\right)$ with $\Theta < \left(\kappa c_0 \wedge c_1\right) \mathcal{S}$, and suppose that the service success probabilities, $\eta_1^\ell \geq \dots \geq \eta_\kappa^\ell$ for $\ell \in \{\mathsf{1}, \mathsf{2}\}$, are such that
$
\eta^\mathsf{1}_k > \eta^\mathsf{2}_k
,
$
for every $1 \leq k \leq \kappa$. Additionally, let $\varrho^{k}$ be any arbitrary non-trivial probability distribution over $0, \dots, k$.

Then, for $\boldsymbol{\theta}^*$ given by Equation~\eqref{thetaStarDef}, the solutions $\boldsymbol{\theta}^\mathsf{LL}, \boldsymbol{\theta}^\mathsf{HP} \in \mathbb{R}_+^{\kappa \times \kappa}$ defined
\begin{align*}
\theta_{k,h}^\mathsf{LL} 
=
\begin{cases}
\varrho^{k^*-1}_{h} \theta^*_{k^*-1} & \text{ if } k = k^*-1,
\\
\varrho^{k^*}_{h}\theta^*_{k^*} & \text{ if } k = k^*,
\\
0 & \text{ otherwise},
\end{cases}
\qquad \text{ and } \qquad
\theta_{k,h}^\mathsf{HP} 
=
\begin{cases}
\theta^*_{k^*-1} & \text{ if } h = k = k^*-1,
\\
\theta^*_{k^*} & \text{ if } h = k = k^*,
\\
0 & \text{ otherwise},
\end{cases}
\end{align*}
are both feasible solutions to the heterogeneous static planning problem, and the objective value of $\boldsymbol{\theta}^\mathsf{HP}$ strictly dominates that of $\boldsymbol{\theta}^\mathsf{LL}$:
\begin{align*}
\sum_{k=1}^\kappa \sum_{h=0}^k
\left(p_{k,h} \eta_k^\mathsf{1} + (1-p_{k,h})\eta_k^\mathsf{2} \right) \theta_{k,h}^\mathsf{LL} 
<
\sum_{k=1}^\kappa \sum_{h=0}^k
\left(p_{k,h} \eta_k^\mathsf{1} + (1-p_{k,h})\eta_k^\mathsf{2} \right) \theta_{k,h}^\mathsf{HP} 
.
\end{align*}

Furthermore, if the Type \textsf{1} success probabilities satisfy
\begin{align}
\eta_j^\mathsf{1}  - \eta_{k^*-1}^\mathsf{1}   \leq \frac{\Delta_{k^*}}{j}\left(k^*-1-j\right)\left(\eta_{k^*-1}^\mathsf{1}  - \eta_{k^*}^\mathsf{1} \right)
,
\label{successProbAssump2}
\end{align}
for each $1 \leq j \leq \kappa$, where $\Delta_k = k \mu_k / (k \mu_k - (k-1)\mu_{k-1})$ and where $k^* \leq \kappa$ is such that $(k^*-1)c_0 \mathcal{S} \leq \Theta < k^* c_0 \mathcal{S}$, then $\boldsymbol{\theta}^\mathsf{HP}$ is an optimal solution to the heterogeneous static planning problem.
\end{proposition}

While we have certainly acknowledged that this analysis is stylized, we think that it identifies an important connection between closure and routing. This insight lies in the contrast of the two parts of this section, or, more specifically, in the contrast between Propositions~\ref{firstSPPprop} and~\ref{secondSPPprop} in light of Lemma~\ref{closureVarLemma}. That is, Lemma~\ref{closureVarLemma} and Proposition~\ref{firstSPPprop} together show that if systematic closure specifies the success probability exactly, then routing policies can treat assignment as equivalent to activity, and standard LL routing will be optimal.\footnote{\edit{Similarly, if all services close naturally, $\eta_k = 1$ for all $k$. This satisfies Equation~\eqref{successProbAssump}, and thus LL is optimal under exclusively natural closure. Unfortunately, our data suggests that this is not the case in this contact center.}} On the other hand, Lemma~\ref{closureVarLemma} \edit{and} Proposition~\ref{secondSPPprop} \edit{together show} that if \edit{the probability of success is} not precisely specified within each level, then the routing policy must address this heterogeneity explicitly, \edit{as done by} LL$+$HP in Section~\ref{sec:insights}.
\edit{This offers an explanation for the superior} performance of \edit{the} LL$+$HP \edit{policy} in our data-driven simulation experiments, \edit{because} data shows that 32\% of the conversations were closed by the system after exactly two hours of inactivity  (i.e., $\tau \not \in \mathcal{T}$). 

\section{Extensions, Discussion, and Conclusion}
\label{sec:conclusion}

As its title indicates, we believe the first-order contribution of this paper is the novel model of the service exchange. We have proposed a bivariate, marked Hawkes cluster model of the customer-agent service interaction. Across the application to data, new theory, data-driven experiments, and synthetic simulations, we hope to have demonstrated the model's practical value and research intrigue. We have also seen that the model and experiments reproduce \edit{behavioral} phenomena seen in both the empirical and theoretical literature on a more granular level. 
Because we have seen as well that the nature of the service success is a function of the closure policy, this work implies that the impact of systematic closure must not be overlooked, and that, on the contrary, closure and routing policies should be considered jointly in managerial strategy. Before concluding, we would like to discuss some limitations to this model, analysis, and experiments, and in turn propose possible extensions to address these.

Starting with the model, while Definition~\ref{modelDef} laid out the bivariate, marked cluster process in general terms, there are still assumptions to be relaxed. For instance, we have followed typical point process convention by assuming independence of the marks, but it is possible that this is too strong of a claim. That is, our conversation model has shown that there is value in recognizing the history dependence of the message timestamps, and there may indeed be value in modeling history dependence \edit{for} the sentiment and word count random variables as well. Auto-regressive models may offer a natural and tractable structure for this dependence among the countable random variable sequence. For a broader level of model extensions, we have maintained a modeling focus at the customer-agent interaction level, with the system-level effects coming in the form of the agent's concurrency, treated as \edit{a} deterministic function that dictates time-varying model parameters. It is of foremost interest to extend this model to the agent and system levels. An agent-level model could perhaps treat an agent with $K$ conversations as a $K+1$ dimensional multivariate Hawkes process, where there would be one agent correspondence rate and then one correspondence rate per customer. Similarly, one can imagine system-level models that incorporate variety in agent skills and in job types. Although this was not present in our data set, we are quite interested in extending to contexts when the impacts and decays vary across the combinations of skills and types.

We also see many opportunities for analysis of this model. Other than the stability condition in Theorem~\ref{stabilityThm}, all of the results we have proven here have centered around one of the model forms with an exponential excitation kernel. Analysis under more general kernels is an intriguing and important open direction. Similarly, one specific open problem is to extend the residual analysis in \edit{Theorem}~\ref{dialplotprop} beyond the univariate case. This result and the associated dial plot figures are built from the cluster decomposition from \citet{daw2023conditional}, which has likewise only been shown for the univariate setting. Replicating this for higher dimensions would enable a host of new questions to be addressed. There is also an opportunity to analyze routing directly in the Hawkes model, building from the results we have shown here in the stylized static planning problem setting setting. Even in the data-driven routing experiment, our policy primarily leveraged point forecasts, but the Hawkes service model should enable distributional forecasts as well. Similarly, the theoretical results on the effects of heterogeneity showed the importance of service closure as an operational decision, and analysis of this novel perspective may yield its own managerial insights.

\edit{The Hawkes cluster model could also allow one to consider the impact of the design of this co-produced service and the impact of that design on the system performance overall. 
Because our stochastic model distinguishes contributions from the customer and the agent, we could change the estimated parameters to reason about alternate service arrangements. 
Conceptually, this exploration would be closely related to models of service design \citep[e.g.,][]{Roels2014,bellos2019should,bellos2021service}, while also existing in a stochastic model of service embedded  in a queueing system.
Much like \citet{Roels2014}, we could explore a spectrum of co-production arrangements by altering the allocation of work between the customer and agent. 
Perturbing away from the data estimates would allow managers to evaluate system performance under different degrees of service co-production.}

We believe that there are also opportunities to translate this model into other service settings. In fact, on some level, tax preparation seems to be the quintessential co-produced service --- it was discussed by both~\citet{chase1978where} and~\citet{Roels2014}, and it occurred to these authors as well. One intriguing aspect of these firms in present-day is that they seem to offer products across the co-production spectrum, and this presents the opportunity to apply this model's lens to additional managerial decisions, such as pricing. \edit{Our framework can also potentially be used to model the company-customer relationship over time, as this is also a co-produced service. In this way, our proposed model may also contribute to the literature on customer lifetime duration and value predictions \citep[e.g.,][]{borle2008customer}. In that application, the $A_i^{x}$'s will be the timestamps of the company-customer interactions, such as purchases \citep{borle2008customer} or communications and engagements between the company  and their customers \citep{ascarza2018some}.} \edit{In practice, this could necessitate modeling and learning parameters on an individual customer-by-customer basis, and learning model parameters may also be valuable on an agent-by-agent basis in practice in some service settings as well \citep{ibrahim2016inter}.}
In general, we are quite interested in applying these models to other data sets where timestamps within service are available, as it may be that many different domains are well-suited to this history-dependent framework.

\printendnotes








\bibliographystyle{informs2014} 
\bibliography{Bibliography.bib} 

%
\begin{APPENDIX}{}

 \section{Proofs, Auxiliary Results, and Technical Lemmas}\label{app:proofs}


\subsection{\edit{Infinitesimal Generator for the Markovian Representation of the Process}}
 
There are four different decay rates in the BHP model form or in the SysBHP for any fixed value of the concurrency, so to analyze  the Hawkes model with Markov process tools we need to treat the correspondence rates as four sub-processes: $\lambda_t^\mathsf{c,c}$, $\lambda_t^\mathsf{c,a}$, $\lambda_t^\mathsf{a,c}$, and $\lambda_t^\mathsf{a,a}$. We will focus on the SysBHP in the following presentation of statements, but this can naturally reduced to the other two model forms by taking $\PP{S_1^x = W_1^x = 1} = 1$ for each $x \in \{\mathsf{c},\mathsf{a}\}$ for the BHP and by additionally assuming $\alpha^{x,y} = \alpha \slash 2$ and $\beta^{x,y} = \beta$ for every $x, y \in \{\mathsf{c},\mathsf{a}\}$ (and fixed $K_t$). Because the word and sentiment sequences are independent from the correspondence processes and because the concurrency is known, the infinitesimal generator can be expressed as follows in Lemma~\ref{dynkins}.

\begin{lemma}[Dynkin's Formula]\label{dynkins} Suppose that $t \in [t_0, t_1)$ for $t_0 < t_1$ in which $K_t$ is unchanged throughout the interval and $\lambda_{t_0}^\mathsf{c,c}$, $\lambda_{t_0}^\mathsf{c,a}$, $\lambda_{t_0}^\mathsf{a,c}$, and $\lambda_{t_0}^\mathsf{a,a}$ are known. Let $\boldsymbol{\lambda}_t = [\lambda_t^\mathsf{c,c}, \lambda_t^\mathsf{c,a}, \lambda_t^\mathsf{a,c}, \lambda_t^\mathsf{a,a}]^\mathsf{T}$ and $\mathbf{N}_t = [N_t^\mathsf{c,c}, N_t^\mathsf{c,a}, N_t^\mathsf{a,c}, N_t^\mathsf{a,a}]^\mathsf{T}$ Then, for a sufficiently regular function $h:\mathbb{R}_+^4 \times \mathbb{N}^4 \to \mathbb{R}$, 
 $$
 \frac{\mathrm{d}}{\mathrm{d}t}\E{h(\boldsymbol{\lambda}_t, \mathbf{N}_t)}
 =
 \E{\mathcal{L} h(\boldsymbol{\lambda}_t, \mathbf{N}_t)}
$$
 where 
 \begin{small}
 \begin{align*}
\mathcal{L} h(\boldsymbol{\lambda}_t, \mathbf{N}_t)
&=
\lambda_t^\mathsf{c,c} 
\mathrm{E}_{S^\mathsf{c},W^\mathsf{c}}\left[
h\left(
\boldsymbol{\lambda}_t + [\alpha^\mathsf{c,c}_\mathsf{1} S^\mathsf{c} + \alpha^\mathsf{c,c}_\mathsf{2}W^\mathsf{c},\, 0,\, \alpha^\mathsf{a,c}_\mathsf{1} S^\mathsf{c} + \alpha^\mathsf{a,c}_\mathsf{2}W^\mathsf{c},\, 0]^\mathsf{T}
, 
\mathbf{N}_t + [1,\, 0,\, 0,\, 0]^\mathsf{T}
\right)
-
h(\boldsymbol{\lambda}_t, \mathbf{N}_t)
\right]
\nonumber
\\
&
+
\lambda_t^\mathsf{c,a} 
\mathrm{E}_{S^\mathsf{c},W^\mathsf{c}}\left[
h\left(
\boldsymbol{\lambda}_t + [\alpha^\mathsf{c,c}_\mathsf{1} S^\mathsf{c} + \alpha^\mathsf{c,c}_\mathsf{2}W^\mathsf{c},\, 0,\, \alpha^\mathsf{a,c}_\mathsf{1} S^\mathsf{c} + \alpha^\mathsf{a,c}_\mathsf{2}W^\mathsf{c},\, 0]^\mathsf{T}
, 
\mathbf{N}_t + [0,\, 1,\, 0,\, 0]^\mathsf{T}
\right)
-
h(\boldsymbol{\lambda}_t, \mathbf{N}_t)
\right]
\nonumber
\\
&
+
\lambda_t^\mathsf{a,c} 
\mathrm{E}_{S^\mathsf{a},W^\mathsf{a}}\left[
h\left(
\boldsymbol{\lambda}_t + \left[0, \frac{\alpha^\mathsf{c,a}_\mathsf{1} S^\mathsf{a} + \alpha^\mathsf{c,a}_\mathsf{2}W^\mathsf{a}}{K_t},\, 0,\, \frac{\alpha^\mathsf{a,a}_\mathsf{1} S^\mathsf{a} + \alpha^\mathsf{a,a}_\mathsf{2}W^\mathsf{a}}{K_t}\right]^\mathsf{T}
, 
\mathbf{N}_t + [0,\, 0,\, 1,\, 0]^\mathsf{T}
\right)
-
h(\boldsymbol{\lambda}_t, \mathbf{N}_t)
\right]
\nonumber
\\
&
+
\lambda_t^\mathsf{c,a} 
\mathrm{E}_{S^\mathsf{c},W^\mathsf{c}}\left[
h\left(
\boldsymbol{\lambda}_t + \left[0, \frac{\alpha^\mathsf{c,a}_\mathsf{1} S^\mathsf{a} + \alpha^\mathsf{c,a}_\mathsf{2}W^\mathsf{a}}{K_t},\, 0,\, \frac{\alpha^\mathsf{a,a}_\mathsf{1} S^\mathsf{a} + \alpha^\mathsf{a,a}_\mathsf{2}W^\mathsf{a}}{K_t}\right]^\mathsf{T}
, 
\mathbf{N}_t + [0,\, 0,\, 0,\, 1]^\mathsf{T}
\right)
-
h(\boldsymbol{\lambda}_t, \mathbf{N}_t)
\right]
\nonumber
\\
&
-
\beta^\mathsf{c,c} \lambda_t^\mathsf{c,c}h_{\partial 1}(\boldsymbol{\lambda}_t, \mathbf{N}_t)
-
\beta^\mathsf{c,a} \lambda_t^\mathsf{c,a}h_{\partial 2}(\boldsymbol{\lambda}_t, \mathbf{N}_t)
-
\frac{\beta^\mathsf{a,c}}{K_t} \lambda_t^\mathsf{a,c}h_{\partial 3}(\boldsymbol{\lambda}_t, \mathbf{N}_t)
-
\frac{\beta^\mathsf{a,a}}{K_t} \lambda_t^\mathsf{a,a}h_{\partial 4}(\boldsymbol{\lambda}_t, \mathbf{N}_t)
,
\end{align*}
\end{small}
with $h_{\delta i}(\cdot)$ as the partial derivative of $h(\cdot)$ with respect to the $i^\text{th}$ coordinate and where $\mathrm{E}_{X,Y}[\cdot]$ is the expectation taken relative to the random variables $X$ and $Y$.
 \end{lemma}
  \proof{Proof.}
  See, e.g., Section 5 of \cite{davis1984piecewise}, and Theorem 5.5 therein for the full regularity conditions on the function $h(\cdot)$.
 \hfill \Halmos 
\endproof
  
As we have discussed, the SysBHP is not a Markov process in general for a state variable vector $[\lambda_t^\mathsf{c,c}, \lambda_t^\mathsf{c,a}, \lambda_t^\mathsf{a,c}, \lambda_t^\mathsf{a,a}]$. This is because updates to the concurrency require the conversation's full history in order to update the agent correspondence rate. In the next subsection, we show that the SysBHP will indeed be Markovian for a larger state space. Furthermore, the Markovian nature of the BHP implies that the SysBHP vector $[\lambda_t^\mathsf{c,c}, \lambda_t^\mathsf{c,a}, \lambda_t^\mathsf{a,c}, \lambda_t^\mathsf{a,a}]$ will be Markovian on intervals on which the concurrency is not changed. As for the regularity conditions on the function $h(\cdot)$, in this paper we are only concerned with the mean. Hence, $h(\cdot)$ is the identity function for one of the sub-processes, and such a function is compactly supported.

\subsection{Existence, Uniqueness, and Stability of the SysBHP}\label{sysMarkov}

While the exponential decay functions of the UHP and BHP model forms will immediately imply that the stochastic process is Markovian when tracking a state vector containing each sub-correspondence rate with a unique decay rate \citep{oakes1975markovian}, this is not true for the SysBHP. While Lemma~\ref{dynkins} used that the four SysBHP conversation-direction correspondence rates are collectively a Markov process on intervals where the concurrency is not changed (because the process is functionally equivalent to the BHP in such cases), this is not true when the concurrency changes. In particular, upon the update of concurrency, the full history of the conversation's epochs, sentiment scores, and word counts would be needed to compute $\lambda_t^{a,y} = \sum_{\ell: A_\ell^y \leq t} (\alpha^{\mathsf{a},y}_\mathsf{1} S_\ell^y + \alpha^{\mathsf{a},y}_\mathsf{2} W_\ell^y) \slash K_t e^{-\beta^{\mathsf{a},y} (t - A_\ell^y) \slash K_{t}}$ for each $y \in \{\mathsf{c},\mathsf{a}\}$, since the change in $K_t$ is discontinuous.

However, it is not necessary to maintain the full history of the conversation when analyzing or simulating the SysBHP; the model form can enjoy the Markov property simply by expanding the state space to include more correspondence rates. Recalling that $K_t \leq \kappa$ for some maximal concurrency $\kappa \in \mathbb{Z}_+$, we actually only need to maintain copies of \emph{possible} correspondence rates for each possible value of the concurrency. In fact, this only needs to be done on the agent side. That is, let us define $\lambda_{t,k}^{a,y} = \sum_{\ell: A_\ell^y \leq t} (\alpha^{\mathsf{a},y}_\mathsf{1} S_\ell^y + \alpha^{\mathsf{a},y}_\mathsf{2} W_\ell^y) \slash k e^{-\beta^{\mathsf{a},y} (t - A_\ell^y) \slash k}$ for each $y \in \{\mathsf{c},\mathsf{a}\}$ and for every $k \in \{1, \dots, \kappa\}$. Together with the customer-side correspondence rates $\lambda_t^\mathsf{c,c}$ and $\lambda_t^\mathsf{c,a}$ (defined as usual), these $\kappa$ copies of the agent-side correspondence rates are sufficient to fully describe the dynamics of the process. That is, let $\lambda_t^\mathsf{c,c}$, $\lambda_t^\mathsf{c,a}$, $\lambda_{t,K_t}^\mathsf{a,c}$, and $\lambda_{t,K_t}^\mathsf{a,a}$ be the intensities that generate new epochs in the process, but at every new epoch all $\lambda_{t,k}^\mathsf{a,c}$ and $\lambda_{t,k}^\mathsf{a,a}$ for every $k$ should be updated along with $\lambda_t^\mathsf{c,c}$ and $\lambda_t^\mathsf{c,a}$. In this way, the state variable vector $[\lambda_t^\mathsf{c,c}, \lambda_t^\mathsf{c,a}, \lambda_{t,1}^\mathsf{a,c}, \lambda_{t,1}^\mathsf{a,a}, \dots, \lambda_{t,\kappa}^\mathsf{a,c}, \lambda_{t,\kappa}^\mathsf{a,a}]$ will be a Markov process. All coordinates will decay at their respective rates, even-indexed coordinates will jump upon a customer message, and odd-indexed coordinates will jump upon an agent message. Customer messages will be driven by $\lambda_t^\mathsf{c,c} + \lambda_{t,K_t}^\mathsf{c,a}$, and agent messages likewise by $\lambda_t^\mathsf{a,c} + \lambda_{t,K_t}^\mathsf{a,a}$. In fact, we can notice that this Markovian construction requires only the present value of $K_t$, and this hints at an associated Markov property for a full queueing system of SysBHP services if the arrival process is also Markov. In that case, the concurrency would be modeled as a stochastic process itself, rather than the deterministic function form we have assumed in this paper.

 \subsection{Proof of Theorem~\ref{stabilityThm}}
 
 \proof{Proof.}
From Theorem 5 of \cite{massoulie1998stability} (or Proposition 2.3 of \cite{abergel2015long} for the former applied to a relevant specific case), the BHP model will be stable if the spectral radius of the ratio matrix 
$$
\mathbf{R} 
=
\begin{bmatrix}
\frac{\alpha^{\mathsf{c,c}}}{\beta^{\mathsf{c,c}}} & \frac{\alpha^{\mathsf{c,a}}}{\beta^{\mathsf{c,a}}} \vspace{.05in}\\ 
\frac{\alpha^{\mathsf{a,c}}}{\beta^{\mathsf{a,c}}} & \frac{\alpha^{\mathsf{a,a}}}{\beta^{\mathsf{a,a}}}
\end{bmatrix}
$$
is strictly less than 1, meaning that both eigenvalues of $\mathbf{R}$ are less than 1 in absolute value. We can note that the eigenvalues of $\mathbf{R}$ will both be strictly less than 1 if and only if the eigenvalues of $\mathbf{R}-\mathbf{I}$ are both less than 0, and likewise, the eigenvalues of $\mathbf{R}$ will both be strictly greater than $-1$ if and only if the eigenvalues of $\mathbf{R}+\mathbf{I}$ are both strictly positive. Now, $\mathsf{det}(\mathbf{R}-\mathbf{I}) = (1 - \frac{\alpha^\mathsf{c,c}}{\beta^\mathsf{c,c}})(1 - \frac{\alpha^\mathsf{a,a}}{\beta^\mathsf{a,a}}) - \frac{\alpha^\mathsf{c,a}}{\beta^\mathsf{c,a}}\frac{\alpha^\mathsf{a,c}}{\beta^\mathsf{a,c}}$ and $\mathsf{trace}(\mathbf{R}-\mathbf{I}) = \frac{\alpha^\mathsf{c,c}}{\beta^\mathsf{c,c}} + \frac{\alpha^\mathsf{a,a}}{\beta^\mathsf{a,a}} - 2$. If
\begin{align}
\frac{\alpha^{\mathsf{c,a}}}{\beta^{\mathsf{c,a}}}\frac{\alpha^{\mathsf{a,c}}}{\beta^{\mathsf{a,c}}}
<
\left(1 - \frac{\alpha^{\mathsf{c,c}}}{\beta^{\mathsf{c,c}}}\right)\left(1 - \frac{\alpha^{\mathsf{a,a}}}{\beta^{\mathsf{a,a}}}\right)
,
\label{bhpStability}
\end{align}
then the determinant of $\mathbf{R} - \mathbf{I}$ will of course be positive, but\edit{, together with the fact that $\bar\alpha^\mathsf{c,c} < \beta^\mathsf{c,c}$ and $\bar\alpha^\mathsf{a,a} < \beta^\mathsf{a,a}$,} \eqref{bhpStability} also implies that the trace will be negative since every $\alpha$ and $\beta$ must be positive. This similarly implies that both the determinant and the trace of $\mathbf{R} + \mathbf{I}$ are positive, and hence \eqref{bhpStability} implies that the bivariate (but not necessarily marked or concurrency dependent) model exists and is stable.

To now extend this to full generality, let us construct an alternate bivariate model (abbreviated ABHP) in which the instantaneous impact in the party $x \in \{\mathsf{c},\mathsf{a}\}$ correspondence rate upon receipt of the $i$th message from party $y \in \{\mathsf{c},\mathsf{a}\}$ is $\bar \alpha^{x,y}$, as opposed to $f^{x,y}(i)$. Likewise, let the ABHP be defined so that the jump size in the ABHP agent correspondence rate is $\bar \alpha^{\mathsf{a},j} \slash K_t$. Let $\tilde{\lambda}_t^\mathsf{c}$ and $\tilde{\lambda}_t^\mathsf{a}$ respectively be the customer and agent correspondence rates for this alternate model, and assume that the ABHP decay functions are the same as in Definition~\ref{modelDef}. Then, the proceeding arguments yield that \eqref{SysBHPStabilityEq} implies that the ABHP is stable. Now, let us suppose further that the Hawkes service model and the ABHP are defined with the same initial values. The processes
$$
\lambda_t^\mathsf{c} - \tilde \lambda_t^\mathsf{c}
\qquad
\text{and}
\qquad
\lambda_t^\mathsf{a} - \tilde \lambda_t^\mathsf{a}
$$
are then zero-mean martingales. Therefore, Doob's martingale convergence theorem yields that since the ABHP is stable given \eqref{SysBHPStabilityEq}, the Hawkes service model is as well. Note that the preceding arguments apply regardless of the value of $K_t$, and this is reflected in the fact  \eqref{SysBHPStabilityEq} does not depend on $K_t$. Equivalently stated, the Hawkes service model will be stable for every possible concurrency value. Because there are almost surely finitely many switches of concurrency, the stability within each value immediately extends to the model overall.

Finally, to show the almost sure finitude of the conversation, let us first identify the limiting mean of the ABHP. From Equation (21) of Hawkes' original work \citep{hawkes1971spectra}, a stationary bivariate Hawkes process with ratio matrix $\mathbf{R}$ (as defined at the beginning of the proof) and baseline intensity vector $\boldsymbol{\nu} \in \mathbb{R}_+^2$ will have steady-state mean intensity equal to $(\mathbf{I} - \mathbf{R})^{-1} \boldsymbol{\nu}$. Since the cluster model has no baseline intensity, we have $\boldsymbol{\nu} = 0$. Thus, $\lim_{t \to \infty} \E{\tilde \lambda_t^\mathsf{c}} = \lim_{t \to \infty} \E{\tilde \lambda_t^\mathsf{a}} = 0$. By construction, $\E{\lambda_t^\mathsf{c}} = \E{\tilde \lambda_t^\mathsf{c}}$ and $\E{\lambda_t^\mathsf{a}} = \E{\tilde \lambda_t^\mathsf{a}}$, so we also have that $\lim_{t \to \infty} \E{ \lambda_t^\mathsf{c}} = \lim_{t \to \infty} \E{ \lambda_t^\mathsf{a}} = 0$. Now, because $\lambda_t^\mathsf{c}$ and $\lambda_t^\mathsf{a}$ are non-negative almost surely, the zero mean implies that $\lim_{t \to \infty} \lambda_t^\mathsf{c} = 0$ 
and
$\lim_{t \to \infty} \lambda_t^\mathsf{a} = 0$ almost surely as well. Since these correspondence rates are the intensities of the counting process for the total number of messages $N_t$, this further implies that with probability 1 there will be only finitely many messages, thus completing the proof.
\hfill \Halmos 
\endproof

To the best of our knowledge, \eqref{bhpStability} is the first closed form stability condition for (marked) bivariate Hawkes processes. Not only can Theorem~\ref{stabilityThm} simplify to the well-known univariate Hawkes process stability condition $\alpha < \beta$, let us note that it also generalizes previously stated stability conditions of bivariate Hawkes processes with more restrictive assumptions. For example, Equation (7) of \citet{bacry2015hawkes} is equivalent to \eqref{SysBHPStabilityEq} above for a bivariate model in which $\alpha^\mathsf{c,c}\slash\beta^\mathsf{c,c} = \alpha^\mathsf{a,a}\slash\beta^\mathsf{a,a}$ and $\alpha^\mathsf{c,a}\slash\beta^\mathsf{c,a} = \alpha^\mathsf{a,c}\slash\beta^\mathsf{a,c}$.

\subsection{Proof of Proposition~\ref{totalMsgPropShort}}

In addition to computing the expected total number of messages as stated in Proposition~\ref{totalMsgPropShort}, we will also find the number of these that are sent by the customer and by the agent. Hence, we prove Proposition~\ref{totalMsgPropShort} by way of proving the following broader claim.

\begin{proposition}\label{totalMsgProp}
Excluding the total number of messages already sent in the observation period up to time $t_0 \geq 0$, the expected number of remaining messages the customer will send is 
\begin{align*}
\E{N_\infty^\mathsf{c} - N_{t_0}^\mathsf{c} \mid \boldsymbol{\lambda}_{t_0}}
&=
\frac{
\left(
1 - \frac{\bar\alpha^\mathsf{a,a}}{\beta^\mathsf{a,a}}
\right)
\left(
\frac{\lambda_{t_0}^{\mathsf{c,c}}}{\beta^{\mathsf{c,c}}} 
+
\frac{\lambda_{t_0}^{\mathsf{c,a}}}{\beta^{\mathsf{c,a}}} 
\right)
+
\frac{\bar\alpha^{\mathsf{c,a}}}{\beta^{\mathsf{c,a}}}
\left(
\frac{\bar \lambda_{t_0}^{\mathsf{a,a}}}{\beta^{\mathsf{a,a}}} 
+
\frac{\bar \lambda_{t_0}^{\mathsf{a,c}}}{\beta^{\mathsf{a,c}}} 
\right)
}{
\left(
1 
- 
\frac{\bar\alpha^\mathsf{c,c}}{\beta^\mathsf{c,c}}
\right)
\left(
1 - \frac{\bar\alpha^\mathsf{a,a}}{\beta^\mathsf{a,a}}
\right)
-
\frac{\bar\alpha^{\mathsf{c,a}}}{\beta^{\mathsf{c,a}}}
\frac{\bar\alpha^{\mathsf{a,c}}}{\beta^{\mathsf{a,c}}}
}
,
\end{align*}
and the total expected number of remaining agent messages is 
\begin{align*}
\E{N_\infty^\mathsf{a} - N_{t_0}^\mathsf{a} \mid \boldsymbol{\lambda}_{t_0}}
&=
\frac{
\frac{\bar\alpha^{\mathsf{a,c}}}{\beta^{\mathsf{a,c}}}
\left(
\frac{\lambda_{t_0}^{\mathsf{c,c}}}{\beta^{\mathsf{c,c}}} 
+
\frac{\lambda_{t_0}^{\mathsf{c,a}}}{\beta^{\mathsf{c,a}}} 
\right)
+
\left(
1 
- 
\frac{\bar\alpha^\mathsf{c,c}}{\beta^\mathsf{c,c}}
\right)
\left(
\frac{\bar \lambda_{t_0}^{\mathsf{a,a}}}{\beta^{\mathsf{a,a}}} 
+
\frac{\bar \lambda_{t_0}^{\mathsf{a,c}}}{\beta^{\mathsf{a,c}}} 
\right)
}{
\left(
1 
- 
\frac{\bar\alpha^\mathsf{c,c}}{\beta^\mathsf{c,c}}
\right)
\left(
1 - \frac{\bar\alpha^\mathsf{a,a}}{\beta^\mathsf{a,a}}
\right)
-
\frac{\bar\alpha^{\mathsf{c,a}}}{\beta^{\mathsf{c,a}}}
\frac{\bar\alpha^{\mathsf{a,c}}}{\beta^{\mathsf{a,c}}}
}
,
\end{align*}
hence the total expected number of messages from the present until the end of the conversation is 
\begin{align*}
\E{N_\infty - N_{t_0} \mid \boldsymbol{\lambda}_{t_0}}
&=
\frac{
\left(
1 + \frac{\bar \alpha^\mathsf{a,c}}{\beta^\mathsf{a,c}} - \frac{\bar\alpha^\mathsf{a,a}}{\beta^\mathsf{a,a}}
\right)
\left(
\frac{\lambda_{t_0}^{\mathsf{c,c}}}{\beta^{\mathsf{c,c}}} 
+
\frac{\lambda_{t_0}^{\mathsf{c,a}}}{\beta^{\mathsf{c,a}}} 
\right)
+
\left(
1
+
\frac{\bar\alpha^{\mathsf{c,a}}}{\beta^{\mathsf{c,a}}}
-
\frac{\bar \alpha^\mathsf{c,c}}{\beta^\mathsf{c,c}}
\right)
\left(
\frac{\bar \lambda_{t_0}^{\mathsf{a,a}}}{\beta^{\mathsf{a,a}}} 
+
\frac{\bar \lambda_{t_0}^{\mathsf{a,c}}}{\beta^{\mathsf{a,c}}} 
\right)
}{
\left(
1 
- 
\frac{\bar\alpha^\mathsf{c,c}}{\beta^\mathsf{c,c}}
\right)
\left(
1 - \frac{\bar\alpha^\mathsf{a,a}}{\beta^\mathsf{a,a}}
\right)
-
\frac{\bar\alpha^{\mathsf{c,a}}}{\beta^{\mathsf{c,a}}}
\frac{\bar\alpha^{\mathsf{a,c}}}{\beta^{\mathsf{a,c}}}
}
,
\end{align*}
where $\bar \lambda_{t_0}^{\mathsf{a},y} = \sum_{\ell: A_\ell^y \leq t_\naught}(\alpha^{\mathsf{a},y}_\mathsf{1} S_\ell^y + \alpha^{\mathsf{a},y}_\mathsf{2} W_\ell^y) e^{-\beta^{\mathsf{a},y} (t_\naught - A_\ell^y) \slash K_{t_\naught}}$ for $y \in \{\mathsf{c},\mathsf{a}\}$.
\end{proposition}
\proof{Proof.}
As $t \to \infty$ in \eqref{nVecSol}, we can observe that
$$
\mathbf{M}\E{\mathbf{N}_\infty - \mathbf{N}_{t_0} \mid \boldsymbol{\lambda}_{t_0}} = - \boldsymbol{\lambda}_{t_0}
,
$$
So, for each $i,j \in \{\mathsf{c},\mathsf{a}\}$, this implies that
$$
\E{N_\infty^{i,j} - N_{t_0}^{i,j} \mid \boldsymbol{\lambda}_{t_0}}
=
\frac{\lambda_0^{i,j}}{\beta^{i,j}} 
+
\frac{\bar\alpha^{i,j}}{\beta^{i,j}}
\left(
\E{N_\infty^{j,\mathsf{c}} - N_{t_0}^{j,\mathsf{c}} \mid \boldsymbol{\lambda}_{t_0}}
+
\E{N_\infty^{j,\mathsf{a}} - N_{t_0}^{j,\mathsf{a}} \mid \boldsymbol{\lambda}_{t_0}}
\right)
.
$$
By noting that the total number of messages sent by one party is the sum of the number they send to the other party and the number they send in follow-up to their own writings, i.e.,\ $\E{N_\infty^i - N_{t_0}^{i} \mid \boldsymbol{\lambda}_{t_0}}
=
\E{N_\infty^{i,\mathsf{c}} - N_{t_0}^{i,\mathsf{c}} \mid \boldsymbol{\lambda}_{t_0}}
+
\E{N_\infty^{i,\mathsf{a}} - N_{t_0}^{i,\mathsf{a}} \mid \boldsymbol{\lambda}_{t_0}}$, this can be re-expressed as
\begin{align*}
\E{N_\infty^i - N_{t_0}^{i} \mid \boldsymbol{\lambda}_{t_0}}
&
=
\frac{\lambda_0^{i,\mathsf{c}}}{\beta^{i,\mathsf{c}}} 
+
\frac{\lambda_0^{i,\mathsf{a}}}{\beta^{i,\mathsf{a}}} 
+
\frac{\bar \alpha^{i,\mathsf{c}}}{\beta^{i,\mathsf{c}}}\E{N_\infty^{\mathsf{c}} - N_{t_0}^\mathsf{c} \mid \boldsymbol{\lambda}_{t_0}}
+
\frac{\bar \alpha^{i,\mathsf{a}}}{\beta^{i,\mathsf{a}}}\E{N_\infty^{\mathsf{a}} - N_{t_0}^\mathsf{a} \mid \boldsymbol{\lambda}_{t_0}}
,
\end{align*}
or simply
$$
\left(1 - \frac{\bar \alpha^\mathsf{c,c}}{\beta^\mathsf{c,c}}\right)\E{N_\infty^\mathsf{c} - N_{t_0}^\mathsf{c} \mid \boldsymbol{\lambda}_{t_0}}
=
\frac{\lambda_0^{\mathsf{c,c}}}{\beta^{\mathsf{c,c}}} 
+
\frac{\lambda_0^{\mathsf{c,a}}}{\beta^{\mathsf{c,a}}} 
+
\frac{\bar \alpha^{\mathsf{c,a}}}{\beta^{\mathsf{c,a}}}\E{N_\infty^{\mathsf{a}} - N_{t_0}^\mathsf{a} \mid \boldsymbol{\lambda}_{t_0}}
,
$$
and
$$
\left(1 - \frac{\bar \alpha^\mathsf{a,a}}{\beta^\mathsf{a,a}}\right)\E{N_\infty^\mathsf{a}- N_{t_0}^\mathsf{a} \mid \boldsymbol{\lambda}_{t_0}}
=
\frac{\bar \lambda_0^{\mathsf{a,a}}}{\beta^{\mathsf{a,a}}} 
+
\frac{\bar \lambda_0^{\mathsf{a,c}}}{\beta^{\mathsf{a,c}}} 
+
\frac{\bar \alpha^{\mathsf{a,c}}}{\beta^{\mathsf{a,c}}}\E{N_\infty^{\mathsf{c}}- N_{t_0}^\mathsf{c} \mid \boldsymbol{\lambda}_{t_0}}
.
$$
Now, substituting the agent's mean message count into the customer equation, we have
\begin{small}
\begin{align*}
\left(
\left(
1 
- 
\frac{\bar\alpha^\mathsf{c,c}}{\beta^\mathsf{c,c}}
\right)
\left(
1 - \frac{\bar\alpha^\mathsf{a,a}}{\beta^\mathsf{a,a}}
\right)
-
\frac{\bar\alpha^{\mathsf{c,a}}}{\beta^{\mathsf{c,a}}}
\frac{\bar\alpha^{\mathsf{a,c}}}{\beta^{\mathsf{a,c}}}
\right)\E{N_\infty^\mathsf{c}- N_{t_0}^\mathsf{c} \mid \boldsymbol{\lambda}_{t_0}}
&=
\left(
\frac{\lambda_0^{\mathsf{c,c}}}{\beta^{\mathsf{c,c}}} 
+
\frac{\lambda_0^{\mathsf{c,a}}}{\beta^{\mathsf{c,a}}} 
\right)
\left(
1 - \frac{\bar\alpha^\mathsf{a,a}}{\beta^\mathsf{a,a}}
\right)
+
\frac{\bar\alpha^{\mathsf{c,a}}}{\beta^{\mathsf{c,a}}}
\left(
\frac{\bar \lambda_0^{\mathsf{a,a}}}{\beta^{\mathsf{a,a}}} 
+
\frac{\bar \lambda_0^{\mathsf{a,c}}}{\beta^{\mathsf{a,c}}} 
\right)
,
\end{align*}
\end{small}
yielding
\begin{align*}
\E{N_\infty^\mathsf{c}- N_{t_0}^\mathsf{c} \mid \boldsymbol{\lambda}_{t_0}}
&=
\frac{
\left(
1 - \frac{\bar\alpha^\mathsf{a,a}}{\beta^\mathsf{a,a}}
\right)
\left(
\frac{\lambda_0^{\mathsf{c,c}}}{\beta^{\mathsf{c,c}}} 
+
\frac{\lambda_0^{\mathsf{c,a}}}{\beta^{\mathsf{c,a}}} 
\right)
+
\frac{\bar\alpha^{\mathsf{c,a}}}{\beta^{\mathsf{c,a}}}
\left(
\frac{\bar \lambda_0^{\mathsf{a,a}}}{\beta^{\mathsf{a,a}}} 
+
\frac{\bar \lambda_0^{\mathsf{a,c}}}{\beta^{\mathsf{a,c}}} 
\right)
}{
\left(
1 
- 
\frac{\bar\alpha^\mathsf{c,c}}{\beta^\mathsf{c,c}}
\right)
\left(
1 - \frac{\bar\alpha^\mathsf{a,a}}{\beta^\mathsf{a,a}}
\right)
-
\frac{\bar\alpha^{\mathsf{c,a}}}{\beta^{\mathsf{c,a}}}
\frac{\bar\alpha^{\mathsf{a,c}}}{\beta^{\mathsf{a,c}}}
}
,
\end{align*}
and
\begin{align*}
\E{N_\infty^\mathsf{a}- N_{t_0}^\mathsf{a} \mid \boldsymbol{\lambda}_{t_0}}
&=
\frac{
\frac{\bar\alpha^{\mathsf{a,c}}}{\beta^{\mathsf{a,c}}}
\left(
\frac{\lambda_0^{\mathsf{c,c}}}{\beta^{\mathsf{c,c}}} 
+
\frac{\lambda_0^{\mathsf{c,a}}}{\beta^{\mathsf{c,a}}} 
\right)
+
\left(
1 
- 
\frac{\bar\alpha^\mathsf{c,c}}{\beta^\mathsf{c,c}}
\right)
\left(
\frac{\bar \lambda_0^{\mathsf{a,a}}}{\beta^{\mathsf{a,a}}} 
+
\frac{\bar \lambda_0^{\mathsf{a,c}}}{\beta^{\mathsf{a,c}}} 
\right)
}{
\left(
1 
- 
\frac{\bar\alpha^\mathsf{c,c}}{\beta^\mathsf{c,c}}
\right)
\left(
1 - \frac{\bar\alpha^\mathsf{a,a}}{\beta^\mathsf{a,a}}
\right)
-
\frac{\bar\alpha^{\mathsf{c,a}}}{\beta^{\mathsf{c,a}}}
\frac{\bar\alpha^{\mathsf{a,c}}}{\beta^{\mathsf{a,c}}}
}
.
\end{align*}
By adding and simplifying these two expressions, we achieve the stated result for the total mean number of messages.
\hfill \Halmos
\endproof

\subsection{Proof of \edit{Theorem}~\ref{dialplotprop}}

\proof{Proof.}
From Theorem 1 of \citet{daw2023conditional}, the joint distribution of the compensator points defined in~\eqref{compEq} is equivalent to that of increasingly sorted differences between a uniformly random parking function and mutually independent standard uniform random variables. That is, $\Lambda_i = (\pi - U)_{(i)}$ for each $1 \leq i \leq N-1$ with $\pi \in \mathbb{Z}_+^{N-1}$ as a uniformly random parking function of length $N-1$ and $U_i \stackrel{\mathsf{iid}}{\sim} \mathsf{Uni}(0,1)$. Because the components of the parking function are integers and because the uniform random variables are contained on $(0,1)$, we can recover the underlying parking function through the ceiling of the compensator points: $\pi_{(i)} = \lceil \Lambda_i \rceil$. Furthermore, all compensator points that share an integer interval will share a parking function value, and thus for every $1 \leq \ell \leq N-1$ the joint distribution of $\Lambda_i - (\ell-1)$ for $i \in \mathcal{I}_\ell$ will be equivalent to the order statistics of $|\mathcal{I}_\ell|$ independent standard uniform random variables. Additionally, these random variables are independent across different values of $\ell$. Finally, we can recognize that this implies that the marginal distribution of $\Lambda_i - \lfloor \Lambda_i \rfloor$ are that of a standard uniform random variable for every $i$, and, moreover, upon randomly shuffling so as to remove the ordering, the random variables become mutually independent and identical. 
\hfill\Halmos
\endproof

 \subsection{Proof of Proposition~\ref{transMean}}
 
 \proof{Proof.}
Because $\E{S_1^\mathsf{c}} = \E{W_1^\mathsf{c}} = \E{S_1^\mathsf{a}} = \E{W_1^\mathsf{a}} = 1$, Lemma~\ref{dynkins} yields that the means of the SysBHP sub-processes satisfy the following system of ordinary differential equations:
 \begin{align*}
\frac{\mathrm{d}}{\mathrm{d}t} \E{\lambda_t^\mathsf{c,c}}
&=
\bar \alpha^\mathsf{c,c} \left(\E{\lambda_t^\mathsf{c,c}} + \E{\lambda_t^\mathsf{c,a}}\right)
-
\beta^\mathsf{c,c}\E{\lambda_t^\mathsf{c,c}}
,
\\
\frac{\mathrm{d}}{\mathrm{d}t} \E{\lambda_t^\mathsf{c,a}}
&=
\bar \alpha^\mathsf{c,a} \left(\E{\lambda_t^\mathsf{a,c}} + \E{\lambda_t^\mathsf{a,a}}\right)
-
\beta^\mathsf{c,a}\E{\lambda_t^\mathsf{c,a}}
,
\\
\frac{\mathrm{d}}{\mathrm{d}t} \E{\lambda_t^\mathsf{a,c}}
&=
\frac{\bar \alpha^\mathsf{a,c}}{K_t} \left(\E{\lambda_t^\mathsf{c,c}} + \E{\lambda_t^\mathsf{c,a}}\right)
-
\frac{\beta^\mathsf{a,c}}{K_t}\E{\lambda_t^\mathsf{a,c}}
,
\\
\frac{\mathrm{d}}{\mathrm{d}t} \E{\lambda_t^\mathsf{a,a}}
&=
\frac{\bar \alpha^\mathsf{a,a}}{K_t} \left(\E{\lambda_t^\mathsf{a,c}} + \E{\lambda_t^\mathsf{a,a}}\right)
-
\frac{\beta^\mathsf{a,a}}{K_t}\E{\lambda_t^\mathsf{a,a}}
.
 \end{align*}
 Letting $\boldsymbol{\lambda}_t = [ \lambda_t^\mathsf{c,c}, \, \lambda_t^\mathsf{c,a}, \, \lambda_t^\mathsf{a,c}, \, \lambda_t^\mathsf{a,a} ]^\mathrm{T}$, this pattern of the source of jumps and self decay gives rise to the linear system
\begin{align*}
\frac{\mathrm{d}}{\mathrm{d}t}
\E{\boldsymbol{\lambda}_t}
&=
\left(
\begin{bmatrix}
\bar \alpha^\mathsf{c,c}  & \bar \alpha^\mathsf{c,c} & & \\
 &  & \bar \alpha^\mathsf{c,a}   & \bar \alpha^\mathsf{c,a}  \\
\frac{\bar \alpha^\mathsf{a,c} }{ K_t } & \frac{\bar \alpha^\mathsf{a,c} }{ K_t } &  & \\
 & & \frac{\bar \alpha^\mathsf{a,a} }{ K_t } & \frac{\bar \alpha^\mathsf{a,a} }{ K_t } \\
\end{bmatrix}
-
\begin{bmatrix}
\beta^\mathsf{c,c} & & & \\
 & \beta^\mathsf{c,a} & & \\
 & & \frac{\beta^\mathsf{a,c}}{K_t} & \\
 & & & \frac{\beta^\mathsf{a,a}}{K_t} \\
\end{bmatrix}
\right)
\E{\boldsymbol{\lambda}_t}
=
\mathbf{M}
\E{\boldsymbol{\lambda}_t}
,
\end{align*}
 thus for $\boldsymbol{ \lambda}_{t_0} = [ \lambda_{t_0}^\mathsf{c,c}, \, \lambda_{t_0}^\mathsf{c,a}, \, \lambda_{t_0}^\mathsf{a,c}, \, \lambda_{t_0}^\mathsf{a,a} ]^\mathrm{T}$ as the vector of known initial values, the vector of mean intensities at time $t$ is given by the solution
\begin{align}
\E{\boldsymbol{\lambda}_t} = e^{\mathbf{M} (t - t_0)} \boldsymbol{ \lambda}_{t_0} 
.
\end{align}

Now, let us define the corresponding sub-counting processes for the number of messages sent by time $t$, collected in the column vector $\mathbf{N}_t = [ N_t^\mathsf{c,c}, \, N_t^\mathsf{c,a}, \, N_t^\mathsf{a,c}, \, N_t^\mathsf{a,a} ]^\mathrm{T}$. Again, through the infinitesimal generator for these Markov processes, we can see that
$$
\frac{\mathrm{d}}{\mathrm{d}t}\E{\mathbf{N}_t}
=
\E{\boldsymbol{\lambda}_t}
= 
e^{\mathbf{M} (t - t_0)} \boldsymbol{ \lambda}_{t_0} 
,
$$
 thus we have that the mean counting process vector is given by
\begin{align}
\E{\mathbf{N}_t}
=
\mathbf{N}_{t_0}
+
\mathbf{M}^{-1}
\left( e^{\mathbf{M} (t-t_0)} - \mathbf{I} \right) \boldsymbol{ \lambda}_{t_0} 
.
\label{nVecSol}
\end{align}
\hfill \Halmos 
\endproof

\subsection{Proof of Proposition~\ref{nextMsgDist}}

\proof{Proof.}
Note that the event that $X \geq x$ is equivalent to the event that there are no arrivals in $[t_0, t_0 + x)$, i.e., $N_{t_0+x} - N_{t_0} = 0$. Because Hawkes processes are conditionally non-stationary Poisson until the next arrival given the process history, this is
\begin{align*}
\PP{N_{t_0+x} - N_{t_0} = 0 \mid \boldsymbol{\lambda}_{t_0}}
&=
e^{-\int_{t_0}^{t_0+x}\left(\lambda_t^\mathsf{c} + \lambda_t^\mathsf{a}\right)\mathrm{d}t}
\\
&
=
e^{-\int_{0}^{x}\left(\lambda_{t_0}^\mathsf{c,c} e^{-\beta^\mathsf{c,c} t} + \lambda_{t_0}^\mathsf{c,a} e^{-\beta^\mathsf{c,a} t} + \frac{\lambda_{t_0}^\mathsf{a,c}}{K_{t_0}} e^{-\frac{\beta^\mathsf{a,c} t}{K_{t_0}}} + \frac{\lambda_{t_0}^\mathsf{a,a}}{K_{t_0}} e^{-\frac{\beta^\mathsf{a,a} t}{K_{t_0}}} \right)\mathrm{d}t}
\\
&
=
e^{-\frac{\lambda_{t_0}^\mathsf{c,c}}{\beta^\mathsf{c,c}} \left(1 - e^{-\beta^\mathsf{c,c} x}\right) - 
\frac{\lambda_{t_0}^\mathsf{c,a}}{\beta^\mathsf{c,a}} \left(1 - e^{-\beta^\mathsf{c,a} x}\right)
-
\frac{\bar \lambda_{t_0}^\mathsf{a,c}}{\beta^\mathsf{a,c}} \left(1 -  e^{-\frac{\beta^\mathsf{a,c} x}{K_{t_0}}} \right)
-
\frac{\bar \lambda_{t_0}^\mathsf{a,a}}{\beta^\mathsf{a,a}} \left(1 -  e^{-\frac{\beta^\mathsf{a,a} x}{K_{t_0}}} \right)
}
.
\end{align*}
\hfill \Halmos
\endproof

One should note that Proposition~\ref{nextMsgDist} uses an agent initial correspondence rate condition that is already adjusted so as to remove the concurrency normalization, since this cancels when divided by the decay rate. This $\bar \lambda_{t_0}^{\mathsf{a},y}$ notation will appear in later statements.

\subsection{Proof of Proposition~\ref{firstSPPprop}}

\proof{Proof.}
Much like how the static planning problem we use here is very similar to that of \citet{Tezcan2014RoutingCustomersb}, the following is indebted to Lemma 1's proof therein.\footnote{In fact, \citet{Tezcan2014RoutingCustomersb} considers many parameter cases and thus that proof is more involved than this one, since our interest here is to instead explore a series of consequences within one particular case.} First, we can see that the routing LP's dual is given by
\begin{align}
\max\quad &\mathcal{S}y_\mathcal{S} + \Theta y_\Theta
\\
\mathrm{s.t.}\quad & \frac{1}{k \mu_k} y_\mathcal{S} + y_\Theta \geq \eta_k \quad \forall\, k \in \{1, \dots, \kappa\}
,
\\
\quad& y_\mathcal{S} \geq 0
.
\end{align}
Now, let $\hat k = \inf\{k \in \mathbb{Z}_+ \mid c_1 \leq c_0 k\}$, and let us first consider the case that $k^* \leq \hat k - 1$. Here, we can see that taking $y_\mathcal{S} = k^*(k^*-1)c_0(\eta_{k^*-1}-\eta_{k^*})$ and $y_\Theta = \eta_{k^*-1} - k^*(\eta_{k^*-1} - \eta_{k^*})$ would produce a dual objective value of 
\begin{align}
\mathcal{S} k^*(k^*-1)c_0(\eta_{k^*-1}-\eta_{k^*}) 
+
\Theta \eta_{k^*-1} - \Theta k^*(\eta_{k^*-1} - \eta_{k^*})
,
\end{align}
and this would be matched by the objective value of the primal solution $\boldsymbol{\theta}^*$, where one can quickly recognize the components of $\boldsymbol{\theta}^*$ in the dual objective value simply by looking at the coefficients of $\eta_{k^*-1}$ and $\eta_{k^*}$. It is clear that $\boldsymbol{\theta}^*$ is primal feasible, since $\frac{1}{ (k^*-1) c_0}(k^*-1)\left(k^* c_0 \mathcal{S} - \Theta\right) + \frac{1}{k^* c_0} k^*\left(\Theta - (k^*-1) c_0 \mathcal{S}\right) = \mathcal{S}$ and $(k^*-1)\left(k^* c_0 \mathcal{S} - \Theta\right) +  k^*\left(\Theta - (k^*-1) c_0 \mathcal{S}\right) = \Theta$, so all that remains is to show dual feasibility of $y_\mathcal{S}$ and $y_\Theta$. By construction, the constraints $\frac{1}{k^* c_0} y_\mathcal{S} + y_\Theta = \frac{1}{k^* \mu_{k^*}} y_\mathcal{S} + y_\Theta = \eta_{k^*}$ and $\frac{1}{(k^*-1) c_0} y_\mathcal{S} + y_\Theta = \frac{1}{(k^*-1) \mu_{k^*-1}} y_\mathcal{S} + y_\Theta = \eta_{k^*-1}$ both hold with equality, and so we are left to inspect the others. We can see from \eqref{successProbAssump} that
\begin{align}
\frac{1}{j c_0} y_\mathcal{S} + y_\Theta
&=
\eta_{k^*-1}
+
\frac{k^*}{j } \left(k^*-1 - j\right)(\eta_{k^*-1}-\eta_{k^*}) 
\geq
\eta_j
,
\end{align}
and so the proposed dual solution meets dual constraints 1 through $\hat k - 1$. To extend this to constraints $\hat k$ and above, we can recognize that $\frac{1}{c_1} \geq \frac{1}{c_\naught k}$ for $k \geq \hat k$ by definition, and thus again by \eqref{successProbAssump} we find that these constraints hold, ensuring dual feasibility.

If $k^* = \hat k$, we have by conjunction with the system stability condition that $(k^*-1)c_0 \mathcal{S} \leq \Theta < c_1 \mathcal{S}$. Here we can take a dual solution of $y_\mathcal{S} = (k^*-1) c_0 c_1 (\eta_{k^*-1} - \eta_{k^*}) \slash (c_1 - (k^*-1)c_0)$ and $y_\Theta = \eta_{k^*-1} - c_1 (\eta_{k^*-1} - \eta_{k^*}) \slash (c_1 - (k^*-1)c_0)$, which would produce a dual objective value of
\begin{align}
c_1 \mathcal{S}
  \frac{(k^*-1) c_0}{c_1 - (k^*-1)c_0}
\left(\eta_{k^*-1} - \eta_{k^*}\right)
+
\Theta
\left(
\eta_{k^*-1} 
- 
\frac{c_1} {c_1 - (k^*-1)c_0}
(\eta_{k^*-1} - \eta_{k^*}) 
\right)
.
\end{align}
We can again quickly observe that $\boldsymbol{\theta}^*$ would match this, where here $\theta^*_{k^* - 1} = (c_1 \mathcal{S} - \Theta)(k^*-1)c_0 / (c_1 - (k^*-1)c_0)$ and $\theta^*_{k^*} = (\Theta - (k^*-1)c_0 \mathcal{S}) c_1 / (c_1 - (k^*-1)c_0)$. It is also again straightforward to verify that $\boldsymbol{\theta}^*$ is primal feasible. For dual feasibility, we can observe that $y_\mathcal{S}/c_1 + y_\Theta = \eta_{k^*}$ and $y_\mathcal{S}/((k^*-1) c_0) + y_\Theta = \eta_{k^* - 1}$ directly from the expressions for $y_\mathcal{S}$ and $y_\Theta$; constraints below $k^* - 1$ will follow from Equation~\eqref{successProbAssump} and constraints above $k^* = \hat k$ will follow simply from the fact that the success probabilities are non-increasing.
\hfill\Halmos\endproof

\subsection{Proof of Lemma~\ref{closureVarLemma}}

\proof{Proof.}
Let us first suppose $\tau \in \mathcal{T}$, with some arbitrary $\zeta \in (0, \alpha^\mathsf{c,c}\slash\beta^\mathsf{c,c} + \alpha^\mathsf{a,c}\slash\beta^\mathsf{a,c}]$ such that $\tau = \inf\{t \geq 0 \mid \boldsymbol{\lambda}_{t} \in L_\zeta\}$. Then, $\tau$ is an almost surely finite stopping time by immediate consequence of Theorem~\ref{stabilityThm}, as the correspondence rates must cross the hyperplane ${\lambda_t^\mathsf{c,c}}\slash{\beta^\mathsf{c,c}}
+
\lambda_t^\mathsf{c,a}\slash{\beta^\mathsf{c,a}}
+
\lambda_t^\mathsf{a,c}\slash{\beta^\mathsf{a,c}}
+
{\lambda_t^\mathsf{a,a}}\slash{\beta^\mathsf{a,a}}
=
\zeta
$. This will be a transient state before being absorbed into the limiting zero correspondence rate state. Furthermore, because the only manner of decrease in the BHP correspondence rates is the continuous decay, we know that the process will not only cross the $L_\zeta$ hyperplane, it must attain a value within it.
For any such point, $\PP{X = \infty \mid \boldsymbol{\lambda}_\tau} = e^{-\zeta}$ with probability one, since $\boldsymbol{\lambda}_\tau^\T(1\slash\boldsymbol{\beta}) = \zeta$ by definition of $\tau$. 

This now suggests the reverse direction. Suppose that $\tau$ is an almost surely finite stopping time and that $\PP{X = \infty \mid \boldsymbol{\lambda}_\tau} = p$ almost surely for some probability $p$. If $p \in [e^{-\alpha^\mathsf{c,c}\slash\beta^\mathsf{c,c} - \alpha^\mathsf{a,c}\slash\beta^\mathsf{a,c}}, 1)$, then immediately $\tau \in \mathcal{T}$ through the mapping $\zeta = -\log(p)$ that we have used above. Hence, we are left to show that there are no other almost surely finite stopping times outside of this set. Let $p < e^{-\alpha^\mathsf{c,c}\slash\beta^\mathsf{c,c} - \alpha^\mathsf{a,c}\slash\beta^\mathsf{a,c}}$. By Proposition~\ref{nextMsgDist}, we can see that the probability that no messages occur after the conversation's initial message is precisely $e^{-\alpha^\mathsf{c,c}\slash\beta^\mathsf{c,c} - \alpha^\mathsf{a,c}\slash\beta^\mathsf{a,c}}$. Furthermore, we can see that, absent any epochs after the initial, the correspondence rates are all strictly decreasing, meaning $\PP{X = \infty \mid \boldsymbol{\lambda}_t}$ will strictly rise. Hence, there is at least a $e^{-\alpha^\mathsf{c,c}\slash\beta^\mathsf{c,c} - \alpha^\mathsf{a,c}\slash\beta^\mathsf{a,c}}$ probability that $\PP{X = \infty \mid \boldsymbol{\lambda}_t}$ will never attain the stopping criterion $p$, and so $\tau$ cannot be almost surely finite.
\hfill\Halmos
\endproof

\subsection{Proof of Proposition~\ref{secondSPPprop}}

\proof{Proof.}
Because the static planning problem constraints are essentially unchanged from the homogeneous to heterogeneous scenarios, the feasibility of $\boldsymbol{\theta}^\mathsf{LL}$ and $\boldsymbol{\theta}^\mathsf{HP}$ follows immediately from the feasibility of $\boldsymbol{\theta}^*$. It is also quick to see that $\boldsymbol{\theta}^\mathsf{HP}$ will outperform $\boldsymbol{\theta}^\mathsf{LL}$:
\begin{align}
&
\sum_{k=1}^\kappa \sum_{h=0}^k
\left(p_{k,h} \eta_k^\mathsf{1} + (1-p_{k,h})\eta_k^\mathsf{2} \right) \theta_{k,h}^\mathsf{LL} 
\\
&=
\sum_{h=0}^{k^*-1}
\left(p_{k^*-1,h} \eta_{k^*-1}^\mathsf{1} + (1-p_{k^*-1,h})\eta_{k^*-1}^\mathsf{2} \right) \varrho^{k^*-1}_{h} \theta^*_{k^*-1}
+
\sum_{h=0}^{k^*}
\left(p_{k^*,h} \eta_{k^*}^\mathsf{1} + (1-p_{k^*,h})\eta_{k^*}^\mathsf{2} \right) \varrho^{k^*}_{h} \theta^*_{k^*}
\\
&<
\eta_{k^*-1}^\mathsf{1}  \theta^*_{k^*-1}
+
 \eta_{k^*}^\mathsf{1} \theta^*_{k^*}
\\
&=
\sum_{k=1}^\kappa \sum_{h=0}^k
\left(p_{k,h} \eta_k^\mathsf{1} + (1-p_{k,h})\eta_k^\mathsf{2} \right) \theta_{k,h}^\mathsf{HP} 
.
\end{align}
To observe the optimality of $\boldsymbol{\theta}^\mathsf{HP}$ when given \eqref{successProbAssump2}, let us first note that $\boldsymbol{\theta}^\mathsf{HP}$ will have objective value
\begin{align}
\frac{\eta_{k^*-1}^\mathsf{1} (k^*-1)c_0}{c_0 + (c_1 - k^* c_0)^+}
\left(\left(k^* c_0 \wedge c_1\right) \mathcal{S} - \Theta\right)  
+
\frac{\eta_{k^*}^\mathsf{1} (k^* c_0 \wedge c_1)}{c_0 + (c_1 - k^* c_0)^+}
\left(\Theta - (k^*-1) c_0 \mathcal{S}\right)
,
\end{align}
and by inspecting the coefficients of $\mathcal{S}$ and $\Theta$ we see that this suggests a dual solution of $y_\mathcal{S} = (k^*-1)c_0\left(k^* c_0 \wedge c_1\right)(\eta^\mathsf{1}_{k^*-1} - \eta^\mathsf{1}_{k^*})/(c_0 + (c_1 - k^* c_0)^+)$ and $y_\Theta = (\eta_{k^*}^\mathsf{1} (k^* c_0 \wedge c_1) - \eta_{k^*-1}^\mathsf{1} (k^*-1)c_0)/(c_0 + (c_1 - k^* c_0)^+)$. It remains to show that this candidate dual solution is feasible, meaning $y_\mathcal{S}/(kc_0 \wedge c_1) + y_\Theta \geq p_{k,h}\eta^\mathsf{1}_{k} + (1-p_{k,h})\eta^\mathsf{2}_k$ for every $k$ and $h$. First, we can quickly find that the constraints are tight at $k=h=k^*-1$ and $k=h=k^*$:
\begin{align}
\frac{1}{(k^*-1)c_0} y_\mathcal{S} + y_\Theta
&=
\frac{\left(k^* c_0 \wedge c_1\right)(\eta^\mathsf{1}_{k^*-1} - \eta^\mathsf{1}_{k^*})}{c_0 + (c_1 - k^* c_0)^+}
+
\frac{\eta_{k^*}^\mathsf{1} (k^* c_0 \wedge c_1) - \eta_{k^*-1}^\mathsf{1} (k^*-1)c_0}{c_0 + (c_1 - k^* c_0)^+}
=
\eta^\mathsf{1}_{k*-1}
,
\\
\frac{1}{(k^*c_0 \wedge c_1)} y_\mathcal{S} + y_\Theta
&=
\frac{(k^*-1)c_0(\eta^\mathsf{1}_{k^*-1} - \eta^\mathsf{1}_{k^*})}{c_0 + (c_1 - k^* c_0)^+}
+
\frac{\eta_{k^*}^\mathsf{1} (k^* c_0 \wedge c_1) - \eta_{k^*-1}^\mathsf{1} (k^*-1)c_0}{c_0 + (c_1 - k^* c_0)^+}
=
\eta^\mathsf{1}_{k*}
.
\end{align}
For the remaining constraints, we can observe that generally for $1 \leq j < \hat k = \inf\{k\in\mathbb{Z}_+ \mid c_1 \leq c_0 k\}$,
\begin{align}
\frac{1}{j c_0} y_\mathcal{S} + y_\Theta
&=
\frac{k^*-1}{j}
\frac{\left(k^* c_0 \wedge c_1\right)(\eta^\mathsf{1}_{k^*-1} - \eta^\mathsf{1}_{k^*})}{c_0 + (c_1 - k^* c_0)^+}
+
\frac{\eta_{k^*}^\mathsf{1} (k^* c_0 \wedge c_1) - \eta_{k^*-1}^\mathsf{1} (k^*-1)c_0}{c_0 + (c_1 - k^* c_0)^+}
\\
&=
\left(
\frac{k^*-1}{j}
-
1
\right)
\frac{\left(k^* c_0 \wedge c_1\right)(\eta^\mathsf{1}_{k^*-1} - \eta^\mathsf{1}_{k^*})}{c_0 + (c_1 - k^* c_0)^+}
+
\eta_{k^*-1}^\mathsf{1}
,
\end{align}
and by recognizing that $\Delta_{k^*} = (k^* c_0 \wedge c_1)/(c_0 + (c_1 - k^* c_0)^+)$, we have via \eqref{successProbAssump2} that
\begin{align}
\frac{1}{j c_0} y_\mathcal{S} + y_\Theta
\geq
\eta^\mathsf{1}_j
.
\end{align}
Because $\eta^\mathsf{1}_j > \eta^\mathsf{2}_j$ for each $j$, $\eta^\mathsf{1}_j \geq p \eta^\mathsf{1}_j + (1-p)\eta^\mathsf{2}_j$ for every $p \in [0,1]$, and thus the dual constraints hold for every $0 \leq h \leq j \leq \hat k - 1$. For $j \geq \hat k$, we can recognize that
\begin{align}
\frac{1}{c_1} y_\mathcal{S} + y_\Theta
\geq
\frac{1}{j c_0} y_\mathcal{S} + y_\Theta
,
\end{align}
by the definition of $\hat k$, and thus by the preceding arguments we again find that these remaining dual constraints are satisfied, and thus $\boldsymbol{\theta}^\mathsf{HP}$ is optimal for the primal.
\hfill\Halmos\endproof

\subsection{Auto-Correlation of the Univariate Correspondence Rate}

As an auxiliary result that demonstrates the pace of the history dependence, in Proposition~\ref{autoCorProp} we derive the auto-correlation of the UHP correspondence rate.

\begin{proposition}\label{autoCorProp}
The UHP correspondence rate has auto-correlation 
\begin{align}
\Corr{\lambda_t, \lambda_{t-u}}
&=
\sqrt{
\frac{
e^{-(\beta - \alpha)u} - e^{-(\beta - \alpha)t}
}{
1 - e^{-(\beta-\alpha)t}
}
}
,
\end{align}
where $t \geq u \geq 0$.
\end{proposition}
\proof{Proof.}
Let us recall that the correlation between two random variables $X$ and $Y$ is defined
\begin{align}
\Corr{X,Y} 
=
\frac{\Cov{X,Y}}{\sqrt{\Var{X}\Var{Y}}}
=
\frac{\E{XY} - \E{X}\E{Y}}{\sqrt{\Var{X}\Var{Y}}}
.
\end{align}
The mean and variance of $\lambda_t$ are readily available in the literature, but, to our surprise, we have not been able to locate an expression for the auto-covariance of $\lambda_t$, nor for $\E{\lambda_t \lambda_{t-u}}$, more specifically. So, let us derive that here. Leveraging the Markov property of $\lambda_t$, we can see through conditional expectation that
\begin{align}
\E{\lambda_t \lambda_{t-u}}
&=
\E{\E{\lambda_t \mid \lambda_{t-u}} \lambda_{t-u}}
=
\E{\lambda_{t-u}^2}e^{-(\beta - \alpha)(t-u)}
.
\end{align}
Now, from well-known expressions for the first and second moments of $\lambda_t$ \citep[e.g., Prop. 2 of][]{daw2017queues}, we cancel terms to find 
\begin{align}
\Corr{\lambda_t,\lambda_{t-u}} 
&=
\frac{\frac{\alpha^2 \lambda_0}{\beta-\alpha}e^{-(\beta-\alpha)t}\left(1-e^{-(\beta-\alpha)(t-u)}\right)}
{\sqrt{\frac{\alpha^2 \lambda_0}{\beta-\alpha}e^{-(\beta-\alpha)t}\left(1-e^{-(\beta-\alpha)t}\right)\frac{\alpha^2 \lambda_0}{\beta-\alpha}e^{-(\beta-\alpha)(t-u)}\left(1-e^{-(\beta-\alpha)(t-u)}\right)}}
\\
&=
\sqrt{
\frac{e^{-(\beta-\alpha)t}\left(1-e^{-(\beta-\alpha)(t-u)}\right)}
{e^{-(\beta-\alpha)(t-u)}\left(1-e^{-(\beta-\alpha)t}\right)}
}
,
\end{align}
which simplifies to the stated result.
\hfill\Halmos\endproof

What we can see here is that if $\alpha$ is near $\beta$, as is the case for the contact center dataset, then the correlation of the conversation's pace will decay quite slow in time. That is, as the time gap $u$ grows,  Proposition~\ref{autoCorProp} shows that the auto-correlation will effectively decay at rate $(\beta - \alpha)\slash 2$.

\section{Description of Estimation and Simulation Procedures}
\label{app:Comp}

Let us now give a brief overview of some of the computational methods we use for these stochastic processes. 
We use two main types of procedures for the Hawkes-based models: parameter estimation and Monte-Carlo simulation. Our estimation procedure is an adaptation of the log-likelihood-based expectation-maximization (EM) algorithm that has enjoyed much success for Hawkes processes \citep[e.g.,][]{lewis2011nonparametric,halpin2012algorithm,Halpin2013} and was originally developed for general branching processes by \citet{veen2008estimation}. The procedure is known to be equivalent to projected gradient ascent for the Hawkes process \citep{lewis2011nonparametric}. Much of its popularity stems from its efficiency, and that is particularly true for our setting thanks to the lack of baseline arrivals in the cluster model. 
For $M$ as the total number of conversations and $N_m$ as the number of messages within conversation $m$, leveraging the cluster structure yields a procedure that is $O\left(\sum_{m=1}^M N_m^2\right)$ rather than $O\left(\left(\sum_{m=1}^M N_m\right)^2\right)$ for the continual point process. On a data set with as many distinct conversations as ours, this is a substantial simplification, even by comparison to standard Hawkes EM implementations. The EM algorithm is also quite interpretable: each iteration calculates the probabilities that each message is in response to every previous message and then re-expresses the process parameters in terms of these probabilities; this underlying branching structure is the missing data targeted by the EM algorithm in \citet{veen2008estimation}. We provide the exact likelihood equations and EM steps in the following subsection.

For simulation, a general Hawkes service model simulation procedure could be achieved through  a combination of two of the most popular Hawkes process simulation algorithms, the Lewis-Shedler-Ogata thinning-based algorithm \citep{lewis1979simulation,ogata1981lewis}  and the compensator inversion technique \citep[originally shown for Hawkes processes by][]{ozaki1979maximum}. Each of these has both strengths and weaknesses in addressing the models we have proposed. The former allows us to handle the non-stationarity induced on the service model from the concurrency changes by simulating in each of the different periods of concurrency values. However, the thinning-based methodology is built on identifying an upper bound for the instantaneous arrival rate of the messages. This means that it is unable to replicate the end-conditions of models such as ours in which the correspondence rates almost surely converge to 0, since the positive upper bounds will always imply some nonzero chance of more messages. By comparison, the exact sampling structure of the inverse transform procedure is very well structured to handle processes that eventually cease; however, it is not designed to handle non-stationarity. Thus, our hybrid simulation procedures uses both: on all concurrency intervals until the last one of the work shift, we use Lewis-Shedler-Ogata, and then for the final concurrency interval, we employ inverse transform sampling to complete the replication.

In the case of the SysBHP model form, we can in fact leverage the Markov property of the alternate construction in Section~\ref{sysMarkov} and simulate in the same style as the Markovian Hawkes process procedure in \citet{dassios2013exact}.

\subsection{Estimation Algorithm}

To estimate the parameters of these processes from data, we use a variant of the expectation-maximization (EM) algorithm. As we will describe in this section, these procedures are highly efficient and easily implementable in practice and are thus quite common in the Hawkes process literature. 
The tractability of these algorithms for exponential kernel Hawkes processes largely lies in the fact that all the supporting calculations within each iteration reduce to solving simple linear equations, which can be found in closed form. 
Nevertheless, other methods of estimation exist in the literature for Hawkes processes, so let us briefly mention a few alternatives. First, the most comparable procedure is maximum likelihood estimation (MLE), as the EM algorithm also relies on the likelihood function. This function was first provided in \cite{ozaki1979maximum}. While EM algorithms do not necessarily have the same level of theoretical guarantees, they do offer considerable computational advantages over the non-linear optimization of direct maximum likelihood estimation on large datasets such as the one we study in this work. By comparison, one could instead use parametric approaches that draw upon advanced optimization techniques, such as in \citet{guo2018consistency}. There are also interesting approaches available for the alternate setting in which there are only a small number of data points available, e.g., in \citet{salehi2019learning}. For an overview and comparison of Hawkes process estimation procedures, see \cite{kirchner2018nonparametric}.

As we have noted, we use the EM algorithm because of its computational simplicity and ease of implementation. This tractability means we can also easily describe the EM approach. Although all the terms can be written in closed form, some become cumbersome. Hence, we reserve some explicit computations for Appendix \ref{app:equations}. Because the three model forms encapsulate one another, we only describe the estimation procedure for the SysBHP in detail. 
Data from one conversation can be considered separately from all other conversations, and this follows from the independence of branches within the Hawkes process models.  The data points we use are the message time stamps and sending parties, meaning the customer or agent. Because system features like the concurrency, sentiment scores, and numbers of words per message are observable in practice, we only seek to estimate the jump size and decay parameters.

To describe the EM algorithm, we begin by first specifying the log-likelihood function for a given conversation. Because the Hawkes process models are stochastic intensity Poisson processes, we can give the log-likelihood in closed form. In the case of the SysBHP model form, this is given by 
\begin{equation}
\mathcal{L}\left(\vartheta \mid \mathcal{D}\right)
=
\sum_{i=1}^{N^\mathsf{c}_\infty} \log\left( \lambda^\mathsf{c}_{{A_i^\mathsf{c}}^-} \right)
+
\sum_{j=1}^{N^\mathsf{a}_\infty} \log\left( \lambda^\mathsf{a}_{{A_j^\mathsf{a}}^-} \right)
-
\int_0^\infty \lambda^\mathsf{c}_t \mathrm{d}t
-
\int_0^\infty \lambda^\mathsf{a}_t \mathrm{d}t
,
\label{likelihoodSimple}
\end{equation}
where $\lambda_t^\mathsf{c}$ and $\lambda_t^\mathsf{a}$ are respectively the customer and agent correspondence rates given in Equations~\eqref{SysBHPdefCustomer} and~\eqref{SysBHPdefAgent} with $\lambda^\mathsf{c}_{{A_i^\mathsf{c}}^-} = \lim_{t \uparrow A_i^c} \lambda^\mathsf{c}_t$ and
$\lambda^\mathsf{a}_{{A_j^\mathsf{a}}^-} = \lim_{t \uparrow A_j^a} \lambda^\mathsf{a}_t$, where 
$\vartheta = \{\alpha^{\mathsf{c,c}}_\mathsf{1}, \alpha^{\mathsf{c,a}}_\mathsf{1}, \alpha^{\mathsf{a,c}}_\mathsf{1}, \alpha^{\mathsf{a,a}}_\mathsf{1}, \alpha^{\mathsf{c,c}}_\mathsf{2}, \alpha^{\mathsf{c,a}}_\mathsf{2}, \alpha^{\mathsf{a,c}}_\mathsf{2}, \alpha^{\mathsf{a,a}}_\mathsf{2}, \beta^{\mathsf{c,c}}, \beta^{\mathsf{c,a}}, \beta^{\mathsf{a,c}}, \beta^{\mathsf{a,a}}\}$ 
is the parameter set, and where $\mathcal{D} = \{(A_1^\mathsf{c}, \dots, A_{N^\mathsf{c}}^\mathsf{c}), (A_1^\mathsf{a}, \dots, A_{N^\mathsf{a}}^\mathsf{a})\}$ is the message timestamps data set for the full conversation. Because the sentiments and word counts are drawn independently from the correspondence processes, we do not include their distributions in the likelihood expressions; these terms will vanish when taking partial derivatives of $\mathcal{L}$ with respect to the parameters in $\theta$. The fully simplified log-likelihood is given in Appendix~\ref{app:equations}.
Note that because the data comprises only completed conversations and all conversations contain finitely many messages, we are using $N^\mathsf{c}_\infty = \lim_{t \to \infty}N_t^\mathsf{c}$ and $N^\mathsf{a}_\infty = \lim_{t \to \infty}N_t^\mathsf{a}$ as the total number of customer and agent messages in the conversation, excluding the initial query. Because conversations are conditionally independent from one another given the concurrency of the agents, we can then note that the log-likelihood function of the full contact center data containing $M \in \mathbb{Z}_+$ conversations, say $\bar{\mathcal{L}}(\vartheta \mid \bar{\mathcal{D}})$, can be obtained from
$$
\bar{\mathcal{L}}(\vartheta \mid \bar{\mathcal{D}})
=
\sum_{m=1}^M \mathcal{L}_m\left(\theta \mid \mathcal{D}_m\right)
,
$$
where $\mathcal{L}_m\left(\vartheta \mid \mathcal{D}_m\right)$ is the log-likelihood for the $m^\text{th}$ conversation as calculated according to Equation~\eqref{lConvDef} and where $\bar{\mathcal{D}} = \bigcup_{m=1}^M \mathcal{D}_m$ is the complete data set.

EM algorithms work by making use of missing data. In our setting, the missing data is the precise conversational dependencies, meaning knowledge of which previous message prompted a given message as response. This is not observable in the data, but we can quantify the probability that one message is in response to another. For example, given the parameters of the SysBHP conversation model and the conversation data, the probability that the $i^\text{th}$ customer message is actually in response to $j^\text{th}$ customer message and is spurred by the sentiment of this message can be calculated via
\begin{align}
p_{i,j}^{\mathsf{c,c-1}}
&=
\frac{1}{\lambda^\mathsf{c}_{{A_i^\mathsf{c}}^-}}\alpha^\mathsf{c,c}_\mathsf{1} S_j^\mathsf{c} e^{-\beta^\mathsf{c,c} (A_i^\mathsf{c} - A_j^\mathsf{c})}  
,
\label{EMprob1}
\end{align}
since this is the amount of excitement generated by the $j^\text{th}$ customer message within the customer message intensity at the time the $i^\text{th}$ customer message was sent. Likewise, the probability that the $i^\text{th}$ message is spurred by the word count of the $j^\text{th}$ message is
\begin{align}
p_{i,j}^{\mathsf{c,c-2}}
&=
\frac{1}{\lambda^\mathsf{c}_{{A_i^\mathsf{c}}^-}}\alpha^\mathsf{c,c}_\mathsf{2} W_j^\mathsf{c} e^{-\beta^\mathsf{c,c} (A_i^\mathsf{c} - A_j^\mathsf{c})}  
,
\nonumber
\end{align}
Similarly, the other response probabilities can thus be calculated as
\begin{align}
p_{i,j}^{\mathsf{c,a-1}}
=
\frac{1}{\lambda^\mathsf{c}_{{A_i^\mathsf{c}}^-}}\alpha^\mathsf{c,a}_\mathsf{1} S_j^\mathsf{a} e^{-\beta^\mathsf{c,a} (A_i^\mathsf{c} - A_j^\mathsf{a})}  
,
&
\qquad
p_{i,j}^{\mathsf{c,a-2}}
=
\frac{1}{\lambda^\mathsf{c}_{{A_i^\mathsf{c}}^-}}\alpha^\mathsf{c,a}_\mathsf{2} W_j^\mathsf{a} e^{-\beta^\mathsf{c,a} (A_i^\mathsf{c} - A_j^\mathsf{a})}  
,
\nonumber
\\
p_{i,j}^{\mathsf{a,c-1}}
=
\frac{1}{\lambda^\mathsf{a}_{{A_i^\mathsf{a}}^-} K_{A_i^\mathsf{a}}}\alpha^\mathsf{a,c}_\mathsf{1}  S_j^\mathsf{c} e^{-\beta^\mathsf{a,c}  (A_i^\mathsf{a} - A_j^\mathsf{c}) \slash K_{A_i^\mathsf{a}}}  
,
&
\qquad
p_{i,j}^{\mathsf{a,c-2}}
=
\frac{1}{\lambda^\mathsf{a}_{{A_i^\mathsf{a}}^-} K_{A_i^\mathsf{a}}}\alpha^\mathsf{a,c}_\mathsf{2}   W_j^\mathsf{c} e^{-\beta^\mathsf{a,c}  (A_i^\mathsf{a} - A_j^\mathsf{c}) \slash K_{A_i^\mathsf{a}}}  
,
\nonumber
\\
p_{i,j}^{\mathsf{a,a-1}}
=
\frac{1}{\lambda^\mathsf{a}_{{A_i^\mathsf{a}}^-} K_{A_i^\mathsf{a}}}\alpha^\mathsf{a,a}_\mathsf{1}   S_j^\mathsf{a} e^{-\beta^\mathsf{a,a}  (A_i^\mathsf{a} - A_j^\mathsf{a}) \slash K_{A_i^\mathsf{a}} }  
,
&
\text{ and}
\quad
p_{i,j}^{\mathsf{a,a-2}}
=
\frac{1}{\lambda^\mathsf{a}_{{A_i^\mathsf{a}}^-} K_{A_i^\mathsf{a}}}\alpha^\mathsf{a,a}_\mathsf{2}   W_j^\mathsf{a} e^{-\beta^\mathsf{a,a}  (A_i^\mathsf{a} - A_j^\mathsf{a}) \slash K_{A_i^\mathsf{a}} }  
.
\label{EMprob2}
\end{align}
Given these response probabilities, one can also then calculate the value of the parameters that are critical points for the full system log-likelihood. By first re-parameterizing the jump sizes in proportion to the decay rate, i.e.~$\hat{\alpha}^{\mathsf{c,c}}_\mathsf{1} = \frac{\alpha^{\mathsf{c,c}}_\mathsf{1}}{\beta^{\mathsf{c,c}}}$, one can in fact give these parameter solutions in closed form, as this change of variable yields that the critical point of each partial derivative is simply found through solving a linear equation. Of course, upon completion of the EM algorithm, one can then obtain the true model jump sizes by simply multiplying $\hat \alpha$ by $\beta$. Because of their length, these expressions are available in the next subsection, Appendix~\ref{app:equations}, so as to not distract from the overarching ideas. This pair of calculations gives us the basis of the iterative EM algorithm, for which we now provide pseudocode in Algorithm~\ref{EM-SysBHP}.

\begin{algorithm}[htb]
\SetAlgoLined
\KwResult{Jump sizes $\vec{\alpha}_*^{(t)}$ and decay rates $\vec{\beta}_*^{(t)}$.}

\textbf{Initialization:} Choose the starting parameters $\vec{\alpha}_*^{(0)}$ and $\vec{\beta}_*^{(0)}$ randomly.

 \While{ $ ||\vec{\alpha}_*^{(t)} - \vec{\alpha}_*^{(t-1)}|| + ||\vec{\beta}_*^{(t)} - \vec{\beta}_*^{(t-1)}|| >\epsilon$
 }{
  \texttt{E-step}: Given the observed data and current parameter estimates $\vec{\alpha}_*^{(t)}$ and $\vec{\beta}_*^{(t)}$, compute the updated response probabilities (each $p_{i,j}^\mathsf{c,c-1}$, $p_{i,j}^\mathsf{c,a-1}$, $p_{i,j}^\mathsf{a,c-1}$, $p_{i,j}^\mathsf{a,a-1}$, $p_{i,j}^\mathsf{c,c-2}$, $p_{i,j}^\mathsf{c,a-2}$, $p_{i,j}^\mathsf{a,c-2}$, and $p_{i,j}^\mathsf{a,a-2}$) within each conversation through Equations~\eqref{EMprob1} and~\eqref{EMprob2}. 
  
  \texttt{M-step}: Using the newly calculated response probabilities and the previous parameter estimates, compute the new parameter estimates $\vec{\alpha}_*^{(t+1)}$ and $\vec{\beta}_*^{(t+1)}$ as the solutions to the linear critical point equations, as given in Equations~\eqref{EMsolve1} through~\eqref{EMsolve8}.
  
   $t \leftarrow t + 1$.
   
 }
 \caption{The SysBHP EM Algorithm}
 \label{EM-SysBHP}
\end{algorithm}

\subsection{Log-Likelihood and EM Algorithm Equations}\label{app:equations}

Here we derive the full log-likelihood for the SysBHP model; the other forms can be simplified from this. Following substitution and simplification from the definition of the correspondence rates and the representation of the log-likelihood in Equation~\eqref{likelihoodSimple}, this log-likelihood can also be expressed
\begin{small}
\begin{align}
\label{lConvDef}
&\mathcal{L}\left(\vartheta \mid \mathcal{D}\right)
=
\sum_{k=1}^{N^\mathsf{c}} \log\Bigg(
\sum_{i=0}^{k-1}
\left(
\alpha^{\mathsf{c},\mathsf{c}}_\mathsf{1} S_i^\mathsf{c}
+
\alpha^{\mathsf{c},\mathsf{c}}_\mathsf{2} W_i^\mathsf{c}
\right)
e^{-\beta^{\mathsf{c},\mathsf{c}}(A_k^\mathsf{c} - A_{i}^{\mathsf{c}})}
+
\sum_{j=1}^{N_{A_k^\mathsf{c}}^{\mathsf{a}}}
\left(
\alpha^{\mathsf{c},\mathsf{a}}_\mathsf{1} S_j^\mathsf{a}
+
\alpha^{\mathsf{c},\mathsf{a}}_\mathsf{2} W_j^\mathsf{a}
\right)
e^{-\beta^{\mathsf{c},\mathsf{a}}(A_k^\mathsf{c} - A_{j}^{\mathsf{a}})}
\Bigg)
\\
&
+
\sum_{k=1}^{N^\mathsf{a}} \log\Bigg(
\sum_{i=0}^{N_{A_k^\mathsf{a}}^{\mathsf{c}}}
\frac{
\alpha^{\mathsf{a},\mathsf{c}}_\mathsf{1} S_i^\mathsf{c}
+
\alpha^{\mathsf{a},\mathsf{c}}_\mathsf{2} W_i^\mathsf{c}
}{
K_{A_k^\mathsf{a}}
}
e^{-\beta^{\mathsf{a},\mathsf{c}}  (A_k^\mathsf{a}-A_{i}^{\mathsf{c}}) \slash K_{A_k^\mathsf{a}}}
+
\sum_{j=1}^{k-1}
\frac{
\alpha^{\mathsf{a},\mathsf{a}}_\mathsf{1} S_j^\mathsf{a}
+
\alpha^{\mathsf{a},\mathsf{a}}_\mathsf{2} W_j^\mathsf{a}
}{
K_{A_k^\mathsf{a}}
}
e^{-\beta^{\mathsf{a},\mathsf{a}}  (A_k^\mathsf{a}-A_{j}^{\mathsf{a}}) \slash K_{A_k^\mathsf{a}}}
\Bigg)
\nonumber
\\
&
- 
\sum_{i=0}^{N^\mathsf{c}} 
\left(
\frac{\alpha^\mathsf{c,c}_\mathsf{1}}{\beta^\mathsf{c,c}} S_i^\mathsf{c}
+
\frac{\alpha^\mathsf{c,c}_\mathsf{2}}{\beta^\mathsf{c,c}} W_i^\mathsf{c}
\right)
-
\sum_{i=0}^{N^\mathsf{c}}  
\left(
\frac{\alpha^\mathsf{a,c}_\mathsf{1}}{\beta^\mathsf{a,c}} S_i^\mathsf{c}
+
\frac{\alpha^\mathsf{a,c}_\mathsf{2}}{\beta^\mathsf{a,c}} W_i^\mathsf{c}
\right)
\sum_{k=1}^\kappa \left(e^{-\beta^\mathsf{a,c} f(K_{\Delta_{k-1}}) (\Delta_{k-1} - A_i^\mathsf{c})^+}  - e^{-\beta^\mathsf{a,c} f(K_{\Delta_{k-1}}) (\Delta_{k} - A_i^\mathsf{c})^+}  \right)
\nonumber
\\
&
-
\sum_{j=1}^{N^\mathsf{a}} 
\left(
\frac{\alpha^\mathsf{c,a}_\mathsf{1}}{\beta^\mathsf{c,a}} S_j^\mathsf{a}
+
\frac{\alpha^\mathsf{c,a}_\mathsf{2}}{\beta^\mathsf{c,a}} W_j^\mathsf{a}
\right)
-
\sum_{j=1}^{N^\mathsf{a}}  
\left(
\frac{\alpha^\mathsf{a,a}_\mathsf{1}}{\beta^\mathsf{a,a}} S_j^\mathsf{a}
+
\frac{\alpha^\mathsf{a,a}_\mathsf{2}}{\beta^\mathsf{a,a}} W_j^\mathsf{a}
\right)
\sum_{k=1}^\kappa \left(e^{-\beta^\mathsf{a,a} f(K_{\Delta_{k-1}}) (\Delta_{k-1} - A_j^\mathsf{a})^+}  - e^{-\beta^\mathsf{a,a} f(K_{\Delta_{k-1}}) (\Delta_{k} - A_j^\mathsf{a})^+}  \right)
,
\nonumber
\end{align}
\end{small}
where $\kappa$ is the total number of successive concurrency values that occurred over the course of this conversation, with $K_t = K_{\Delta_{k-1}}$ on $t \in [\Delta_{k-1}, \Delta_k)$ for each $k \leq \kappa$.

Taking the subscript $*$ for the roots of the first derivative of the log-likelihood with respect to each parameter, we can express the jump sizes in terms of the response probabilities as

\begin{equation}
\hat{\alpha}^{\mathsf{c,c}}_{\mathsf{1},*}
=
\frac{
\sum_{m=1}^M \sum_{k=1}^{N_{\infty,m}^\mathsf{c}} \sum_{i=0}^{k-1} p_{k,i,m}^\mathsf{c,c-1} 
}{
\sum_{m=1}^M  \sum_{i=0}^{N^\mathsf{c}_{\infty,m}} S_{i,m}^\mathsf{c}
}
,
\quad
\hat{\alpha}^{\mathsf{c,c}}_{\mathsf{2},*}
=
\frac{
\sum_{m=1}^M \sum_{k=1}^{N_{\infty,m}^\mathsf{c}} \sum_{i=0}^{k-1} p_{k,i,m}^\mathsf{c,c-2} 
}{
\sum_{m=1}^M  \sum_{i=0}^{N^\mathsf{c}_{\infty,m}} W_{i,m}^\mathsf{c}
}
,
\label{EMsolve1}
\end{equation}
\begin{equation}
\hat{\alpha}^\mathsf{c,a}_{\mathsf{1},*}
=
\frac{
\sum_{m=1}^M
\sum_{k=1}^{N^\mathsf{c}_{\infty,m}} 
  \sum_{j=1}^{N_{A_{k,m}^\mathsf{c},m}^{\mathsf{a}}}
p^\mathsf{c,a-1}_{k,j,m}
}{
\sum_{m=1}^M
\sum_{j=1}^{N^\mathsf{a}_{\infty,m}} S_{j,m}^\mathsf{a}
}
,
\quad
\hat{\alpha}^\mathsf{c,a}_{\mathsf{2},*}
=
\frac{
\sum_{m=1}^M
\sum_{k=1}^{N^\mathsf{c}_{\infty,m}} 
  \sum_{j=1}^{N_{A_{k,m}^\mathsf{c},m}^{\mathsf{a}}}
p^\mathsf{c,a-2}_{k,j,m}
}{
\sum_{m=1}^M
\sum_{j=1}^{N^\mathsf{a}_{\infty,m}} W_{j,m}^\mathsf{a}
}
,
\label{EMsolve2}
\end{equation}
\begin{equation*}
\hat{\alpha}^\mathsf{a,c}_{\mathsf{1},*}
=
\frac{
\sum_{m=1}^M
\sum_{k=1}^{N_{\infty,m}^\mathsf{a}}
\sum_{i=0}^{N^\mathsf{c}_{A_{k,m}^\mathsf{a},m}} p_{k,i,m}^\mathsf{a,c-1}
}
{
\sum_{m=1}^M
\sum_{i=0}^{N^\mathsf{c}_{\infty,m}}  S_{i,m}^\mathsf{c} \sum_{k=1}^{\kappa_m} \bigg(
e^{-\beta^\mathsf{a,c} \slash K_{\Delta_{k-1,m}} (\Delta_{k-1,m} - A_{i,m}^\mathsf{c})^+}
 - 
 e^{-\beta^\mathsf{a,c} \slash K_{\Delta_{k-1,m}} (\Delta_{k,m} - A_{i,m}^\mathsf{c})^+}  
\bigg)
}
,
\end{equation*}
\begin{equation}
\hat{\alpha}^\mathsf{a,c}_{\mathsf{2},*}
=
\frac{
\sum_{m=1}^M
\sum_{k=1}^{N_{\infty,m}^\mathsf{a}}
\sum_{i=0}^{N^\mathsf{c}_{A_{k,m}^\mathsf{a},m}} p_{k,i,m}^\mathsf{a,c-2}
}
{
\sum_{m=1}^M
\sum_{i=0}^{N^\mathsf{c}_{\infty,m}}   W_{i,m}^\mathsf{c}
\sum_{k=1}^{\kappa_m} \bigg(
e^{-\beta^\mathsf{a,c} \slash K_{\Delta_{k-1,m}} (\Delta_{k-1,m} - A_{i,m}^\mathsf{c})^+}
 - 
 e^{-\beta^\mathsf{a,c} \slash K_{\Delta_{k-1,m}} (\Delta_{k,m} - A_{i,m}^\mathsf{c})^+}  
\bigg)
}
,
\label{EMsolve3}
\end{equation}
and
\begin{equation*}
\hat{\alpha}^\mathsf{a,a}_{\mathsf{1},*}
=
\frac{ 
\sum_{m=1}^M
\sum_{k=1}^{N^\mathsf{a}_{\infty,m}}
\sum_{j=1}^{k-1}
p_{k,j,m}^\mathsf{a,a-1}
}{
\sum_{m=1}^M
\sum_{j=1}^{N^\mathsf{a}_{\infty,m}}  S_{j,m}^\mathsf{a}
\sum_{k=1}^{\kappa_m} \left(e^{-\beta^\mathsf{a,a}\slash K_{\Delta_{k-1,m}} (\Delta_{k-1,m} - A_{j,m}^\mathsf{a})^+}  - e^{-\beta^\mathsf{a,a} \slash K_{\Delta_{k-1,m}} (\Delta_{k,m} - A_{j,m}^\mathsf{a})^+}  \right)
}
,
\end{equation*}
\begin{equation}
\hat{\alpha}^\mathsf{a,a}_{\mathsf{2},*}
=
\frac{ 
\sum_{m=1}^M
\sum_{k=1}^{N^\mathsf{a}_{\infty,m}}
\sum_{j=1}^{k-1}
p_{k,j,m}^\mathsf{a,a-2}
}{
\sum_{m=1}^M
\sum_{j=1}^{N^\mathsf{a}_{\infty,m}}   W_{j,m}^\mathsf{a}
\sum_{k=1}^{\kappa_m} \left(e^{-\beta^\mathsf{a,a} \slash K_{\Delta_{k-1,m}} (\Delta_{k-1,m} - A_{j,m}^\mathsf{a})^+}  - e^{-\beta^\mathsf{a,a} \slash K_{\Delta_{k-1,m}} (\Delta_{k,m} - A_{j,m}^\mathsf{a})^+}  \right)
}
.
\label{EMsolve4}
\end{equation}
Likewise, the decay rates are given by
\begin{equation}
\beta^\mathsf{c,c}_*
=
\frac{ 
\sum_{m=1}^M
\sum_{k=1}^{N_{\infty,m}^\mathsf{c}}
\sum_{i=0}^{k-1} 
(
p_{k,i,m}^\mathsf{c,c-1} + p_{k,i,m}^\mathsf{c,c-2} 
)
}{
\sum_{m=1}^M
\sum_{k=1}^{N_{\infty,m}^\mathsf{c}}
\sum_{i=0}^{k-1} 
(
p_{k,i,m}^\mathsf{c,c-1} + p_{k,i,m}^\mathsf{c,c-2} 
)
\left( A_{k,m}^\mathsf{c} - A_{i,m}^\mathsf{c} \right)
}
,
\label{EMsolve5}
\end{equation}
\begin{equation}
\beta^\mathsf{c,a}_*
=
\frac{
\sum_{m=1}^M \sum_{k=1}^{N^\mathsf{c}_{\infty,m}} 
  \sum_{j=1}^{N_{A_{k,m}^\mathsf{c},m}^{\mathsf{a}}}
  (
p^\mathsf{c,a-1}_{k,j,m} + p^\mathsf{c,a-2}_{k,j,m}
)
}{
\sum_{m=1}^M \sum_{k=1}^{N^\mathsf{c}_{\infty,m}} 
  \sum_{j=1}^{N_{A_{k,m}^\mathsf{c},m}^{\mathsf{a}}}
  (
p^\mathsf{c,a-1}_{k,j,m} + p^\mathsf{c,a-2}_{k,j,m}
)
\left(A_{k,m}^\mathsf{c} - A_{j,m}^\mathsf{a}\right)
}
,
\label{EMsolve6}
\end{equation}
\begin{align}
\beta^\mathsf{a,c}_*
&=
\Bigg(
\sum_{m=1}^M
 \sum_{k=1}^{N_{\infty,m}^\mathsf{a}} 
\sum_{i=0}^{N^\mathsf{c}_{A_{k,m}^\mathsf{a},m}}
(
p_{k,i,m}^\mathsf{a,c-1} + p_{k,i,m}^\mathsf{a,c-2}
)
\Bigg)
\Bigg\slash
\Bigg(
 \sum_{m=1}^M
 \sum_{k=1}^{N_{\infty,m}^\mathsf{a}} 
 \sum_{i=0}^{N^\mathsf{c}_{A_{k,m}^\mathsf{a},m}} 
(
p_{k,i,m}^\mathsf{a,c-1} + p_{k,i,m}^\mathsf{a,c-2}
)
\frac{ A_{k,m}^\mathsf{a} - A_{i,m}^\mathsf{c}}{K_{A_{k,m}^\mathsf{a}}}
 \nonumber
\\
&
\quad
-
\sum_{m=1}^M
\sum_{i=0}^{N_{\infty,m}^\mathsf{c}}  
(
\hat{\alpha}^{\mathsf{a,c}}_\mathsf{1}  S_{i,m}^\mathsf{c} + \hat{\alpha}^{\mathsf{a,c}}_\mathsf{2}  W_{i,m}^\mathsf{c} 
)
\sum_{k=1}^{\kappa_m} \bigg(
e^{-\beta^\mathsf{a,c} \frac{(\Delta_{k-1,m} - A_{i,m}^\mathsf{c})^+}{K_{\Delta_{k-1,m}}}} 
\frac{(\Delta_{k-1,m} - A_{i,m}^\mathsf{c})^+}{K_{\Delta_{k-1,m}}}
 \nonumber
\\
&
\quad
 - 
 e^{-\beta^\mathsf{a,c} \frac{(\Delta_{k,m} - A_{i,m}^\mathsf{c})^+}{K_{\Delta_{k-1,m}}}}  
\frac{(\Delta_{k,m} - A_{i,m}^\mathsf{c})^+}{K_{\Delta_{k-1,m}}}
\bigg)
\Bigg)
,
\label{EMsolve7}
\end{align}
and
\begin{align}
\beta^\mathsf{a,a}_*
&=
\Bigg(
\sum_{m=1}^M
\sum_{k=1}^{N^\mathsf{a}_{\infty,m}}
\sum_{j=1}^{k-1}
(
p_{k,j,m}^\mathsf{a,a-1}
+
p_{k,j,m}^\mathsf{a,a-2}
)
\Bigg)
\Bigg\slash
\Bigg(
\sum_{m=1}^M
\sum_{k=1}^{N^\mathsf{a}_{\infty,m}}
\sum_{j=1}^{k-1}
(
p_{k,j,m}^\mathsf{a,a-1}
+
p_{k,j,m}^\mathsf{a,a-2}
)
\frac{ A_{k,m}^\mathsf{a} - A_{j,m}^\mathsf{a}}{K_{A_{k,m}^\mathsf{a}}}
   \nonumber
\\
&
\quad
-
\sum_{m=1}^M
\sum_{j=1}^{N^\mathsf{a}_{\infty,m}}
(
\hat{\alpha}^{\mathsf{a,a}}_\mathsf{1} S_{j,m}^\mathsf{a} + \hat{\alpha}^{\mathsf{a,a}}_\mathsf{2} W_{j,m}^\mathsf{a}
 ) 
\sum_{k=1}^{\kappa_m} 
\bigg(
e^{-\beta^\mathsf{a,a}  \frac{(\Delta_{k-1,m} - A_{j,m}^\mathsf{a})^+}{K_{\Delta_{k-1,m}}} }
\frac{(\Delta_{k-1,m} - A_{j,m}^\mathsf{a})^+}{K_{\Delta_{k-1,m}}}
  \nonumber
\\
&
\quad
- 
 e^{-\beta^\mathsf{a,a} \frac{(\Delta_{k,m} - A_{j,m}^\mathsf{a})^+}{K_{\Delta_{k-1,m}}} }  
\frac{(\Delta_{k,m} - A_{j,m}^\mathsf{a})^+}{K_{\Delta_{k-1,m}}}
\bigg)
\Bigg)
.
\label{EMsolve8}
\end{align}
With these quantities in hand, one can directly compute all steps of Algorithm~\ref{EM-SysBHP}.

\subsection{Benchmark Service Models: Static and Stage-Dependent Phase-Type Models}
\label{sec:model_phase}
We compare our suggested model to  service models that are inspired by well-known activity-based service models.
The classic service literature views the service process as a series of tasks, using phase-type distributions as the stochastic models for the service progression \citep{Mandelbaum1998}. With this in mind, our benchmark model views the service as a repeating task Markov chain model, where the time between messages is exponentially distributed. This is similar in nature to re-entry service models, such as the Erlang-R, used to describe service duration in hospitals and contact centers \citep[see][]{YomTov2014ErlangR,campello2017}. In that model, the number of messages is geometric with mean $1/(1-p)$ and the service duration is viewed as the sum of i.i.d.\ exponentials generated via a phase-type model. Following this, our basic model assumes i.i.d.\ exponential response times with rate $\mu$ and a geometrically distributed number of messages, where all random variables are independent. We denote this model as the \textit{sum of exponentials-static} (SES) model.

However, there is no reason to believe that the conversation phases progress with constant rate or satisfy a universal mean rate. Indeed, our data reveals that the number of messages in a conversation fits to a negative-binomial distribution rather than a geometric (see Figure \ref{fig:GapsBinomial}) and that response time may vary with the conversation state (see Table \ref{tbl:ParametersGammaTrain}). The latter was also confirmed by  \citet{Altman2019EmotionalLoad}, showing that the conversation's stage predicts agent response time. We cannot know in real time what stage the conversation has reached, but nevertheless, we can know how many messages the conversation has had so far. Therefore, our second benchmark service model generalizes the SES model by allowing for phase-varying rates for both the transition probability and the response time between messages. We call this model \textit{sum of exponentials-dynamic} (SED). Here, the exponential random variables are independent but non-identical, and we draw the number of gaps from a negative-binomial distribution. 
In this way, both SES and SED may be viewed as Coxian phase-type constructions of the service duration, where SES has the same rate and absorption probability in every phase but SED allows the rates and probabilities to vary between phases.\footnote{We also considered gamma distribution variants of these models, i.e.,~SGS and SGD, but their performance was not different enough from SES and SED to merit their inclusion.} 

It is important to note, however, that both the SES and SED models are still path-independent. Although the SED rates may change from one phase to the next, the response time random variables are still independent from one another and from the total number of messages. So, while SES and SED may be well-represented in the literature, these stochastic processes cannot capture the dependencies between messages.

\section{Additional Figures, Tables, and Descriptions of Data and Experiment Outcomes}
\label{app:outof}

In this final section of the appendix, we provide additional depictions and descriptions of our results. Let us briefly detail each one. In Figure~\ref{fig:Arrivals}, we show that customer arrivals follow a standard service pattern through the hours of the day, and that this pattern is largely consistent over the course of the month. Although initial customer arrivals are outside our modeling scope, this provides a valuable frame of reference, particularly so relative to Figures~\ref{fig:durations} and~\ref{fig:gaptimes}. To inspect time-of-day effects within the conversations, Figure~\ref{fig:durations} shows the mean service duration by hour of day and by primary (e.g., most held) agent concurrency in the conversation, and Figure~\ref{fig:gaptimes} shows the mean time between messages by hour of day and by agent concurrency at the start of the gap. We see here that the average duration and gap time across concurrencies show only mild hourly variation relative to the arrival rate Figure~\ref{fig:Arrivals}, and the concurrency-conditional means show even less. Furthermore, we see that larger values of the concurrency have larger impacts on these means, supporting the style of $\mu_k$ in Propositions~\ref{firstSPPprop} and~\ref{secondSPPprop}.

Then, Figure~\ref{fig:GapsBinomial} and Table~\ref{tbl:ParametersGammaTrain} provide details and context for the stage-dependent benchmark model (SED) discussed in Section~\ref{sec:model}. Table~\ref{tbl:ParametersGammaTrain} shows the rates of the exponential random variables in each stage, while Figure~\ref{fig:GapsBinomial} motivates the parametric form of the distribution for the number of stages, meaning the number of gaps or messages in the conversation. Next, Figures~\ref{fig:starplotsall1} and~\ref{fig:starplotsall2} show the broader collection of dial plots, as referred to in Section~\ref{sec:residual}. Figure~\ref{fig:starplotsall1} shows the dial plots for conversations with total number of messages ranging from $N=3$ to $N=18$, and Figure~\ref{fig:starplotsall2} then contains $N=19$ to $N=31$. Finally, Tables~\ref{tbl:CompAllMIwait} through~\ref{tbl:CompAllSDOwait} contain detailed results of the data-driven routing experiment from Section~\ref{sec:insights}. Table~\ref{tbl:CompAllMIwait} shows the mean inner wait for each policy in each parameter setting, and similarly Table~\ref{tbl:CompAllMOwait} shows each mean outer wait. Tables~\ref{tbl:CompAllSDIwait} and~\ref{tbl:CompAllSDOwait} then show the standard deviation of mean and outer wait, respectively, and Table~\ref{tbl:CompAllPOwait} displays the fraction of customers who experience outer wait in each setting. In all five tables, the value in parenthesis represents the percent improvement of the various LL$+$HP policies over the standard LL policy.

\begin{figure}[tbh]
\centering
\subfigure[All days]{
\includegraphics[width=0.30\textwidth]{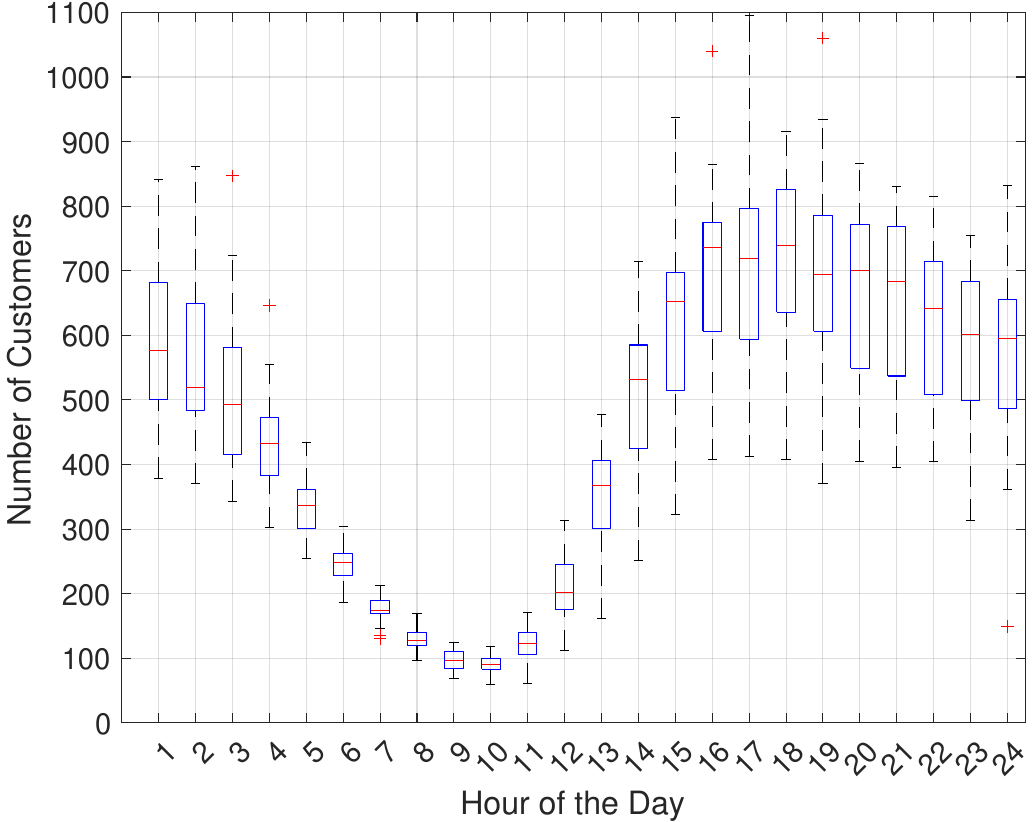} \label{fig:ArrivalAll}
} 
\subfigure[Train set]{
\includegraphics[width=0.30\textwidth]{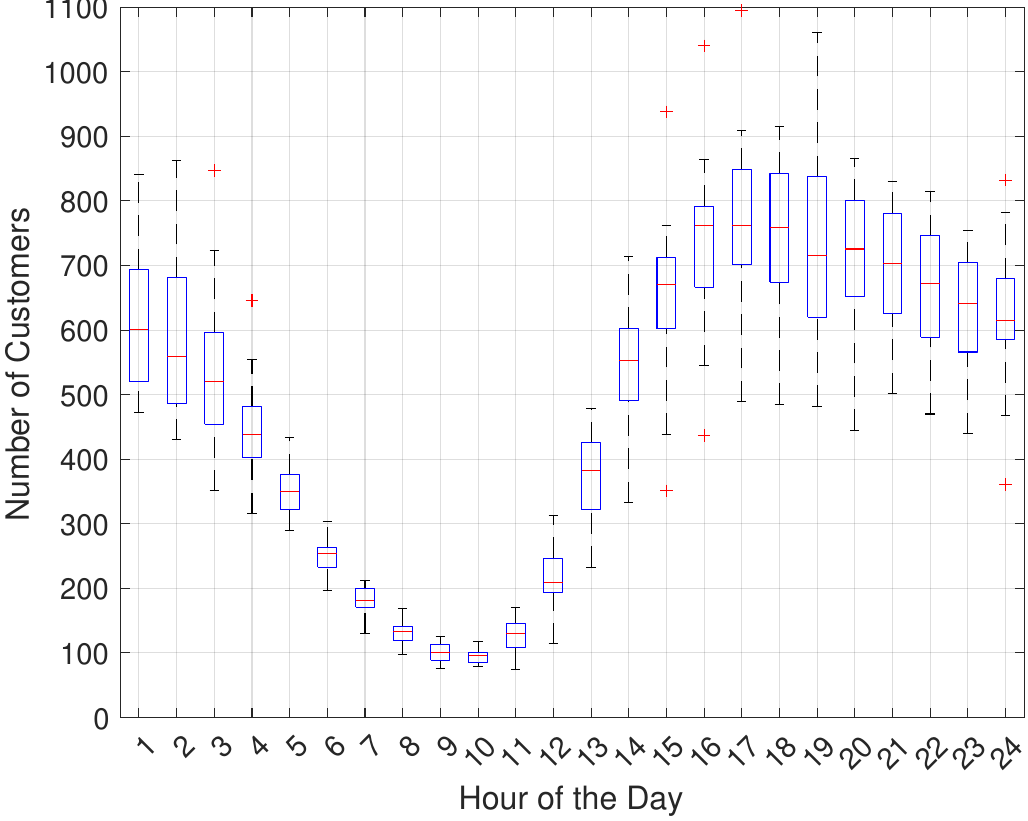} \label{fig:ArrivalTrain}
} 
\subfigure[Test set]{
\includegraphics[width=0.30\textwidth]{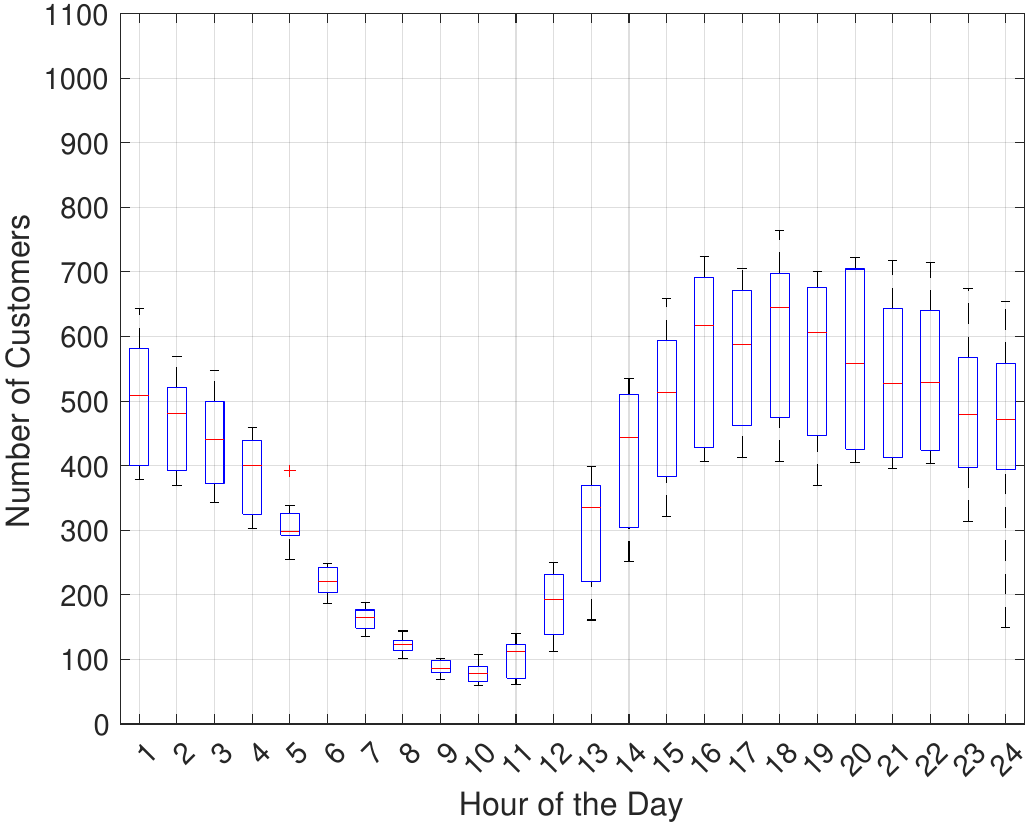} \label{fig:ArrivalHour}
}  \label{fig:ArrivalTest}
\caption{Number of conversations per hour. May 2017.}
\label{fig:Arrivals}
\end{figure}

\begin{figure}[htbp]
    \centering
    \includegraphics[width=\textwidth]{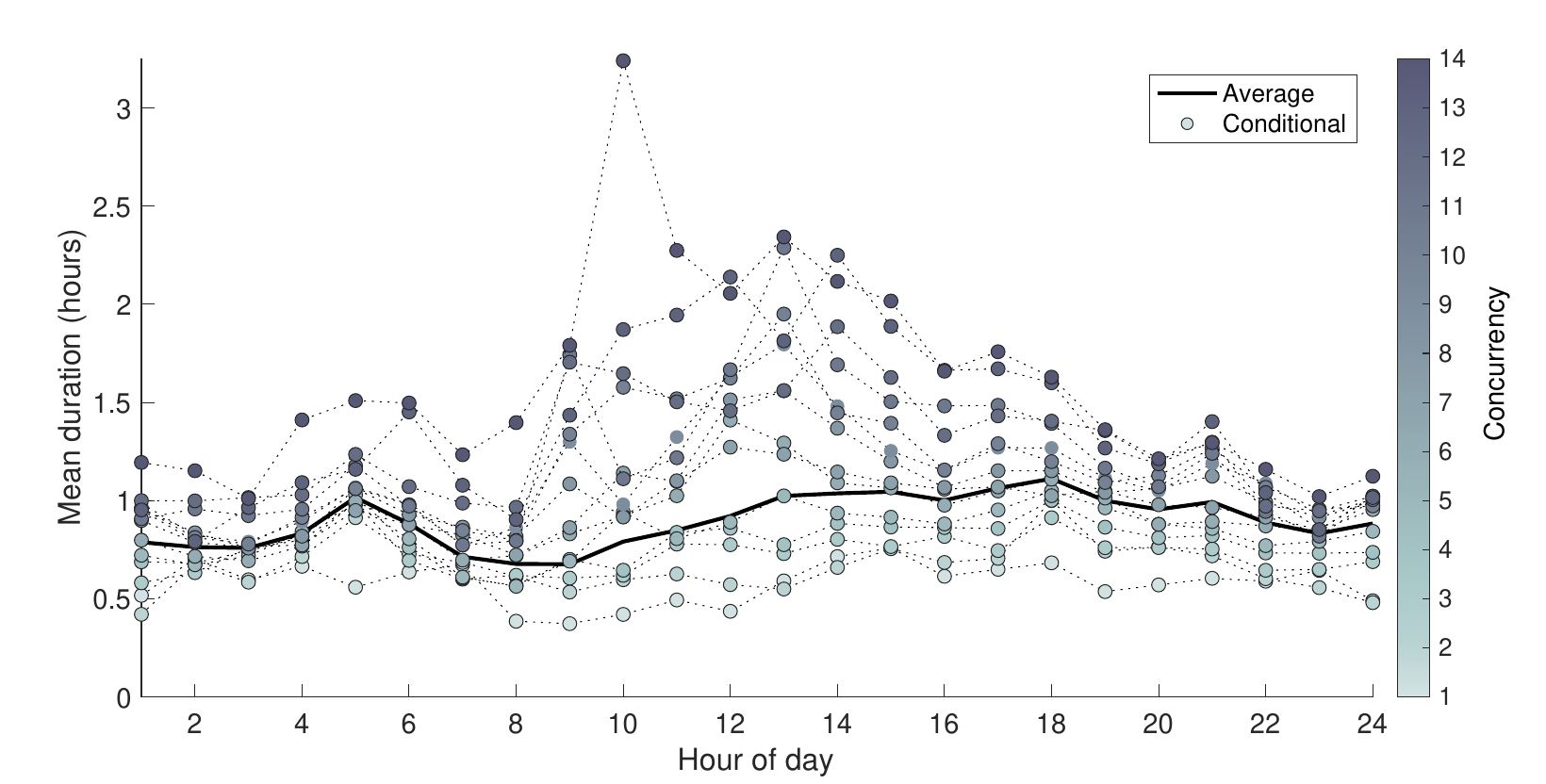}
    \caption{Average service duration by hour of day and primary agent concurrency during the conversation.}
    \label{fig:durations}
\end{figure}

\begin{figure}[htbp]
    \centering
    \includegraphics[width=\textwidth]{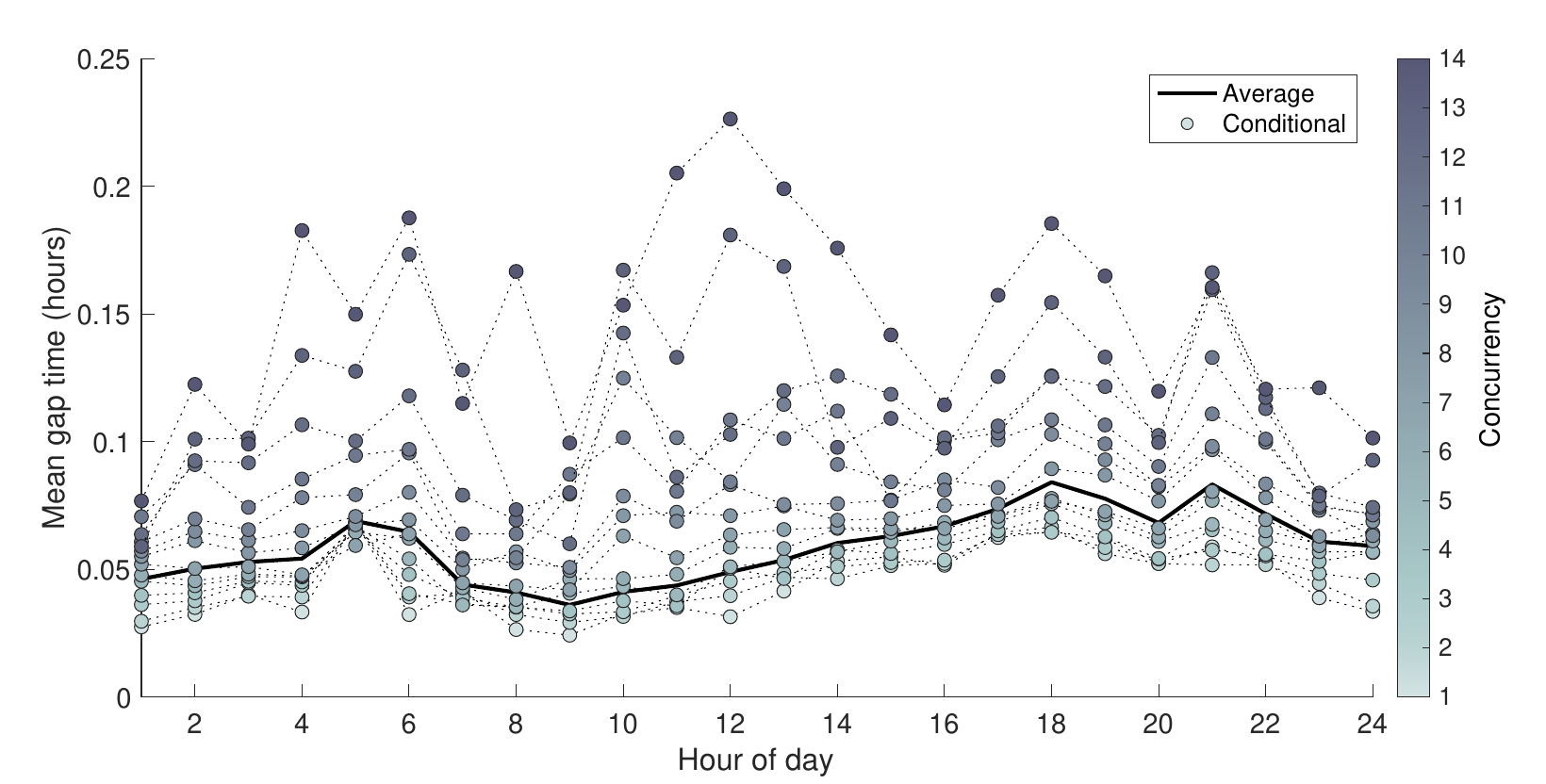}
    \caption{Average gap time (time between messages) by hour of day and initial agent concurrency in the gap.}
    \label{fig:gaptimes}
\end{figure}

\begin{figure}[htbp]
    \centering
    \includegraphics[width=0.55\textwidth]{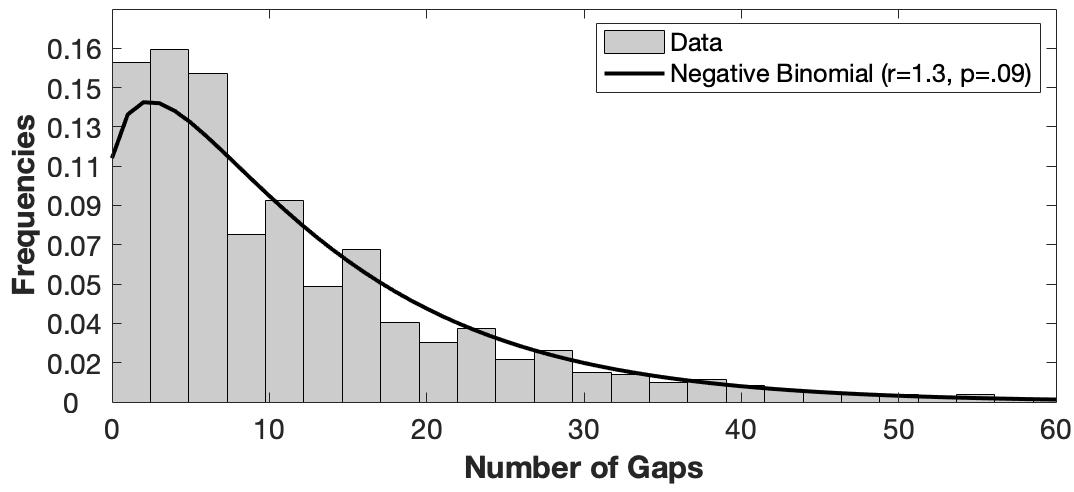}
    \caption{Negative binomial distribution fit on number of gaps, May 2017.}
    \label{fig:GapsBinomial}
\end{figure}

\begin{table}[htbp]
 \centering
  \caption{Estimated Parameters for SED Model  (Training set May 1-23 2017).}
  \begin{scriptsize}
  \begin{tabular}{ll} \toprule
  Gap number & Mean (\textit{standard error})  \\
  \midrule
  1 & 	$ 1/\mu_{1}=0.183$ (\textit{0.001}) \tabularnewline
  2  &	$ 1/\mu_{2}=0.085$ (\textit{0.002}) \tabularnewline
  3  &
	$ 1/\mu_{3}=0.095$ (\textit{0.002}) \tabularnewline
 4   &
	$ 1/\mu_{4}=0.092$
	(\textit{0.002})
	\tabularnewline
 5   &
 $ 1/\mu_{5}=0.081$ (\textit{0.002}) \tabularnewline
  6  &
 $  1/\mu_{6}=0.073$ (\textit{0.002}) \tabularnewline
 7   &
  $ 1/\mu_{7}=0.065$ 
  (\textit{0.002})
  \tabularnewline
 8   & 
 $ 1/\mu_{8}=0.062$ (\textit{0.002})
 \tabularnewline
 9   &
	$ 1/\mu_{9}=0.059$ (\textit{0.002}) \tabularnewline
  10  &
	$ 1/\mu_{10}=0.058$ (\textit{0.002}) \tabularnewline
 11   &
 $ 1/\mu_{11}=0.056$ (\textit{0.002}) \tabularnewline
 12   &
	$ 1/\mu_{12}=0.054$ (\textit{0.002}) \tabularnewline
  13  &
	$ 1/\mu_{13}=0.052$ (\textit{0.002}) \tabularnewline
 $>$14   &
	$ 1/\mu_{>14}=0.046$ (\textit{0.002})
 \tabularnewline
  \bottomrule
  \end{tabular}
  \end{scriptsize}
 \label{tbl:ParametersGammaTrain}
\end{table}


\begin{figure}[htbp]
\centering
\subfigure[$N=3$]{
\includegraphics[width=0.225\textwidth]{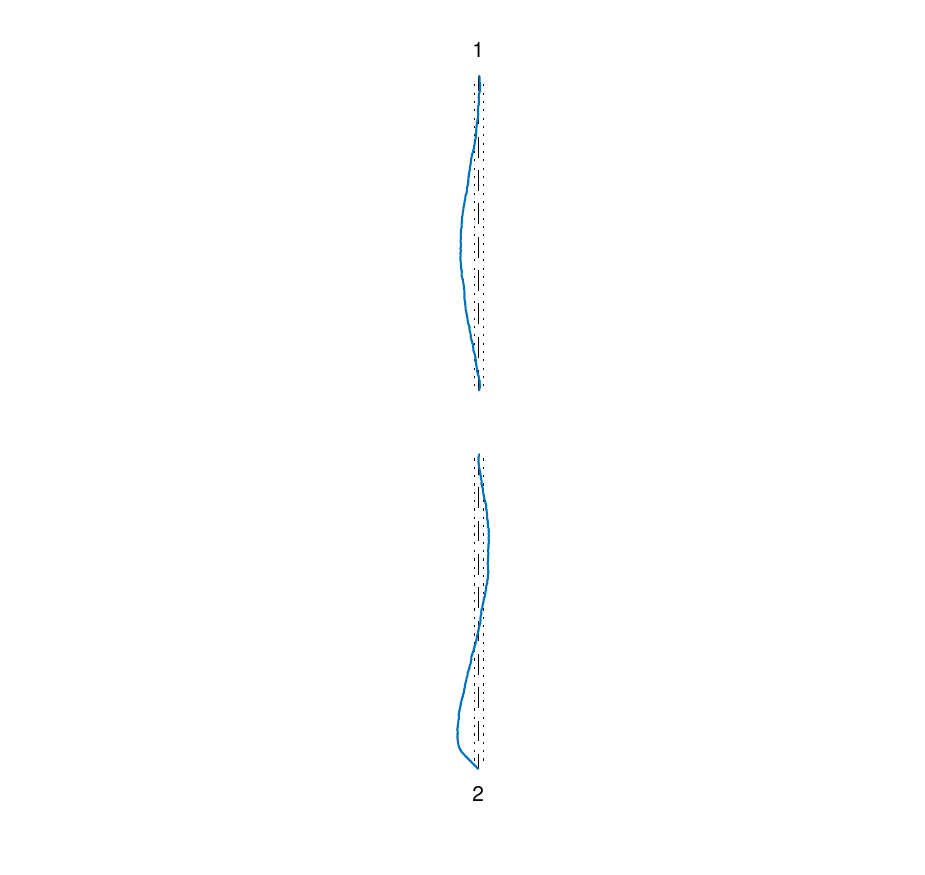} 
} 
\subfigure[$N=4$]{
\includegraphics[width=0.225\textwidth]{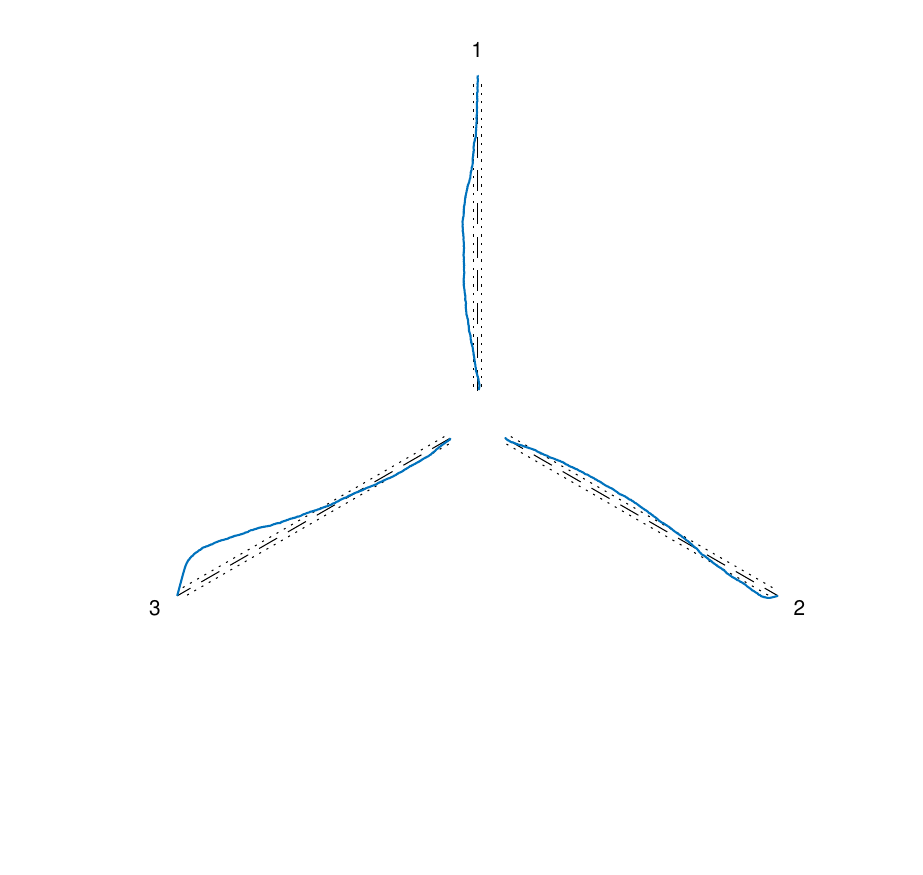} 
} 
\subfigure[$N=5$]{
\includegraphics[width=0.225\textwidth]{StartPlot_N_5.eps} 
} 
\subfigure[$N=6$]{
\includegraphics[width=0.225\textwidth]{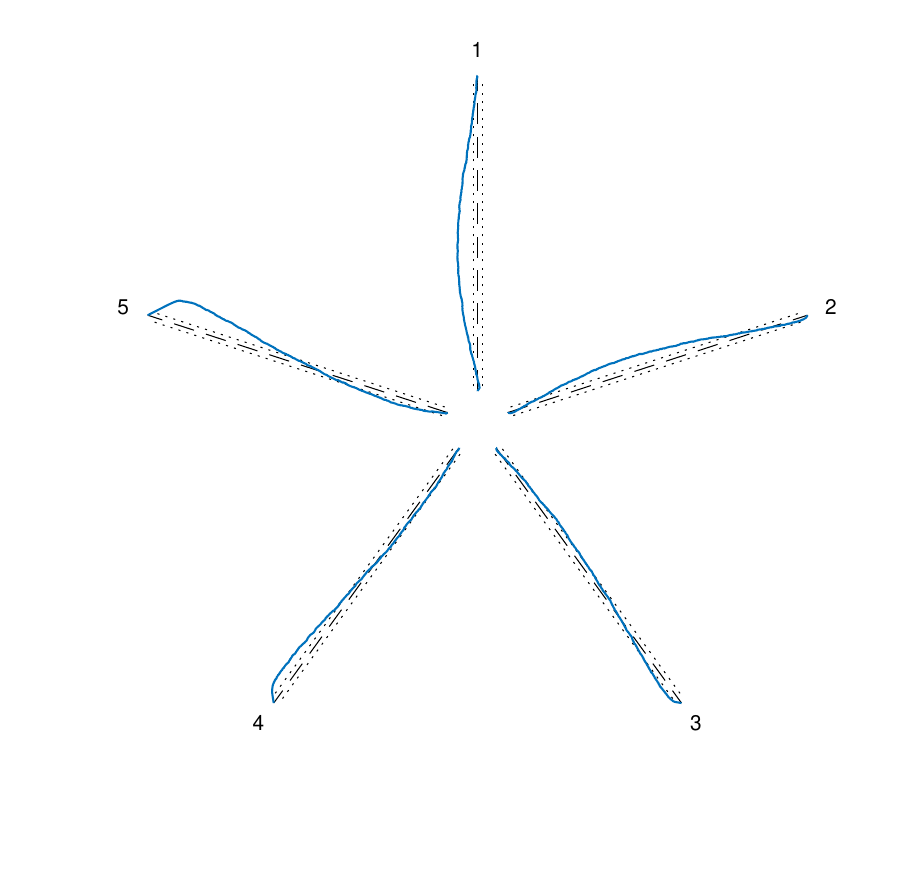} 
} 
\subfigure[$N=7$]{
\includegraphics[width=0.225\textwidth]{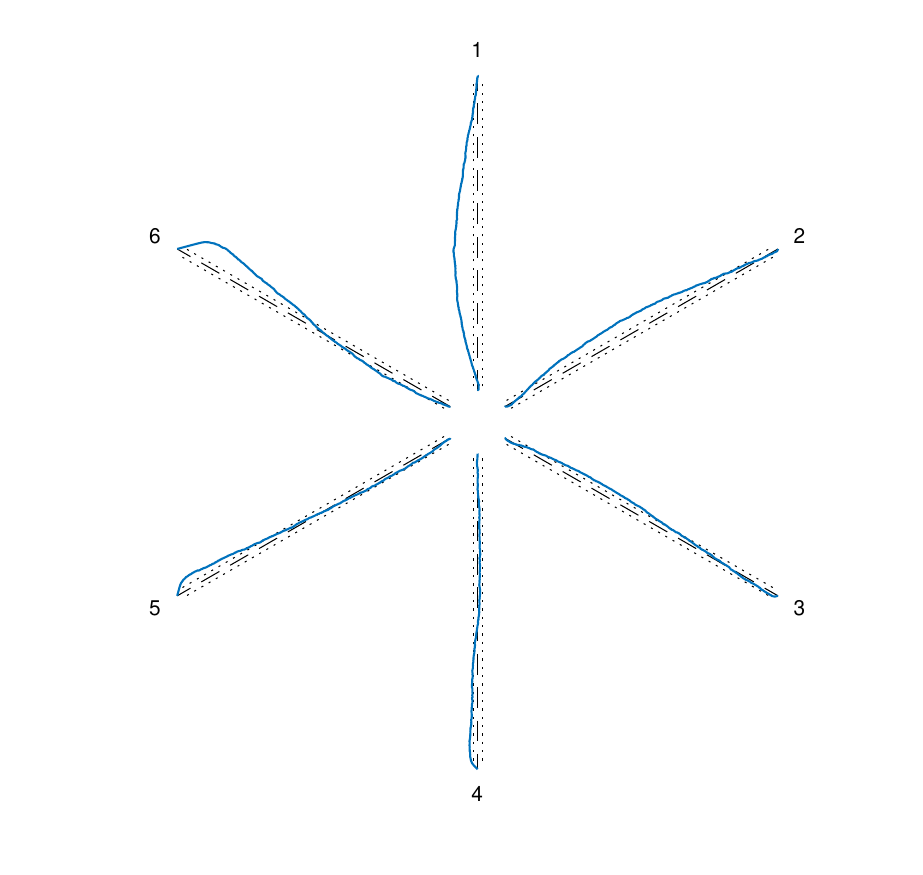} 
} 
\subfigure[$N=8$]{
\includegraphics[width=0.225\textwidth]{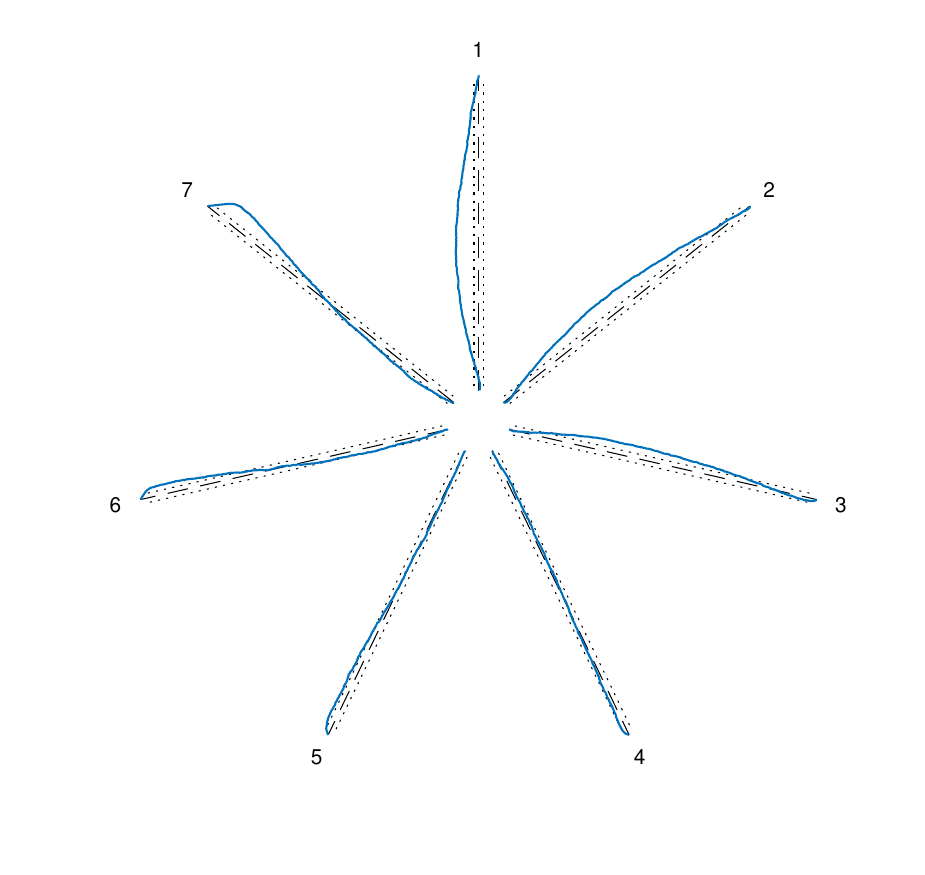} 
} 
\subfigure[$N=9$]{
\includegraphics[width=0.225\textwidth]{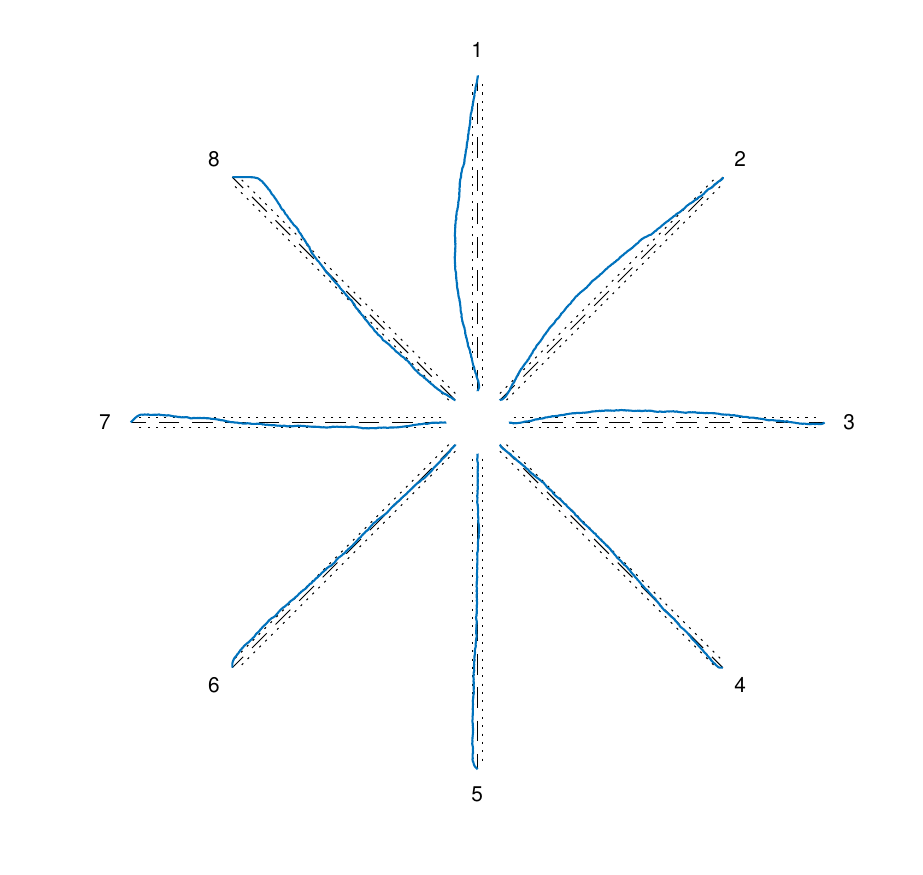} 
} 
\subfigure[$N=10$]{
\includegraphics[width=0.225\textwidth]{StartPlot_N_10.eps} 
} 
\subfigure[$N=11$]{
\includegraphics[width=0.225\textwidth]{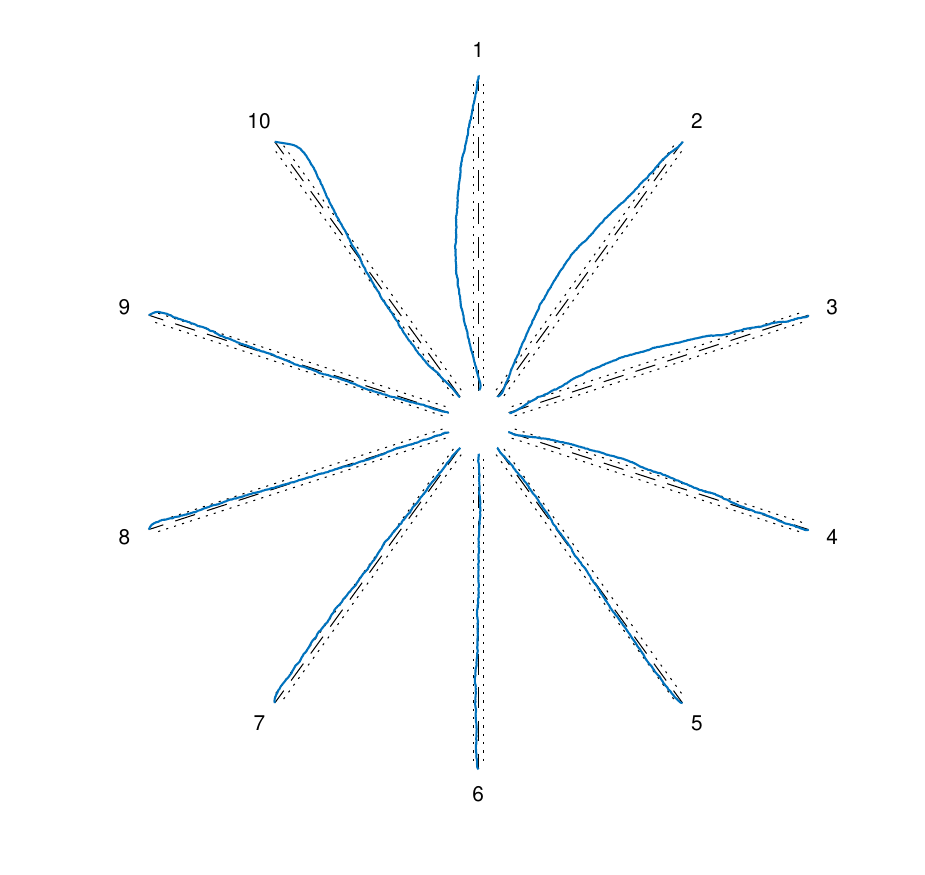} 
} 
\subfigure[$N=12$]{
\includegraphics[width=0.225\textwidth]{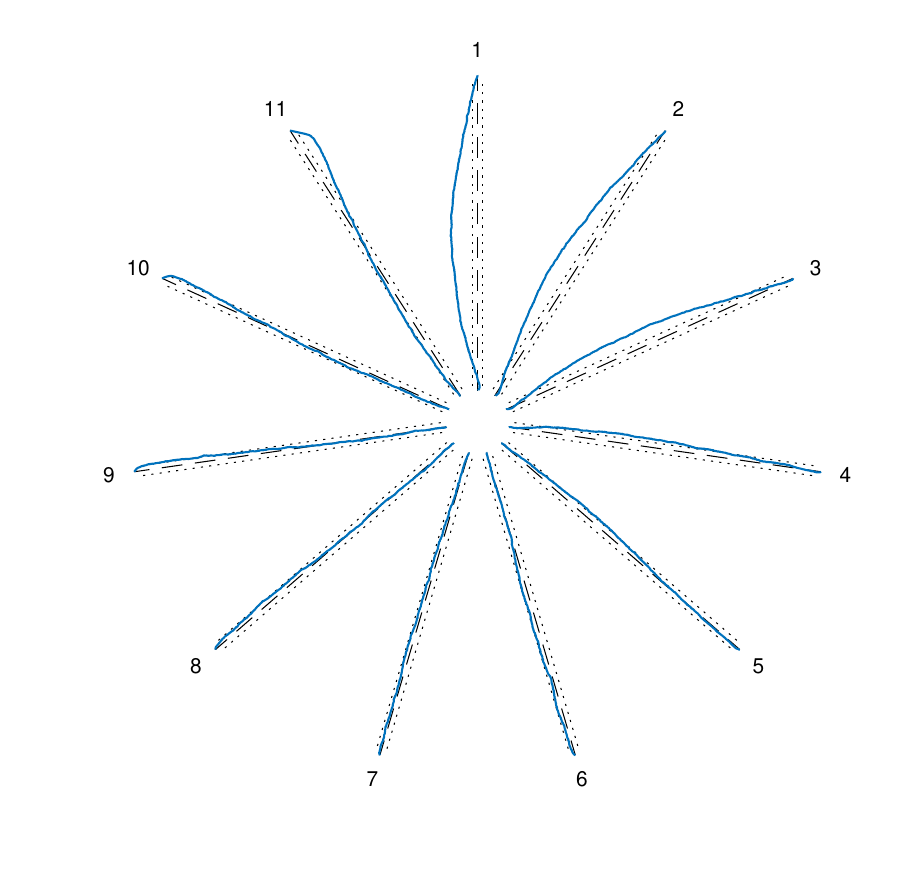} 
} 
\subfigure[$N=13$]{
\includegraphics[width=0.225\textwidth]{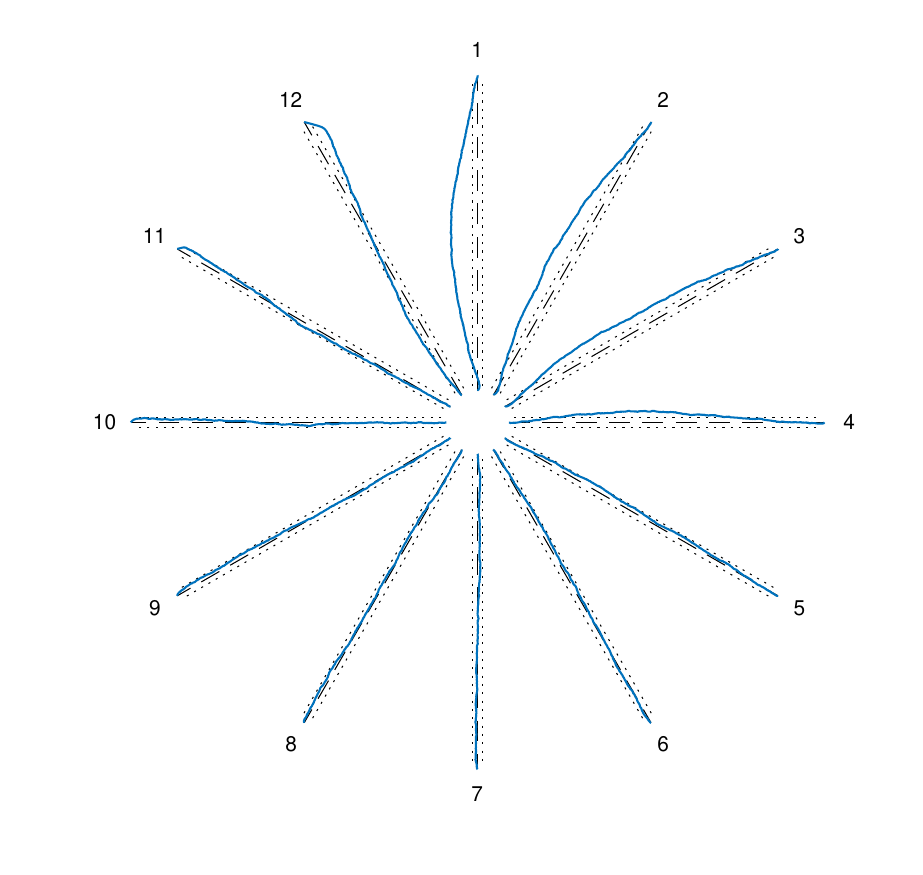} 
} 
\subfigure[$N=14$]{
\includegraphics[width=0.225\textwidth]{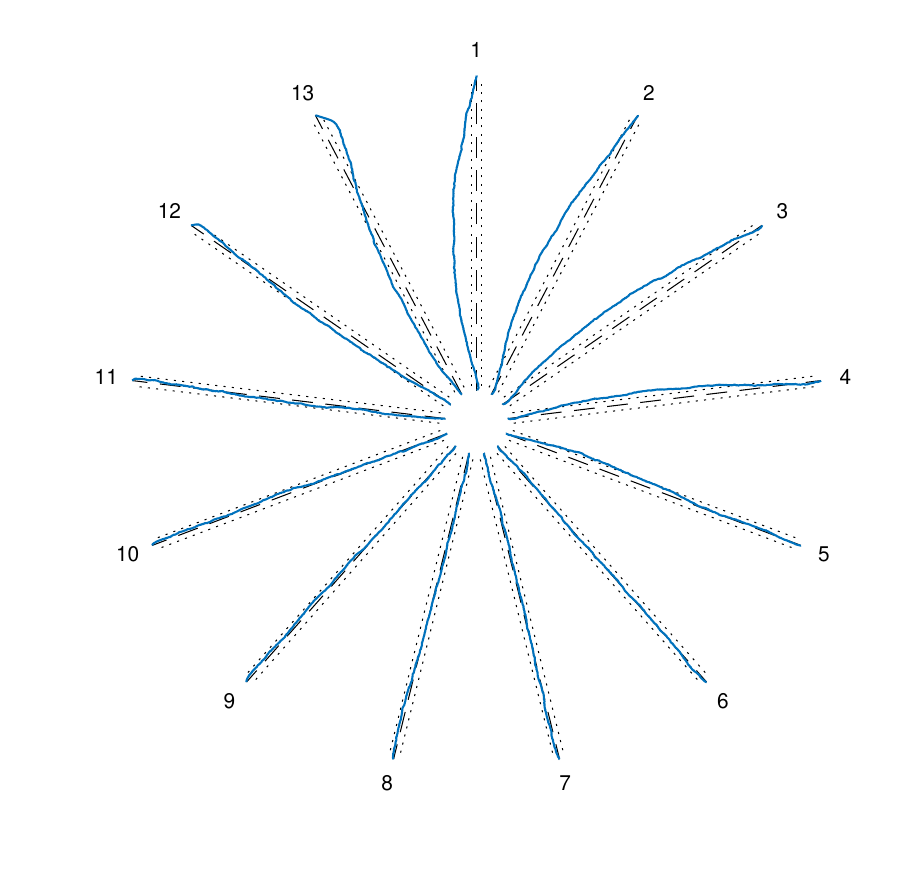} 
}
\subfigure[$N=15$]{
\includegraphics[width=0.225\textwidth]{StartPlot_N_15.eps} 
} 
\subfigure[$N=16$]{
\includegraphics[width=0.225\textwidth]{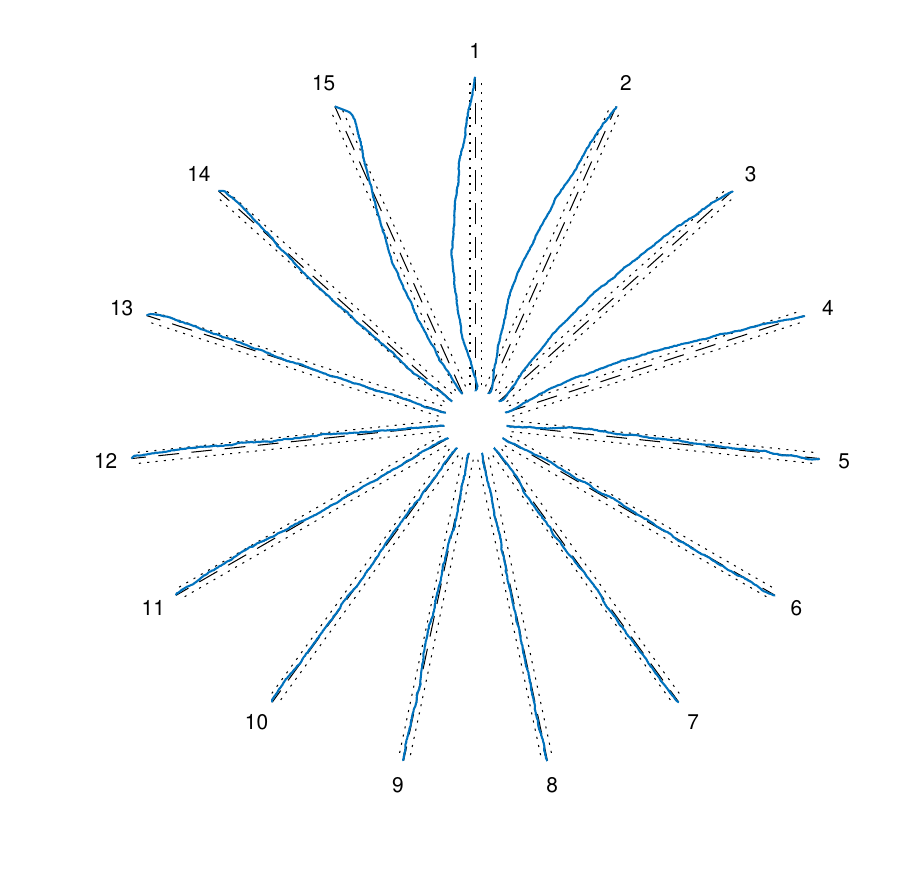} 
} 
\subfigure[$N=17$]{
\includegraphics[width=0.225\textwidth]{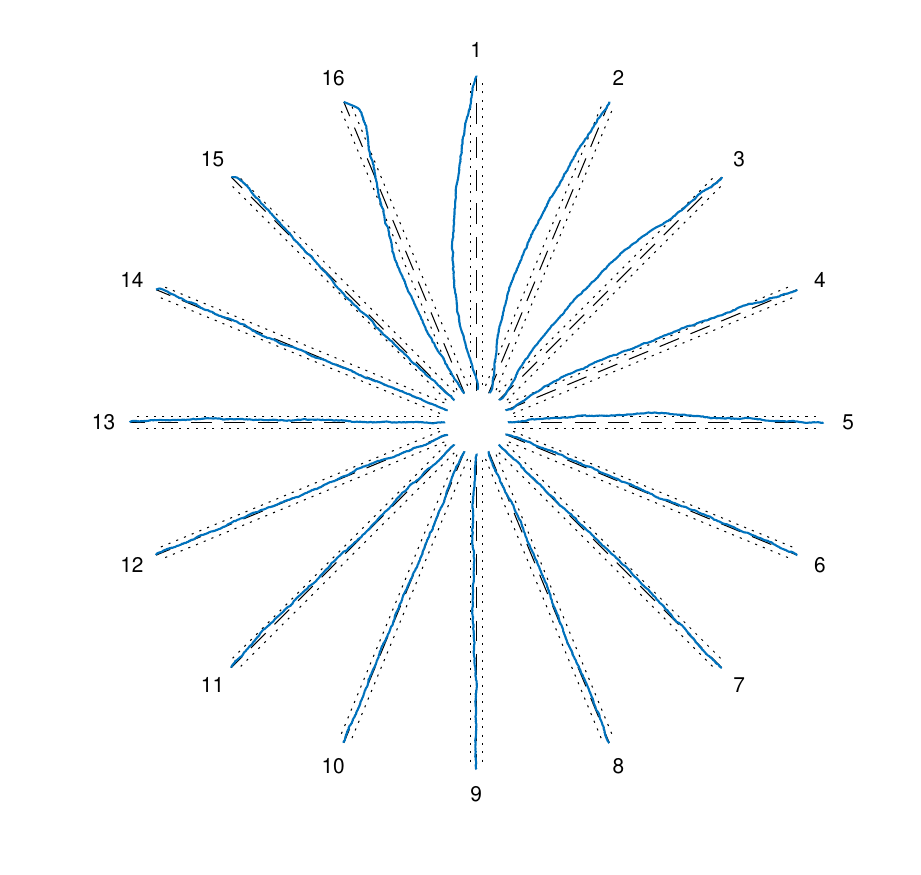} 
} 
\subfigure[$N=18$]{
\includegraphics[width=0.225\textwidth]{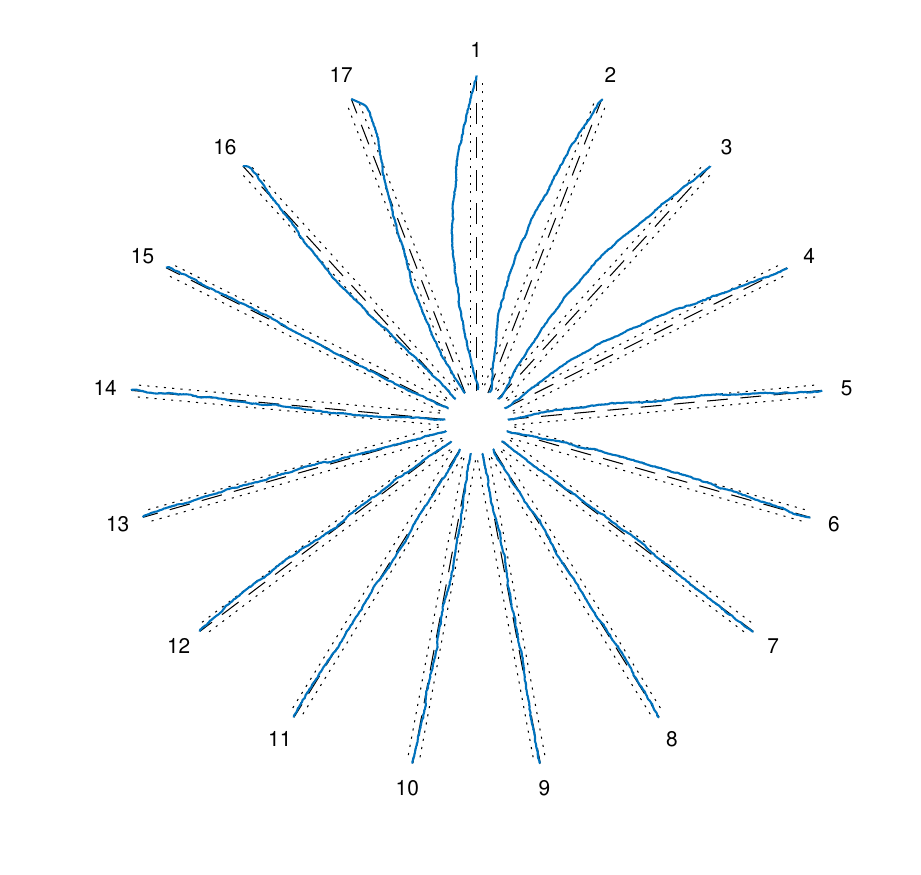} 
} 
\caption{All residual analysis ``dial plots'' of conditionally uniform UHP transformed empirical CDFs compared with true standard uniform CDFs and 99\% error bounds for $N = 3$ through $18$. Out-of-sample test. }
\label{fig:starplotsall1}
\end{figure}

\begin{figure}[htbp]
\centering
\subfigure[$N=19$]{
\includegraphics[width=0.225\textwidth]{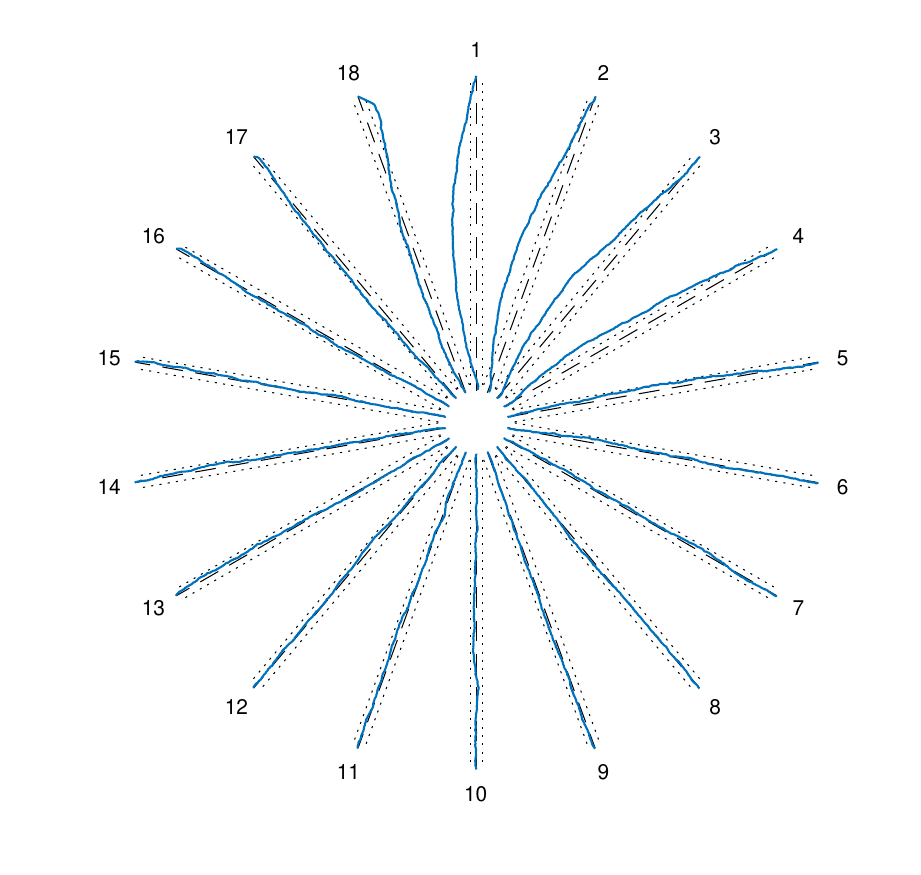} 
} 
\subfigure[$N=20$]{
\includegraphics[width=0.225\textwidth]{StartPlot_N_20.eps} 
} 
\subfigure[$N=21$]{
\includegraphics[width=0.225\textwidth]{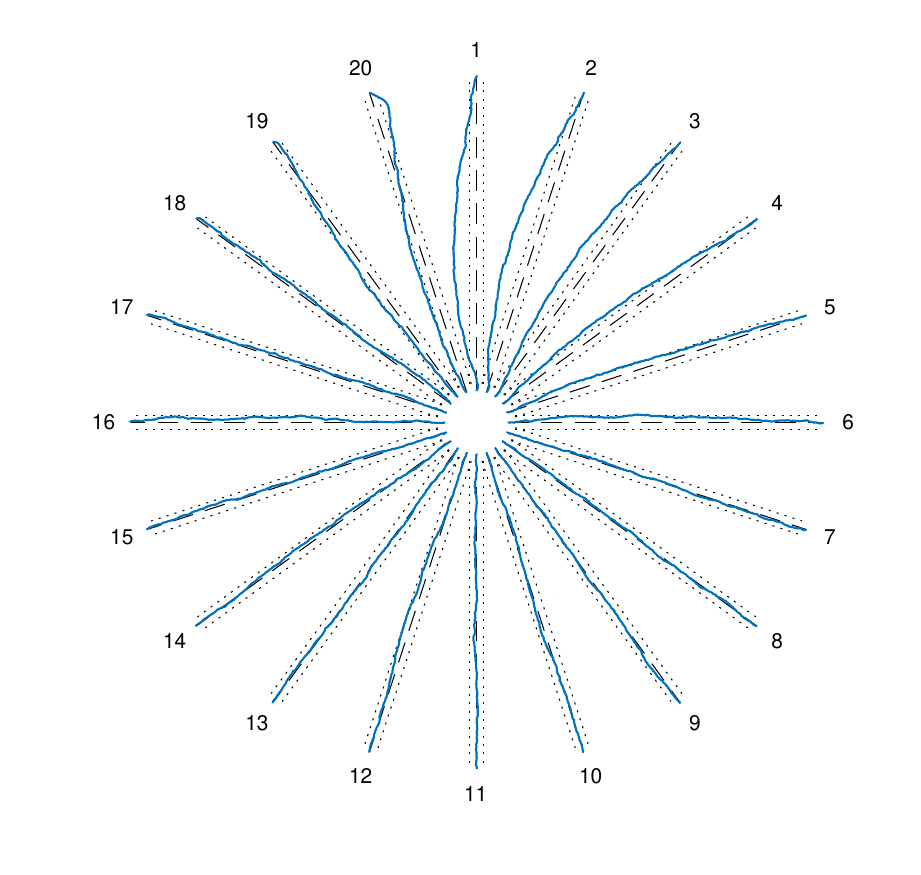} 
} 
\subfigure[$N=22$]{
\includegraphics[width=0.225\textwidth]{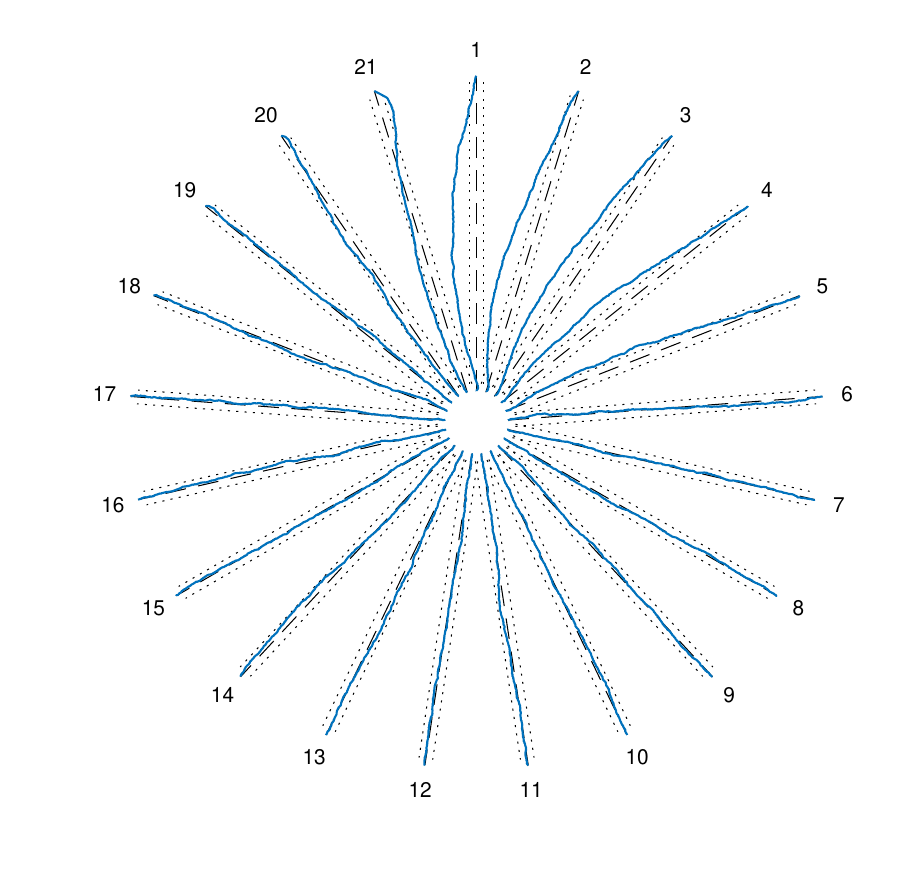} 
} 
\subfigure[$N=23$]{
\includegraphics[width=0.225\textwidth]{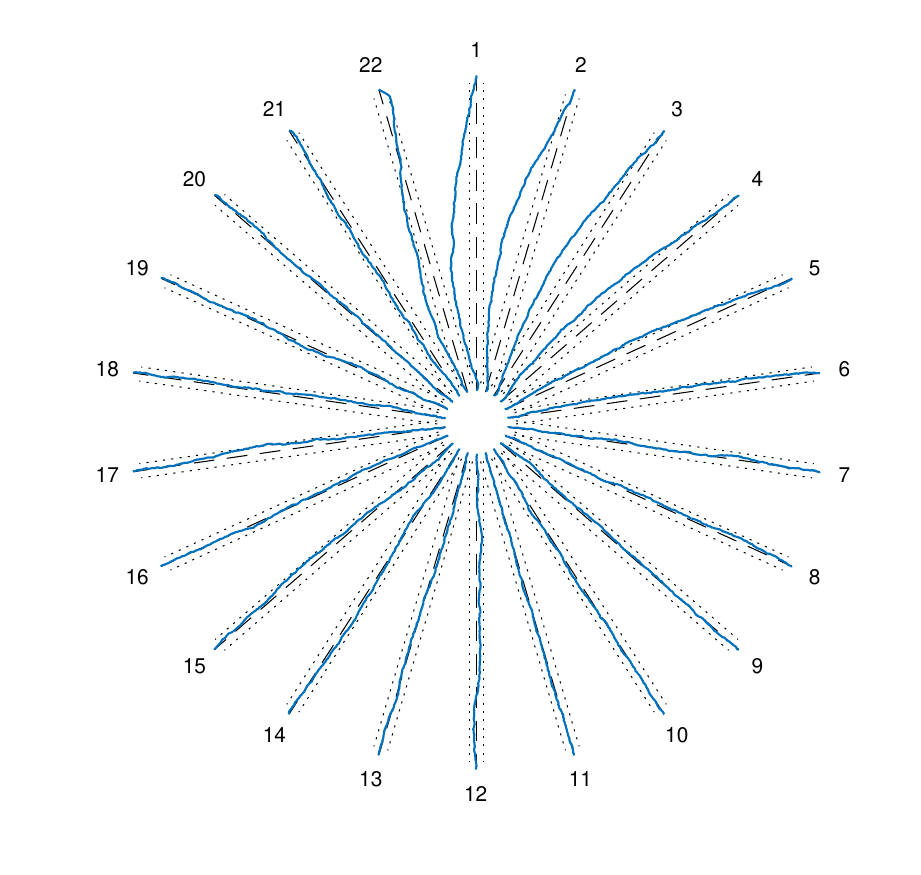} 
} 
\subfigure[$N=24$]{
\includegraphics[width=0.225\textwidth]{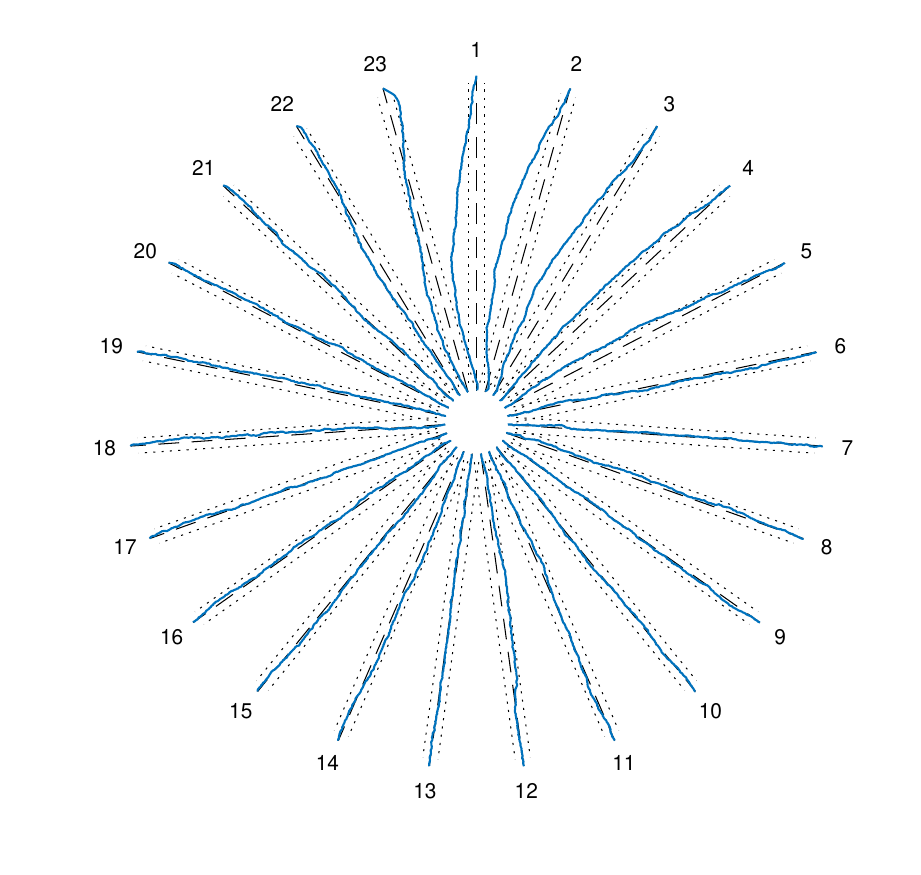} 
} 
\subfigure[$N=25$]{
\includegraphics[width=0.225\textwidth]{StartPlot_N_25.eps} 
} 
\subfigure[$N=26$]{
\includegraphics[width=0.225\textwidth]{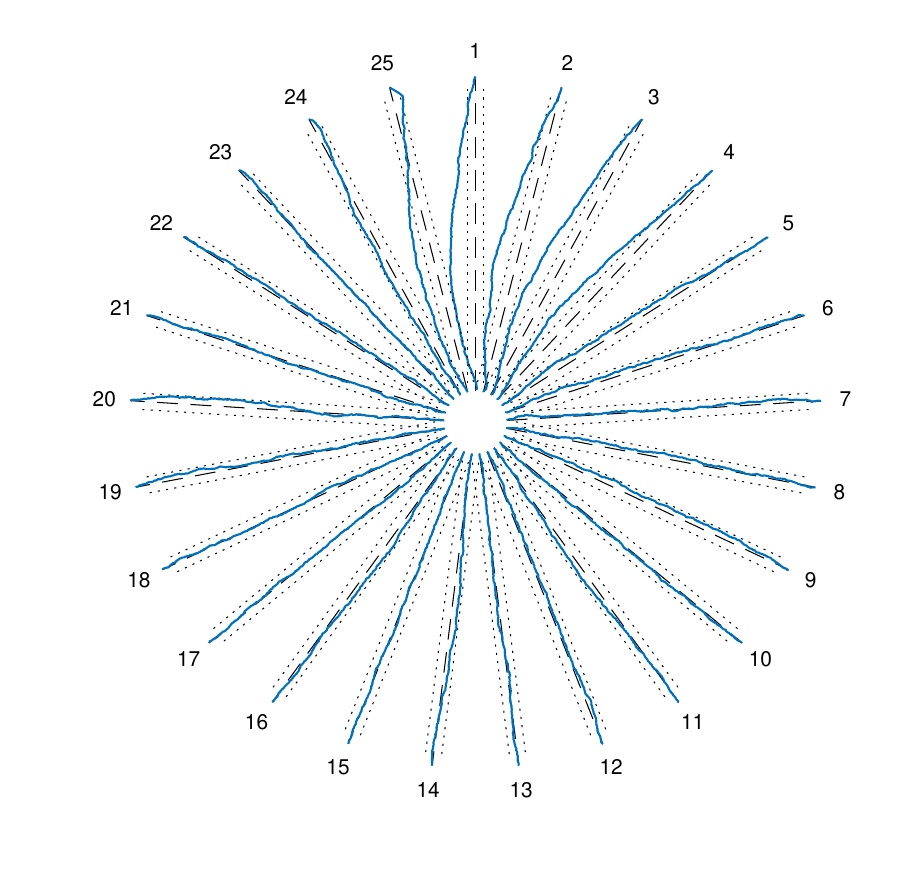} 
} 
\subfigure[$N=27$]{
\includegraphics[width=0.225\textwidth]{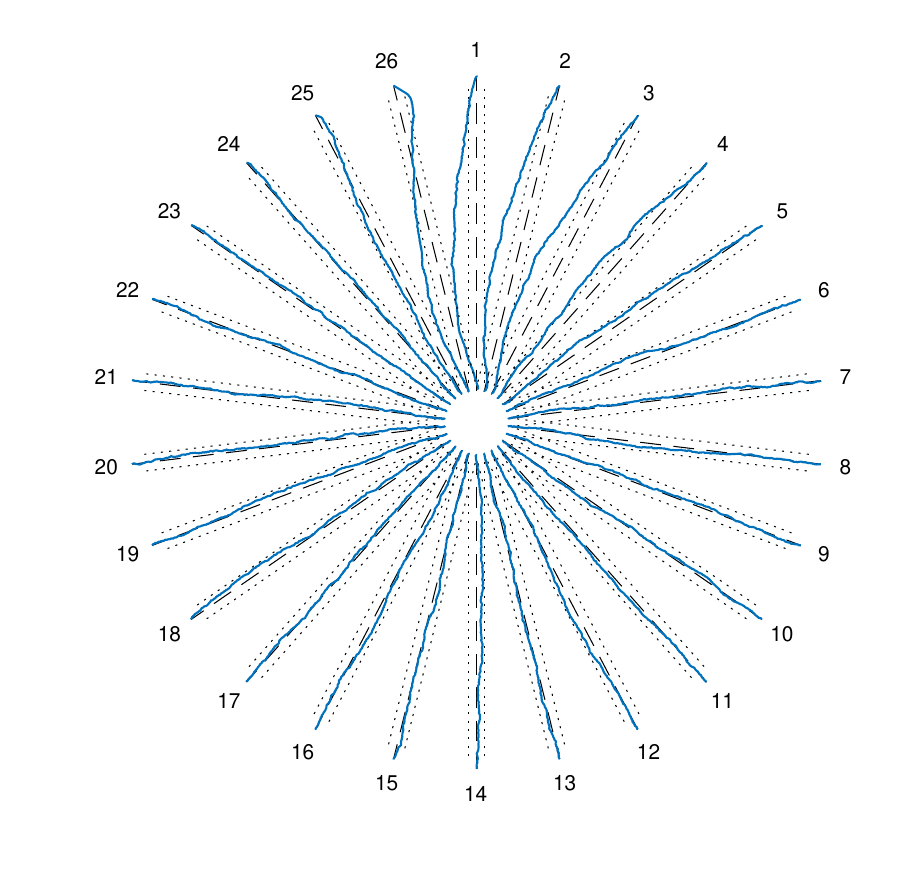} 
} 
\subfigure[$N=28$]{
\includegraphics[width=0.225\textwidth]{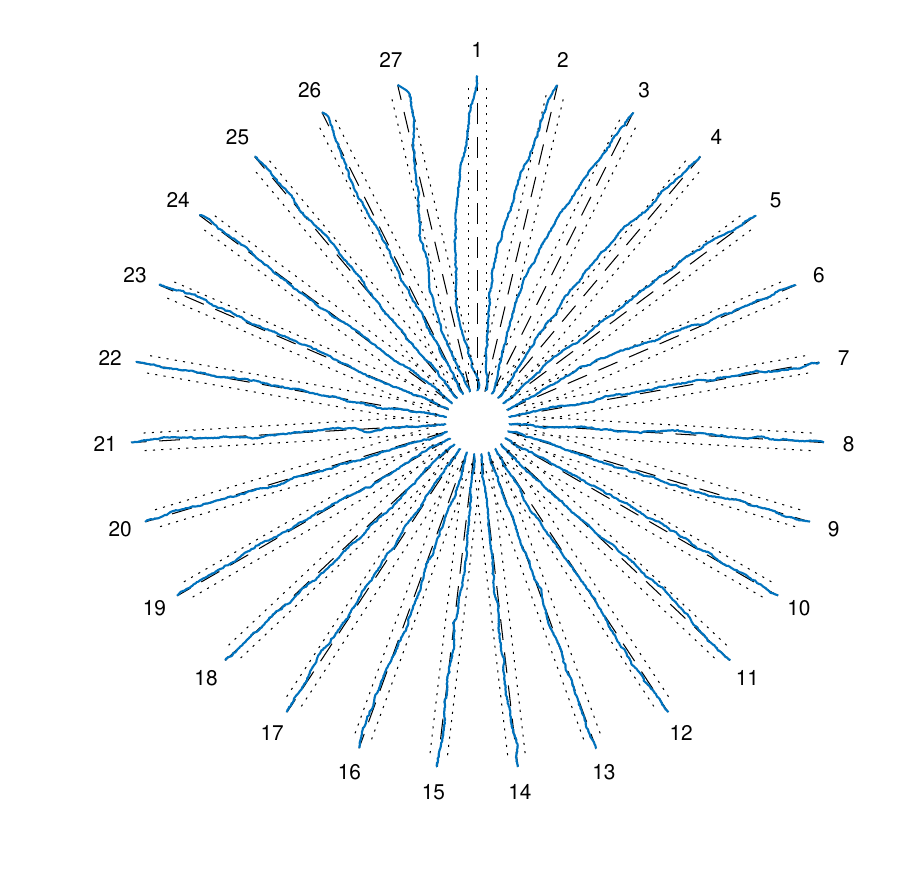} 
} 
\subfigure[$N=29$]{
\includegraphics[width=0.225\textwidth]{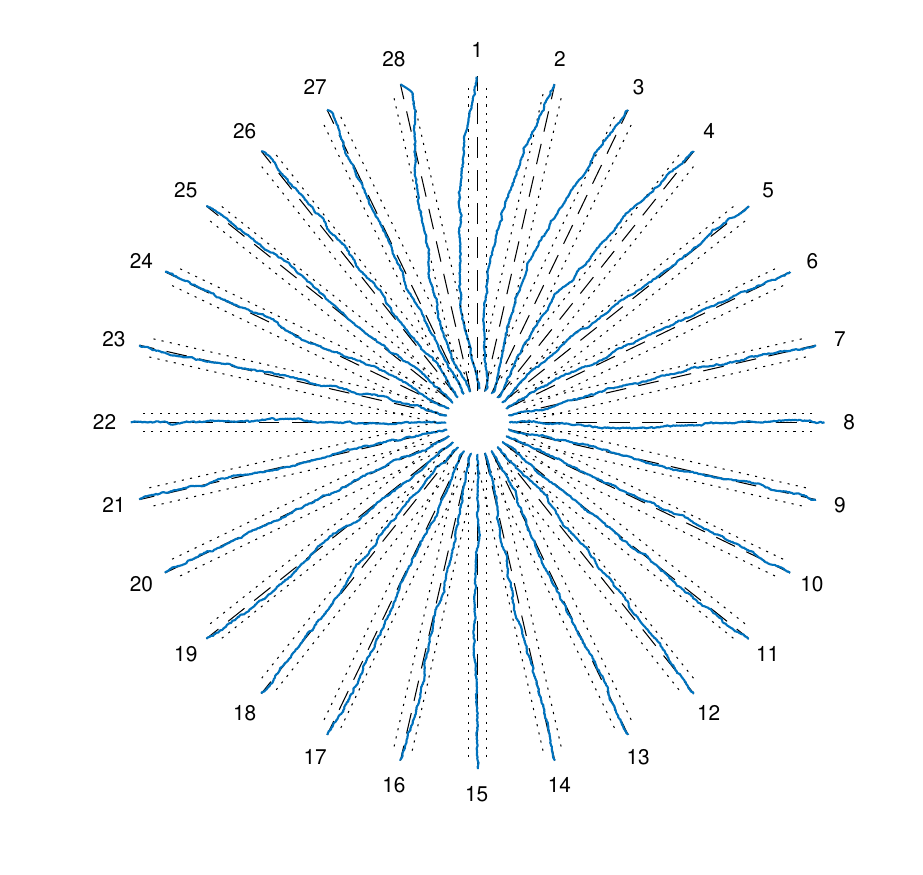} 
} 
\subfigure[$N=30$]{
\includegraphics[width=0.225\textwidth]{StartPlot_N_30.eps} 
} 
\subfigure[$N=31$]{
\includegraphics[width=0.225\textwidth]{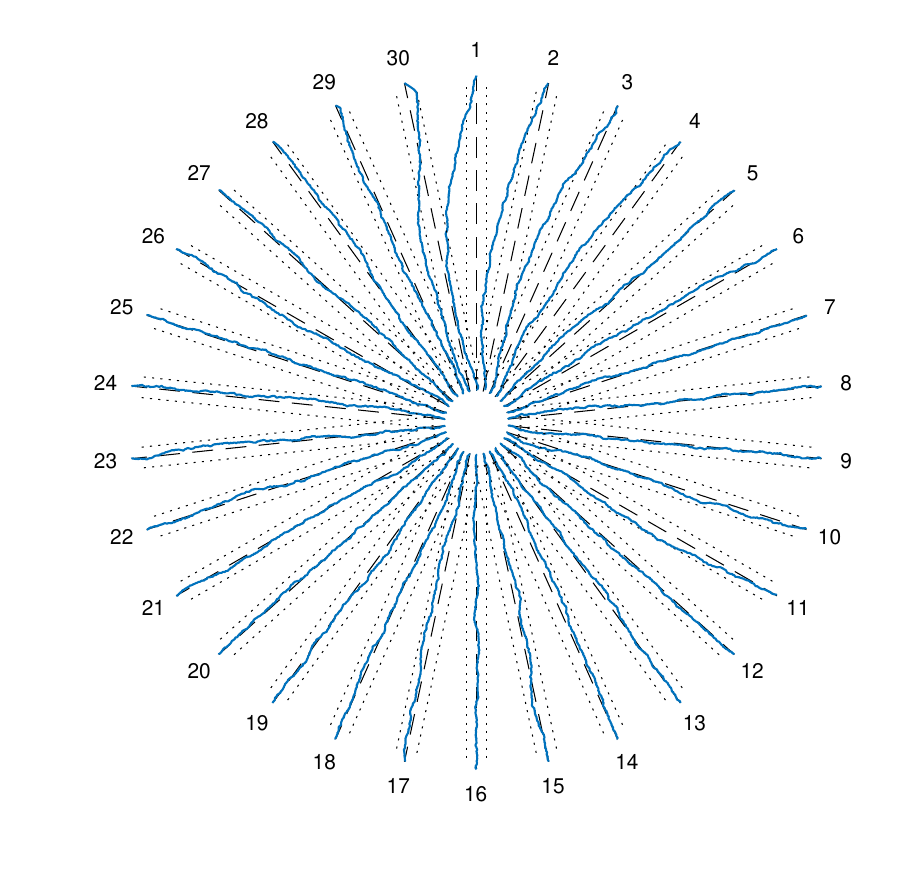} 
} 
\subfigure[$N=32$]{
\includegraphics[width=0.225\textwidth]{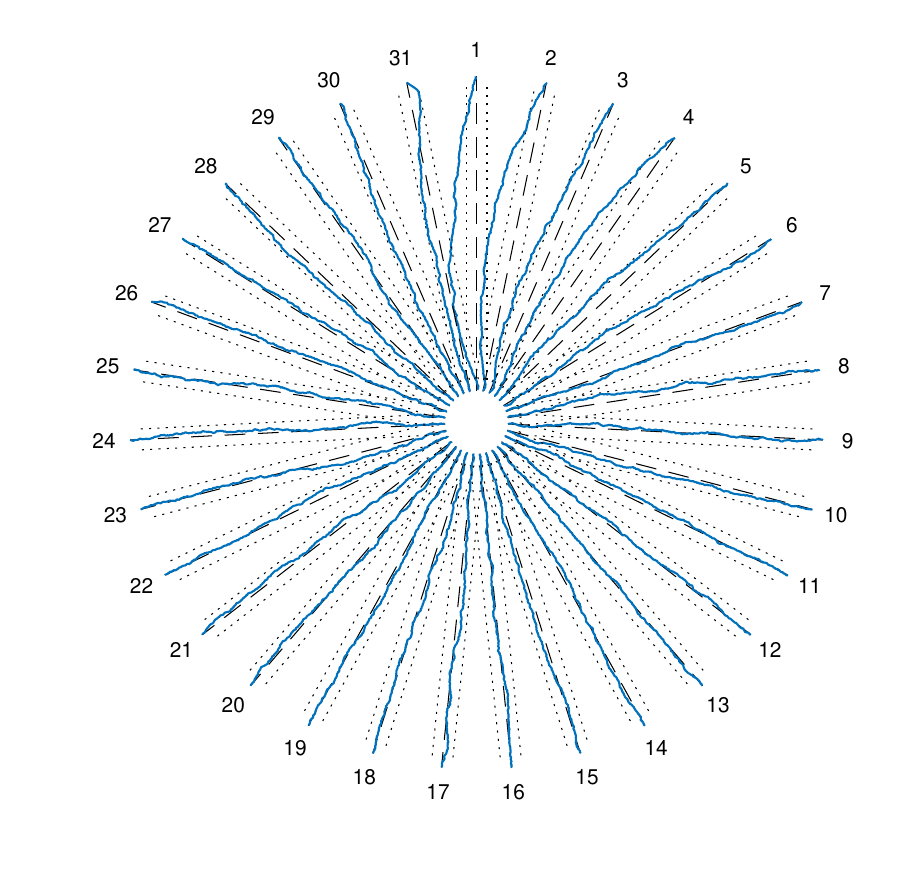} 
} 
\subfigure[$N=33$]{
\includegraphics[width=0.225\textwidth]{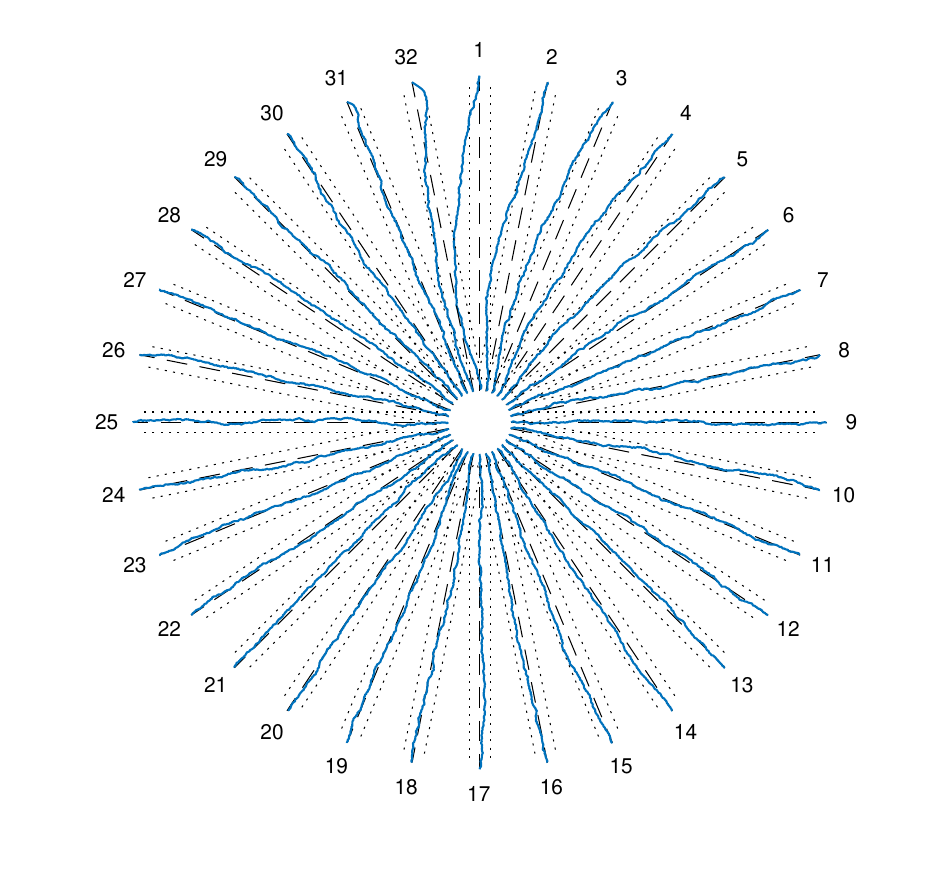} 
} 
\caption{All residual analysis ``dial plots'' of conditionally uniform UHP transformed empirical CDFs compared with true standard uniform CDFs and 99\% error bounds for $N = 19$ through $33$. Out-of-sample test. }
\label{fig:starplotsall2}
\end{figure}

\begin{table}[htbp]
 \centering
  \caption{Mean Inner Wait (and Percent Improvement) for the Lightest Load Routing Policies with and without Hawkes Projections.}
  \begin{scriptsize}
  \hspace*{-0.4in}
  \begin{tabular}{cccccccccccccc} \toprule
    $\mathcal{S}$ & $\kappa$ && LL && UHP && \multicolumn{3}{c}{BHP} && \multicolumn{3}{c}{SysBHP}
    \tabularnewline
        &  &  & & & & & $\delta = 0.5$ & $\delta = 5$ & $\delta = \infty$ && $\delta = 0.5$ & $\delta = 5$ & $\delta = \infty$
    \tabularnewline
  \cmidrule{1-2}
  \cmidrule{4-4}
  \cmidrule{6-6}
  \cmidrule{8-10}
  \cmidrule{12-14}
125 
 & 5 && 3.86 && 3.71 (3.8\%) && 3.68 (4.6\%) & 3.69 (4.3\%) & 3.69 (4.4\%) && 3.70 (4.0\%) & 3.68 (4.6\%) & 3.74 (3.2\%) 
 \tabularnewline
 & 10 && 9.51 && 9.13 (4.0\%) && 9.03 (5.1\%) & 9.09 (4.5\%) & 9.10 (4.4\%) && 9.00 (5.4\%) & 9.05 (4.9\%) & 9.13 (4.0\%) 
 \tabularnewline
 & 15 && 14.26 && 13.74 (3.7\%) && 13.66 (4.2\%) & 13.69 (4.0\%) & 13.72 (3.8\%) && 13.58 (4.8\%) & 13.61 (4.6\%) & 13.71 (3.9\%)  \tabularnewline
 & 20 && 17.45 && 16.85 (3.5\%) && 16.72 (4.2\%) & 16.80 (3.7\%) & 16.94 (2.9\%) && 16.60 (4.9\%) & 16.59 (4.9\%) & 16.72 (4.2\%)   \tabularnewline
    \cmidrule{1-2}
  \cmidrule{4-4}
  \cmidrule{6-6}
  \cmidrule{8-10}
  \cmidrule{12-14}
135
 & 5 && 3.77 && 3.60 (4.5\%) && 3.60 (4.4\%) & 3.58 (5.1\%) & 3.60 (4.5\%) && 3.59 (4.7\%) & 3.61 (4.2\%) & 3.61 (4.2\%) 
 \tabularnewline
 & 10 && 8.87 && 8.44 (4.8\%) && 8.32 (6.2\%) & 8.37 (5.6\%) & 8.40 (5.3\%) && 8.32 (6.2\%) & 8.34 (5.9\%) & 8.42 (5.0\%) 
 \tabularnewline
 & 15 && 12.91 && 12.42 (3.8\%) && 12.27 (4.9\%) & 12.32 (4.6\%) & 12.33 (4.5\%) && 12.22 (5.3\%) & 12.27 (4.9\%) & 12.30 (4.8\%)  \tabularnewline
 & 20 && 15.06 && 14.54 (3.4\%) && 14.30 (5.0\%) & 14.33 (4.8\%) & 14.38 (4.5\%) && 14.28 (5.2\%) & 14.34 (4.8\%) & 14.34 (4.8\%)  \tabularnewline
  \bottomrule
    \end{tabular}
    \hspace*{-0.4in}
  \end{scriptsize}
 \label{tbl:CompAllMIwait}
\end{table}

\begin{table}[htbp]
 \centering
  \caption{Mean Outer Wait (and Percent Improvement) for the Lightest Load Routing Policies with and without Hawkes Projections.}
  \begin{scriptsize}
  \hspace*{-0.7in}
  \begin{tabular}{cccccccccccccc} \toprule
    $\mathcal{S}$ & $\kappa$ && LL && UHP && \multicolumn{3}{c}{BHP} && \multicolumn{3}{c}{SysBHP}
    \tabularnewline
        &  &  & & & & & $\delta = 0.5$ & $\delta = 5$ & $\delta = \infty$ && $\delta = 0.5$ & $\delta = 5$ & $\delta = \infty$
    \tabularnewline
  \cmidrule{1-2}
  \cmidrule{4-4}
  \cmidrule{6-6}
  \cmidrule{8-10}
  \cmidrule{12-14}
125 
 & 5 && 185.25 && 182.73 (1.4\%) && 182.68 (1.4\%) & 182.72 (1.4\%) & 182.47 (1.5\%) && 179.86 (2.9\%) & 179.10 (3.3\%) & 180.25 (2.7\%) 
 \tabularnewline
 & 10 && 72.95 && 70.47 (3.4\%) && 68.92 (5.5\%) & 70.12 (3.9\%) & 70.33 (3.6\%) && 68.99 (5.4\%) & 69.58 (4.6\%) & 69.68 (4.5\%)  
 \tabularnewline
 & 15 && 51.29 && 48.49 (5.5\%) && 48.02 (6.4\%) & 48.39 (5.7\%) & 48.80 (4.9\%) && 47.15 (8.1\%) & 47.85 (6.7\%) & 47.87 (6.7\%)  \tabularnewline
 & 20 && 40.37 && 37.90 (6.1\%) && 36.42 (9.8\%) & 37.72 (6.6\%) & 38.20 (5.4\%) && 36.56 (9.4\%) & 36.83 (8.8\%) & 37.19 (7.9\%)  \tabularnewline
   \cmidrule{1-2}
  \cmidrule{4-4}
  \cmidrule{6-6}
  \cmidrule{8-10}
  \cmidrule{12-14}
135
 & 5 && 152.81 && 150.62 (1.4\%) && 149.82 (2.0\%) & 150.25 (1.7\%) & 150.21 (1.7\%) && 147.51 (3.5\%) & 147.95 (3.2\%) & 147.38 (3.6\%) 
 \tabularnewline
 & 10 && 55.89 && 53.19 (4.8\%) && 52.06 (6.8\%) & 52.38 (6.3\%) & 52.64 (5.8\%) && 52.21 (6.6\%) & 51.88 (7.2\%) & 52.44 (6.2\%) 
 \tabularnewline
 & 15 && 37.52 && 35.33 (5.8\%) && 33.87 (9.7\%) & 34.03 (9.3\%) & 34.39 (8.4\%) && 33.90 (9.7\%) & 34.29 (8.6\%) & 34.26 (8.7\%)  \tabularnewline
 & 20 && 30.50 && 28.47 (6.6\%) && 26.99 (11.5\%) & 27.18 (10.9\%) & 27.15 (11.0\%) && 27.36 (10.3\%) & 27.63 (9.4\%) & 27.26 (10.6\%)   \tabularnewline
  \bottomrule
    \end{tabular}
    \hspace*{-0.7in}
  \end{scriptsize}
 \label{tbl:CompAllMOwait}
\end{table}

\begin{table}[htbp]
 \centering
  \caption{Probability of Outer Wait (and Percent Improvement) for the Lightest Load Routing Policies with and without Hawkes Projections.}
  \begin{scriptsize}
  \hspace*{-0.3in}
  \begin{tabular}{cccccccccccccc} \toprule
    $\mathcal{S}$ & $\kappa$ && LL && UHP && \multicolumn{3}{c}{BHP} && \multicolumn{3}{c}{SysBHP}
    \tabularnewline
        &  &  & & & & & $\delta = 0.5$ & $\delta = 5$ & $\delta = \infty$ && $\delta = 0.5$ & $\delta = 5$ & $\delta = \infty$
    \tabularnewline
  \cmidrule{1-2}
  \cmidrule{4-4}
  \cmidrule{6-6}
  \cmidrule{8-10}
  \cmidrule{12-14}
125 
 & 5 && 0.678 && 0.671 (1.0\%) && 0.670 (1.2\%) & 0.668 (1.4\%) & 0.669 (1.3\%) && 0.665 (1.9\%) & 0.665 (1.9\%) & 0.665 (1.8\%) 
 \tabularnewline
 & 10 && 0.349 && 0.344 (1.7\%) && 0.345 (1.2\%) & 0.344 (1.4\%) & 0.339 (2.9\%) && 0.330 (5.4\%) & 0.337 (3.7\%) & 0.336 (3.8\%)  
 \tabularnewline
 & 15 && 0.225 && 0.222 (1.5\%) && 0.220 (2.2\%) & 0.223 (0.8\%) & 0.222 (1.3\%) && 0.218 (3.1\%) & 0.218 (3.1\%) & 0.216 (3.9\%) 
 \tabularnewline
 & 20 && 0.116 && 0.111 (4.3\%) && 0.110 (5.2\%) & 0.116 (0.4\%) & 0.112 (3.4\%) && 0.107 (8.1\%) & 0.110 (5.3\%) & 0.109 (6.0\%)  \tabularnewline
    \cmidrule{1-2}
  \cmidrule{4-4}
  \cmidrule{6-6}
  \cmidrule{8-10}
  \cmidrule{12-14}
135
 & 5 && 0.640 && 0.635 (0.9\%) && 0.632 (1.3\%) & 0.633 (1.2\%) & 0.633 (1.2\%) && 0.628 (2.0\%) & 0.629 (1.8\%) & 0.630 (1.7\%) 
 \tabularnewline
 & 10 && 0.290 && 0.287 (1.1\%) && 0.285 (1.9\%) & 0.284 (2.1\%) & 0.287 (1.1\%) && 0.286 (1.4\%) & 0.284 (2.2\%) & 0.284 (2.0\%) 
 \tabularnewline
 & 15 && 0.154 && 0.151 (1.6\%) && 0.150 (2.5\%) & 0.149 (3.3\%) & 0.152 (1.0\%) && 0.150 (2.8\%) & 0.145 (5.5\%) & 0.145 (5.9\%) 
 \tabularnewline
 & 20 && 0.074 && 0.071 (4.3\%) && 0.068 (8.4\%) & 0.070 (6.1\%) & 0.070 (5.8\%) && 0.068 (8.2\%) & 0.069 (6.3\%) & 0.068 (8.3\%) 
 \tabularnewline
  \bottomrule
    \end{tabular}
    \hspace*{-0.45in}
  \end{scriptsize}
 \label{tbl:CompAllPOwait}
\end{table}

\begin{table}[htbp]
 \centering
  \caption{Standard Deviation (and Percent Improvement) of Inner Wait for the Lightest Load Routing Policies with and without Hawkes Projections.}
  \begin{scriptsize}
  \hspace*{-0.3in}
  \begin{tabular}{cccccccccccccc} \toprule
    $\mathcal{S}$ & $\kappa$ && LL && UHP && \multicolumn{3}{c}{BHP} && \multicolumn{3}{c}{SysBHP}
    \tabularnewline
        &  &  & & & & & $\delta = 0.5$ & $\delta = 5$ & $\delta = \infty$ && $\delta = 0.5$ & $\delta = 5$ & $\delta = \infty$
    \tabularnewline
  \cmidrule{1-2}
  \cmidrule{4-4}
  \cmidrule{6-6}
  \cmidrule{8-10}
  \cmidrule{12-14}
125 
 & 5 && 17.98 && 17.71 (1.5\%) && 17.55 (2.4\%) & 17.70 (1.5\%) & 17.69 (1.6\%) && 17.66 (1.7\%) & 17.64 (1.9\%) & 17.68 (1.6\%) 
 \tabularnewline
 & 10 && 27.49 && 26.69 (2.9\%) && 26.53 (3.5\%) & 26.40 (4.0\%) & 26.94 (2.0\%) && 26.45 (3.8\%) & 26.53 (3.5\%) & 26.67 (3.0\%) 
 \tabularnewline
 & 15 && 33.20 && 32.46 (2.2\%) && 32.32 (2.6\%) & 32.45 (2.3\%) & 32.98 (0.7\%) && 32.60 (1.8\%) & 32.35 (2.6\%) & 32.71 (1.5\%)  \tabularnewline
 & 20 && 37.62 && 36.63 (2.6\%) && 36.50 (3.0\%) & 36.57 (2.8\%) & 37.30 (0.9\%) && 36.94 (1.8\%) & 36.68 (2.5\%) & 36.55 (2.9\%)  \tabularnewline
    \cmidrule{1-2}
  \cmidrule{4-4}
  \cmidrule{6-6}
  \cmidrule{8-10}
  \cmidrule{12-14}
135
 & 5 && 17.82 && 17.28 (3.0\%) && 17.52 (1.7\%) & 17.49 (1.9\%) & 17.43 (2.2\%) && 17.44 (2.1\%) & 17.46 (2.0\%) & 17.50 (1.8\%) 
 \tabularnewline
 & 10 && 26.39 && 25.49 (3.4\%) && 25.30 (4.1\%) & 25.37 (3.9\%) & 25.71 (2.6\%) && 25.54 (3.3\%) & 25.37 (3.9\%) & 25.67 (2.7\%) 
 \tabularnewline
 & 15 && 31.93 && 31.04 (2.8\%) && 30.83 (3.4\%) & 30.70 (3.8\%) & 30.98 (3.0\%) && 30.82 (3.5\%) & 31.10 (2.6\%) & 30.98 (3.0\%)  \tabularnewline
 & 20 && 34.95 && 34.16 (2.2\%) && 33.71 (3.5\%) & 33.60 (3.9\%) & 33.91 (3.0\%) && 34.03 (2.6\%) & 34.15 (2.3\%) & 34.06 (2.6\%)  \tabularnewline
  \bottomrule
    \end{tabular}
    \hspace*{-0.4in}
  \end{scriptsize}
 \label{tbl:CompAllSDIwait}
\end{table}

\begin{table}[htbp]
 \centering
  \caption{Standard Deviation (and Percent Improvement) of Outer Wait for the Lightest Load Routing Policies with and without Hawkes Projections.}
  \begin{scriptsize}
  \hspace*{-0.7in}
  \begin{tabular}{cccccccccccccc} \toprule
    $\mathcal{S}$ & $\kappa$ && LL && UHP && \multicolumn{3}{c}{BHP} && \multicolumn{3}{c}{SysBHP}
    \tabularnewline
        &  &  & & & & & $\delta = 0.5$ & $\delta = 5$ & $\delta = \infty$ && $\delta = 0.5$ & $\delta = 5$ & $\delta = \infty$
    \tabularnewline
  \cmidrule{1-2}
  \cmidrule{4-4}
  \cmidrule{6-6}
  \cmidrule{8-10}
  \cmidrule{12-14}
125 
 & 5 && 140.35 && 138.24 (1.5\%) && 138.97 (1.0\%) & 138.45 (1.4\%) & 138.17 (1.6\%) && 135.83 (3.2\%) & 135.27 (3.6\%) & 136.35 (2.9\%) 
 \tabularnewline
 & 10 && 35.58 && 34.40 (3.3\%) && 34.22 (3.8\%) & 34.75 (2.3\%) & 34.48 (3.1\%) && 32.99 (7.3\%) & 33.76 (5.1\%) & 33.72 (5.2\%) 
 \tabularnewline
 & 15 && 17.84 && 17.15 (3.9\%) && 17.10 (4.1\%) & 17.44 (2.3\%) & 17.40 (2.5\%) && 16.90 (5.3\%) & 16.81 (5.8\%) & 16.77 (6.0\%)  \tabularnewline
 & 20 && 7.84 && 7.20 (8.1\%) && 7.02 (10.5\%) & 7.59 (3.2\%) & 7.50 (4.3\%) && 6.96 (11.2\%) & 7.10 (9.5\%) & 7.08 (9.7\%)  \tabularnewline
    \cmidrule{1-2}
  \cmidrule{4-4}
  \cmidrule{6-6}
  \cmidrule{8-10}
  \cmidrule{12-14}
135
 & 5 && 113.93 && 112.64 (1.1\%) && 111.97 (1.7\%) & 112.17 (1.5\%) & 111.98 (1.7\%) && 109.97 (3.5\%) & 110.17 (3.3\%) & 110.17 (3.3\%) 
 \tabularnewline
 & 10 && 23.87 && 23.09 (3.3\%) && 22.66 (5.1\%) & 22.82 (4.4\%) & 23.11 (3.2\%) && 22.89 (4.1\%) & 22.61 (5.3\%) & 22.82 (4.4\%) 
  \tabularnewline
 & 15 && 9.80 && 9.41 (4.1\%) && 9.09 (7.3\%) & 9.04 (7.8\%) & 9.33 (4.8\%) && 9.02 (8.0\%) & 9.00 (8.2\%) & 8.88 (9.4\%)  \tabularnewline
 & 20 && 4.18 && 3.86 (7.7\%) && 3.57 (14.6\%) & 3.67 (12.3\%) & 3.69 (11.6\%) && 3.63 (13.1\%) & 3.76 (10.0\%) & 3.65 (12.8\%)    \tabularnewline
  \bottomrule
    \end{tabular}
    \hspace*{-0.7in}
  \end{scriptsize}
 \label{tbl:CompAllSDOwait}
\end{table}

\clearpage\printendnotes

\end{APPENDIX}
\end{document}